\documentclass[fleqn,usenatbib]{mnras}

\usepackage{newtxtext,newtxmath}
\usepackage{orcidlink}

\usepackage[T1]{fontenc}

\DeclareRobustCommand{\VAN}[3]{#2}
\let\VANthebibliography\thebibliography
\def\thebibliography{\DeclareRobustCommand{\VAN}[3]{##3}\VANthebibliography}

\usepackage{graphicx}	
\usepackage{amsmath}	
\usepackage{pdflscape}

\title[SQUALO II: Molecular outflows]{SQUALO II: Probing Accretion Through CO Molecular Outflows in Massive Star-forming Clumps}

\author[S. Mercimek et al.]{
S. Mercimek$^{1,2}$\thanks{E-mail: seyma.mercimek@manchester.ac.uk}\orcidlink{0000-0002-0742-7934},
G. A. Fuller$^{1,2,3}$\orcidlink{0000-0001-8509-1818},
A. Traficante$^{4}$\orcidlink{0000-0003-1665-6402},
T. Pillai$^{5}$\orcidlink{0000-0003-2133-4862},
A. Nucara$^{4,6}$\orcidlink{0009-0005-9192-5491},
W. -A. Chen$^{1,2}$\orcidlink{0000-0002-8336-2837},
D. Walker$^{1}$\orcidlink{0000-0001-7330-8856},
\newauthor
L. Testi$^{7}$\orcidlink{0000-0003-1859-3070},
M. Benedettini$^{4}$\orcidlink{0000-0002-3597-7263}
\\
$^{1}$ Jodrell Bank Centre for Astrophysics, University of Manchester, Manchester M13 9PL, UK \\
$^{2}$Cluster of Excellence "Our Dynamic Universe" (DYNAVERSE),  Institute of Physics I, University of Cologne, Zülpicher Str. 77, 50937 Cologne, Germany\\
$^{3}$ Physikalisches Institut der Universität zu Köln, Zülpicher Str. 77, 50937 Köln, Germany \\
$^{4}$ INAF – Istituto di Astrofisica e Planetologia Spaziali, Via Fosso del
Cavaliere 100, 00133 Roma, Italy\\
$^{5}$ Haystack Observatory, Massachusetts Institute of Technology, 99
Millstone Rd., Westford, MA 01886, USA\\
$^{6}$ Dipartimento di Fisica, Università di Roma Tor Vergata, Via della
Ricerca Scientifica 1, 00133 Roma, Italy\\
$^{7}$Alma Mater Studiorum – Università di Bologna, Dipartimento di
Fisica e Astronomia “Augusto Righi”, Via Gobetti 93/2, 40129,
Bologna, Italy\\
}
\date{Accepted 2026 November 05. Received 2026 November 05; in original form 2026 August 11}
\pubyear{\the\year{}}

\begin{document}
\label{firstpage}
\pagerange{\pageref{firstpage}--\pageref{lastpage}}
\maketitle

\begin{abstract}
Molecular outflows probe protostellar accretion and mass transfer from clump scales to forming stars. We present an ALMA study of CO ($J=2$--1) outflows toward 13 massive clumps from the SQUALO sample, spanning a broad evolutionary range and exhibiting clump-scale infall. We identify 51 high-velocity CO lobes associated with 31 outflow systems and derive their physical and dynamical properties assuming local thermodynamic equilibrium. Outflow-inferred accretion rates are typically $10^{-5}$--$10^{-3}\ M_\odot,{\rm yr}^{-1}$, consistent with active massive star formation. The maximum outflow velocity shows little dependence on candidate driving-core mass, indicating that it is not a reliable tracer of protostellar mass in clustered environments. Outflow momentum flux correlates positively with clump surface density, whereas accretion rate shows a tentative, statistically nonsignificant decrease with increasing virial parameter. Total outflow-inferred accretion rate shows a tentative positive relation with clump-scale infall rate but remains systematically lower, suggesting that the detected outflows represent only part of the large-scale mass flux. The clump $L/M$ ratio moderately anticorrelates with projected outflow length, although projection, observational limitations, and the embedded source population prevent a unique evolutionary interpretation. Together with the continuum-derived fragmentation properties, our results show that compact cores embedded within infalling clumps drive the outflows. The tentative relation between clump-scale infall and protostellar accretion suggests a possible connection across these scales, although our observations do not directly trace mass transfer from filaments to individual protostars.

\end{abstract}




\section{Introduction}

High-mass stars ($M \gtrsim 8 M_\odot$) play a fundamental role in shaping the interstellar medium and driving the evolution of galaxies through their intense radiative output, ionizing feedback, stellar winds, and eventual supernova explosions \citep{Zhang2005}. Despite their importance, the physical mechanisms governing their formation remain a subject of active debate. The earliest stages of massive star formation occur deeply embedded within dense molecular clouds, evolve rapidly, and are typically located at large distances, making direct observations of protostellar accretion particularly challenging \citep[e.g.,][]{Zinnecker2007,Tan2014, Motte2018, Beuther2025}. As a result, indirect tracers have become essential tools for investigating how massive protostars accumulate mass during their embedded phases.

Among these tracers, molecular outflows provide one of the most powerful probes of ongoing accretion. In magneto-centrifugal launching models, outflows are driven by accretion through circumstellar discs, establishing a direct physical connection between mass ejection and protostellar growth \citep{Blandford1982}. Observations have shown that massive protostars drive powerful molecular outflows whose masses, momenta, and mechanical luminosities exceed those of their low-mass counterparts by orders of magnitude \citep{Beuther2002a,Beuther2002b}. Because outflow properties are closely linked to the underlying accretion process, quantities such as outflow mass, momentum, outflow force, and mechanical luminosity provide indirect measures of the rate at which material is being accreted onto forming stars \citep{Arce2007,Duarte-Cabral2013,Maud2015}. Consequently, molecular outflows offer a unique window into the mass assembly process during the deeply embedded stages of star formation.

Large surveys of massive star-forming regions have demonstrated that outflow activity evolves systematically with the physical and evolutionary properties of their parent clumps. Studies of Galactic plane samples have revealed correlations between outflow energetics and tracers such as bolometric luminosity, dust temperature, and the clump luminosity-to-mass ratio ($L/M$), a widely adopted indicator of clump evolution \citep{Molinari2008,Molinari2016}. These correlations suggest that molecular outflows trace the evolution of protostellar accretion across different stages of massive star formation \citep{Li2018,Yang2018,Yang2022}. More recently, \citet{Towner2024} found that protocluster-integrated SiO outflow properties correlate most strongly with the total mass contained in compact cores, rather than with the bulk properties of the parent clump. They found no significant dependence on the clump luminosity-to-mass ratio, suggesting that protocluster outflow activity does not follow a simple evolutionary sequence described by L/M. Instead, their results indicate that the collective outflow activity is more closely connected to the assembled core population. Although L/M generally increases during clump evolution, it is not a unique evolutionary tracer because it also depends on external heating and dust temperature, instantaneous star-formation efficiency, gas dispersal, protostellar accretion, and the sampling of the stellar initial mass function, as well as on systematic uncertainties in the derived clump masses \citet{Ma2013}.

At the same time, a growing body of observational evidence suggests that massive star formation cannot be understood solely in terms of isolated collapsing cores. Many massive clumps exhibit signatures of large-scale gravitational collapse, including widespread infall motions, global velocity gradients, and filamentary structures funnelling material toward their centres \citep[e.g.,][]{Schneider2010,Csengeri2014,Peretto2013,Sanhueza2019}. Infall rates measured on parsec scales frequently reach values of $\sim10^{-4}$--$10^{-3} M_\odot {\rm yr}^{-1}$ \citep[e.g.,][]{Wyrowski2016}, comparable to those required to sustain the growth of massive protostars. These observations support a hierarchical scenario in which material is continuously transferred from filaments to clumps and from clumps to dense cores \citep[e.g.,][]{Ballesteros-Paredes2011,Wang2020, Traficante2020}. In this framework, massive stars form within a dynamically evolving environment that remains connected to a larger-scale mass reservoir throughout much of the star-formation process \citep{Motte2018,Avison2021}.

If continued clump-scale infall contributes to protostellar growth, the large-scale mass supply may regulate the current accretion activity of the embedded protostars. Because the outflow momentum flux provides an indirect measure of protostellar accretion, this connection can be tested by comparing the cumulative outflow-inferred accretion rate within each clump with its independently measured clump-scale infall rate. Relationships between outflow properties and other clump-scale quantities, such as surface density and virial parameter, provide complementary tests of whether the parent-clump environment influences protostellar accretion. Although recent surveys have identified relationships between outflow properties and clump-scale observables \citep{Yang2018,Yang2022,Towner2024}, direct tests of this connection remain limited because most samples are selected based on general indicators of star formation rather than confirmed signatures of ongoing clump-scale accretion.

The Star formation in QUiescent And Luminous Objects (SQUALO) project was specifically designed to investigate the connection between clump-scale accretion and core-scale star formation. The first paper of the series, SQUALO I \citep{Traficante2023}, presented ALMA 1.3 mm continuum observations of a sample of massive clumps spanning more than three orders of magnitude in $L/M$, all exhibiting clear signatures of parsec-scale infall. The survey identified and characterised the embedded core populations, providing measurements of their masses, sizes, densities, and fragmentation properties. These continuum results were interpreted together with the physical and dynamical properties of the parent clumps derived in previous studies, including their gravitational stability and infall characteristics. Collectively, these analyses revealed centrally concentrated and gravitationally bound clumps hosting rich populations of embedded cores, consistent with active mass assembly, and established the core-scale framework required to investigate how large-scale accretion is connected to ongoing star formation.

In this paper, we extend the SQUALO study by presenting the first systematic ALMA CO ($J=2$--1) survey of molecular outflows toward the same sample of clumps. The 13 targets were selected to exhibit clear signatures of clump-scale infall, with inferred clump accretion rates of approximately $10^{-4}$--$10^{-2}\,M_\odot\,{\rm yr}^{-1}$, while spanning a broad evolutionary sequence from deeply embedded 70-$\mu$m-quiet clumps to more evolved protostellar and H\,{\sc ii} region stages. We derive the physical and dynamical properties of the molecular outflows and investigate how they relate to the physical, dynamical, and evolutionary properties of the parent clumps, together with the embedded core populations identified in SQUALO I and the higher-resolution ALMAGAL continuum observations. Our primary goal is to determine whether molecular outflows retain signatures of the large-scale accretion flows identified in SQUALO I, thereby providing new observational constraints on the role of clump-fed accretion in high-mass star formation.

The paper is organised as follows. Section \ref{2} describes the observations and data reduction. Section \ref{3} presents the identification of the molecular outflows and the derivation of their physical and dynamical properties. Section \ref{4} discusses the outflow properties in the context of the embedded cores, the parent clumps, and the wider filamentary environment, and compares the SQUALO sample with previous observational studies. Finally, Section \ref{5} summarises the main conclusions.

\begin{table*}
\centering
     \begin{tabular}{lccccccccc}
        \hline
Clump   & d$^{a}$ & L$\rm_{cl}$$^{a}$ & M$\rm_{cl}$$^{a}$ & L$\rm_{cl}$/M$\rm_{cl}$$^{b}$&V$\rm_{LSR}$$^{c}$ &  $\dot{M}_{cl}$$^{d}$   & rms$^{e}$   & Beam size$^{e}$  & Beam PA$^{e}$  \\ 
 & (kpc) &(L$_{\odot}$)  &(M$_{\odot}$)  &(L$_{\odot}$/M$_{\odot}$)  & (km s$^{-1}$)  & (10$^{-3}$M$_{\odot}$yr$^{-1}$)  
  & (mJy beam$\rm ^{-1}$) & ($\arcsec \times \arcsec$) & ($\rm ^{\circ})$   \\
\hline
HIGALBM31.9462+0.0759 &   5.5 &133 &1392 &0.1 &+94.7 &8.2 &9.8 &1.7$\times$1.3 & 87.3\\
HIGALBM24.0116+0.4897$^{1}$  &5.2 &158 & 1334 &0.1 & +94.8&4.0 &7.8 & 1.6$\times$1.2 & 87.9 \\
HIGALBM28.1957--0.0724 & 5.4 &594 &1598 & 0.4 &+98.8 &0.7 & 9.8 & 1.7$\times$1.3 & 87.3\\
HIGALBM338.9260+0.6340$^{2}$ &  4.2 &10280 & 2689 &3.8 &--65.1 &62.1 &8.6 &1.2$\times$1.1 & 75.2\\
HIGALBM332.6045--0.1674 &  3.1 &1405 & 177 &7.9 & --48.3 & 1.6 & 5.9  & 1.3$\times$1.1 & --68.6\\
HIGALBM341.2149--0.2359$^{3}$ &  3.4 &7210 & 700 &10.3 & --45.7 &2.3& 5.3 & 1.2$\times$1.1 & 81.0\\
HIGALBM327.3918+0.1996$^{4}$ &5.2 &14146 &1121  &12.6 &--90.4 & 4.7& 5.9&1.3$\times$1.2 & --55.0 \\
HIGALBM343.7560--0.1629  & 2.0 &6778  & 461&14.7 &--26.2 &3.8 & 6.0&1.2$\times$1.1 &87.2 \\
HIGALBM344.1032--0.6609   & 2.0 & 3586&230  &15.6 & --27.9&8.9 &6.1 &1.2$\times$1.1 &87.3 \\
HIGALBM343.5212--0.5172$^{5}$    &  3.0 &7039 & 347 & 20.3& --34.6& 3.5& 6.9&1.2$\times$1.1 &86.8 \\
HIGALBM331.1314--0.2438$^{6}$     &5.0 &54845 & 2467 &22.2 &--89.4 &11.8 &7.8 &1.3$\times$1.1 &--64.4 \\
HIGALBM327.4022+0.4449$^{7}$  &  4.6 &73841 & 2157 & 34.2&--80.8 &27.3 &5.0 &1.3$\times$1.2 &--53.5 \\
HIGALBM344.2210--0.5932      & 2.0 &21645 & 202 &107.0 &--27.0 &1.7 &7.1 &1.2$\times$1.1 &87.2 \\
     
        \hline
      \end{tabular}
       \caption{Properties of the 13 Hi-GAL clumps targeted in the SQUALO survey. $^{1-7}$ identify the seven SQUALO clumps that overlap with the ALMAGAL survey; the corresponding ALMAGAL source identifiers are 1=107025, 2=839678, 3=850946, 4=G327.393+00.199, 5=G343.5213-00.5171, 6=799475, 7=783350). $^{a}$ Properties of the clump are taken from \citet{Elia2021}. $^{b}$ Luminosity-to-mass ratios are taken from \citet{Traficante2023}, $^{c}$ Systemic velocities are adopted from the GRS \citep{Jackson2006}, MALT90 \citep{Jackson2013}, the Mopra Southern Galactic Plane CO Survey \citep{Burton2013}, and THRUMMS \citep{Barnes2015}, as compiled by \citet{Traficante2023}. $^{d}$ Clump accretion rates are adopted from \citet{Traficante2017} and \citet{Traficante2018a}, as compiled by \citet{Traficante2023}. $^{e}$ Observational properties of the final primary-beam-corrected SQUALO CO (2--1) cubes used in this work. Here, $\sigma_{ch}$ is the rms noise per 2.54 km s$^{-1}$ channel, measured in line-free channels of the primary-beam-corrected cubes. The beam size and position angle refer to the synthesized beam of the same cubes. The synthesized beam is characterized by its major and minor axes and position angle (PA).}
\label{sources}
    \end{table*}

\section{OBSERVATIONS}
\label{2}
Continuum and molecular-line observations, including the spectral setup, central frequencies, bandwidths, and channel widths, are described in detail by \citet{Traficante2023}. In this work, we focus on the CO ($J=2$--1) transition at a rest frequency of 230.538 GHz.

The CO ($J=2$--1) line was observed in a dedicated spectral window with a native channel width of $\approx$1.27\,km\,s$^{-1}$. For each target, the calibrated 12\,m and 7\,m visibilities were regridded to the LSRK frame with the CASA task \texttt{mstransform}, trimmed to the frequency range common to all execution blocks, and concatenated into a single measurement set. The continuum was subtracted in the $uv$-plane with \texttt{uvcontsub}, fitting a zeroth-order polynomial to line-free channels for each source, which were identified from the visibility spectra.

The 12\,m and 7\,m data were jointly imaged with \texttt{tclean} (CASA 6.2) using the \texttt{mosaic} gridder to account for the different primary beams of the two arrays. We adopted Briggs weighting (robust = 0.5), a pixel size of 0.2\arcsec, and multiscale
deconvolution with scales of 0, 6 and 18 pixels. The cubes have a channel width of 2.54\,km\,s$^{-1}$, cover $\pm$120\,km\,s$^{-1}$ around the systemic velocity of each clump, and were restored with a common beam across all channels. The clean mask was created
automatically in each channel with the \texttt{auto-multithresh} algorithm \citep{Kepley2020}, using \texttt{sidelobethreshold} = 2.0, \texttt{noisethreshold} = 4.25, \texttt{lownoisethreshold} = 2.0 and \texttt{minbeamfrac} = 0.15, with only positive emission masked. Because multiscale deconvolution of bright, extended CO emission is prone to divergence, the number of minor-cycle iterations between major cycles was limited to 10 (\texttt{cycleniter} = 10). This forces frequent major cycles, which we found to be necessary to avoid divergence. To recover the more extended emission, the primary-beam-corrected cubes were combined with the ALMA Total Power (TP) observations using the CASA task \texttt{feather}. The TP cubes, which were produced by the ALMA single-dish pipeline, were first regridded on to the spatial and spectral grid of the interferometric cubes, and no additional scaling was applied to the TP data when feathering (\texttt{sdfactor} = 1).

In this work we analyse the final array-combined, continuum-subtracted, primary-beam-corrected CO ($J=2$--1) cubes. The angular resolution, spectral resolution, rms noise, and other observational properties of the CO ($J=2$--1) data used in this work are summarized in Table~\ref{sources}. Coordinates of the sources are available in the first SQUALO study \citep{Traficante2023}.

When available, we used the 1.3 mm continuum images from the ALMAGAL survey (ALMA Cycle 7 Large Program; 2019.1.00195.L), with an angular resolution of $\sim0.4''$, to aid the identification of the most likely outflow-driving cores. The higher angular resolution of the ALMAGAL data allowed multiple compact continuum sources to be resolved in several regions where the SQUALO continuum observations identified only a single compact structure. In cases where the SQUALO cores were not further resolved by the ALMAGAL observations, we retained the original SQUALO core identification to ensure a consistent analysis across the full sample. The ALMAGAL data (see \citealt{Molinari2025} for details) were used solely for core identification and outflow association. All outflow properties and kinematic measurements presented in this paper were derived exclusively from the SQUALO CO ($J=2$--1) observations. Seven of the 13 SQUALO clumps are also included in the ALMAGAL survey and therefore have complementary ALMAGAL continuum observations. These overlapping targets are identified by superscript numbers in Table 1, where their corresponding ALMAGAL source identifiers are provided.

The availability of complementary ALMAGAL data for each SQUALO target is indicated in Table \ref{sources} with their ALMAGAL name in parenthesis.

To investigate the large-scale environment surrounding the targets, we used Spitzer/IRAC three-colour images from the GLIMPSE survey \citep{Benjamin2003}. The RGB images were constructed using the 8.0, 5.8, and 4.5 $\mu$m bands assigned to the red, green, and blue channels, respectively. These images were used to examine the relationship between the outflows and the surrounding filamentary structures and infrared environment. The IRAC instrument is described by \citet{Fazio2004}.

\begin{figure*}
\centering
\includegraphics[width=16cm]{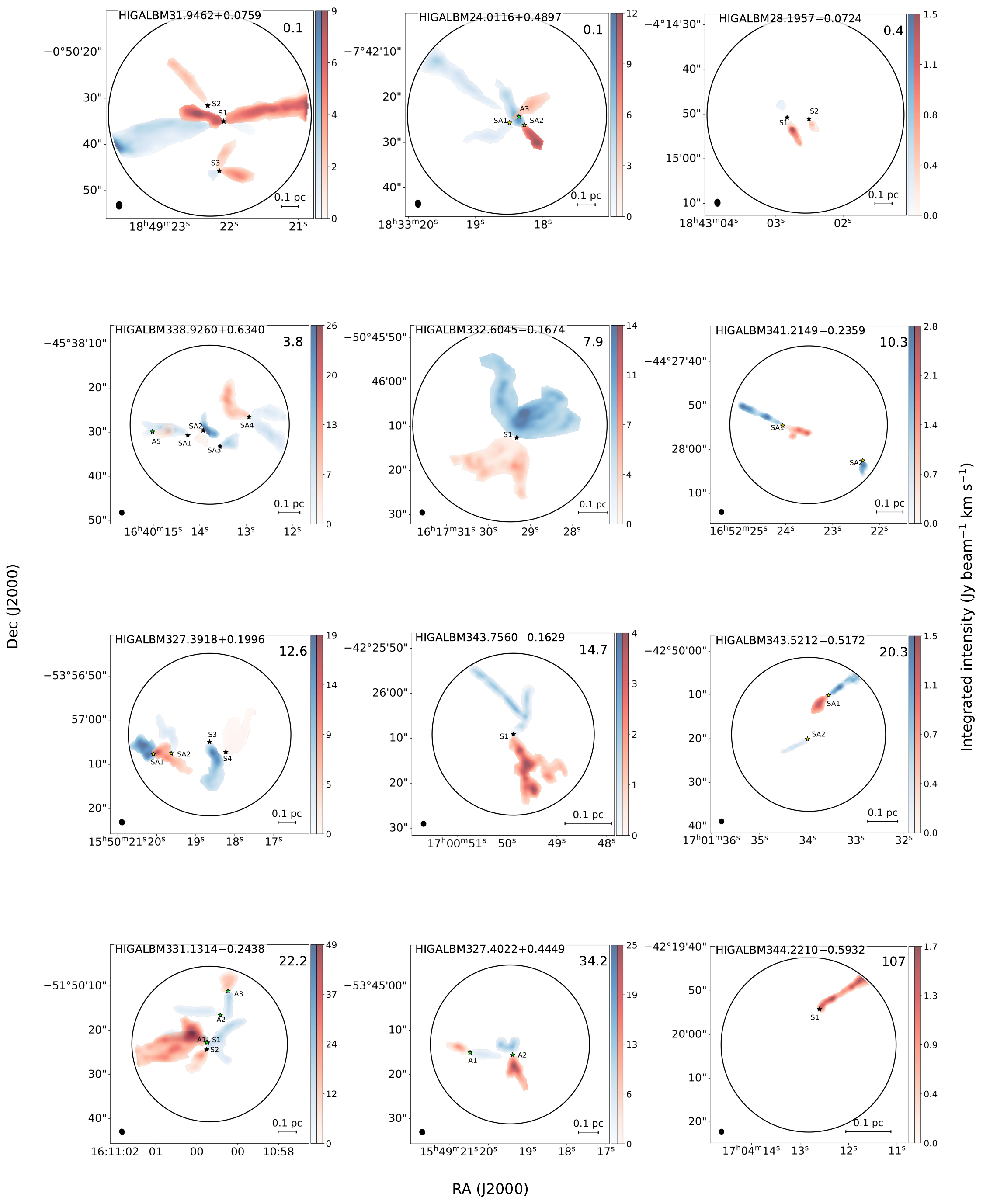}
\caption{Velocity-integrated CO (2–1) outflow maps toward the SQUALO sources. Blue- and red-shifted emission are shown in blue and red, respectively. The integrated intensity maps were constructed by summing emission above a 5$\sigma$ threshold within polygon masks enclosing the outflow lobes, thereby suppressing noise and highlighting only significant outflow emission. Stars mark the positions of compact fragments identified from high-resolution continuum observations. Black fragments labeled with "S" are identified in the SQUALO data, green fragments labeled with "A" are identified in the ALMAGAL data, and yellow fragments labeled with "SA" are detected in both datasets (see Table~\ref{flows}). The source name is shown at the top of each panel, while the number in the upper-right corner indicates the clump luminosity-to-mass ratio. The black circle outlines the region analyzed for each source. The synthesized beam is shown in the lower-left corner, and a 0.1 pc scale bar is displayed in the lower-right corner of each panel. The color bars indicate the integrated CO (2–1) intensity in units of Jy beam$^{-1}$ km s$^{-1}$. Integrated velocity channels for outflow lobes are reported in Table \ref{flows}.
}
\label{CO_MAPS}
\end{figure*}

\section{Results}
\label{3}
\subsection{Outflow identification}
\label{identification}
Molecular outflow activity was identified through high-velocity CO emission offset from the systemic velocity of each source. We inspected the channel maps with \(5\sigma_{\rm ch}\) contours and, starting from \(V_{\rm LSR}\), examined successive channels toward the blue- and red-shifted velocities. For each outflow lobe, we set the inner boundary at the first channel where the emission was spatially distinguishable from the extended ambient component and plausibly associated with a compact continuum core identified in the SQUALO data or, where available, the ALMAGAL data. Where emission could be distinguished from ambient or unrelated components, the outer boundary was set at the most extreme channel with spatially coherent emission above \(5\sigma_{\rm ch}\). 

For ambiguous or blended features, we assessed the spatial and kinematic coherence of the emission and retained it only when it could be plausibly associated with an outflow; otherwise, it was excluded from the outflow calculations. We selected the velocity intervals individually for each outflow lobe, accounting for differences in ambient linewidth and emission complexity.
The adopted interval for each lobe is listed in Table 2. Given the channel spacing of \(2.54\ \mathrm{km\,s^{-1}}\), the uncertainty in each velocity boundary is at least one channel. We integrated the emission over these intervals to create preliminary moment-0 maps and drew polygonal apertures around the significant lobe emission. We used the apertures to extract the spectra. The PV diagrams were extracted along linear cuts following the longest axes of the lobes. For the final displayed moment-0 maps, we integrated the CO emission above \(5\sigma_{\rm ch}\) within the adopted velocity intervals and polygonal apertures.
 Isolated high-velocity features lacking a plausible driving source were not classified as molecular outflows. Integrated intensity maps reveal 51 individual outflow lobes associated with continuum cores in 12 sources of the sample. No molecular outflow activity is detected toward HIGALBM344.1032--0.6609.

For sources exhibiting both blue- and red-shifted lobes, the outflow axis was defined by the line connecting the intensity-weighted centroids of the two components. In cases where only a single lobe was detected, the outflow orientation was estimated from the morphology of the high-velocity emission and its relation to the continuum source. The absence of a counter-lobe in these systems may result from sensitivity limitations, confusion with ambient cloud emission, or projection effects that make the counter-lobe difficult to distinguish from the surrounding molecular gas.

Based on the morphology of the integrated intensity maps, the identified outflows were classified into three categories: (i) clearly bipolar systems with spatially separated blue- and red-shifted lobes, (ii) monopolar systems with only one detected high-velocity lobe, and (iii) complex systems exhibiting overlapping or multiple outflow components. Figure \ref{CO_MAPS} presents the integrated high-velocity CO ($J=2$--1) emission toward the SQUALO sources with detected molecular outflow activity. The blue- and red-shifted emission components are shown in blue and red, respectively, relative to the systemic velocity of each source. The maps were constructed using the source- and lobe-specific LSR velocity intervals reported in Table \ref{flows}. Within these intervals, the integration was restricted to emission exceeding \(5\sigma_{\rm ch}\) within the polygonal aperture defined for each lobe, where \(\sigma_{\rm ch}\) is the channel rms reported in Table \ref{sources}. The colour wedges give the CO integrated intensity in \(\mathrm{Jy\,beam^{-1}\,km\,s^{-1}}\); the blue and red components share the same intensity scale within each panel, although the scale varies between sources.

Figure~\ref{CO_MAPS} illustrates the diversity of the observed morphologies, ranging from well-defined bipolar structures to monopolar and complex systems with multiple overlapping lobes. In several regions, multiple continuum sources are present within the same field, making the identification of the driving source non-trivial. Candidate driving sources were assigned based on the spatial distribution of the high-velocity gas, supported by the higher-angular-resolution ALMAGAL continuum data when available (7 of the 13 targets). The integrated intensity maps and associated position--velocity analysis provide the basis for the derivation of the outflow properties presented in the following sections.

\begin{landscape}
\begin{figure}
    \centering
    \includegraphics[
        width=0.95\linewidth,
        height=0.90\textheight,
        keepaspectratio
    ]{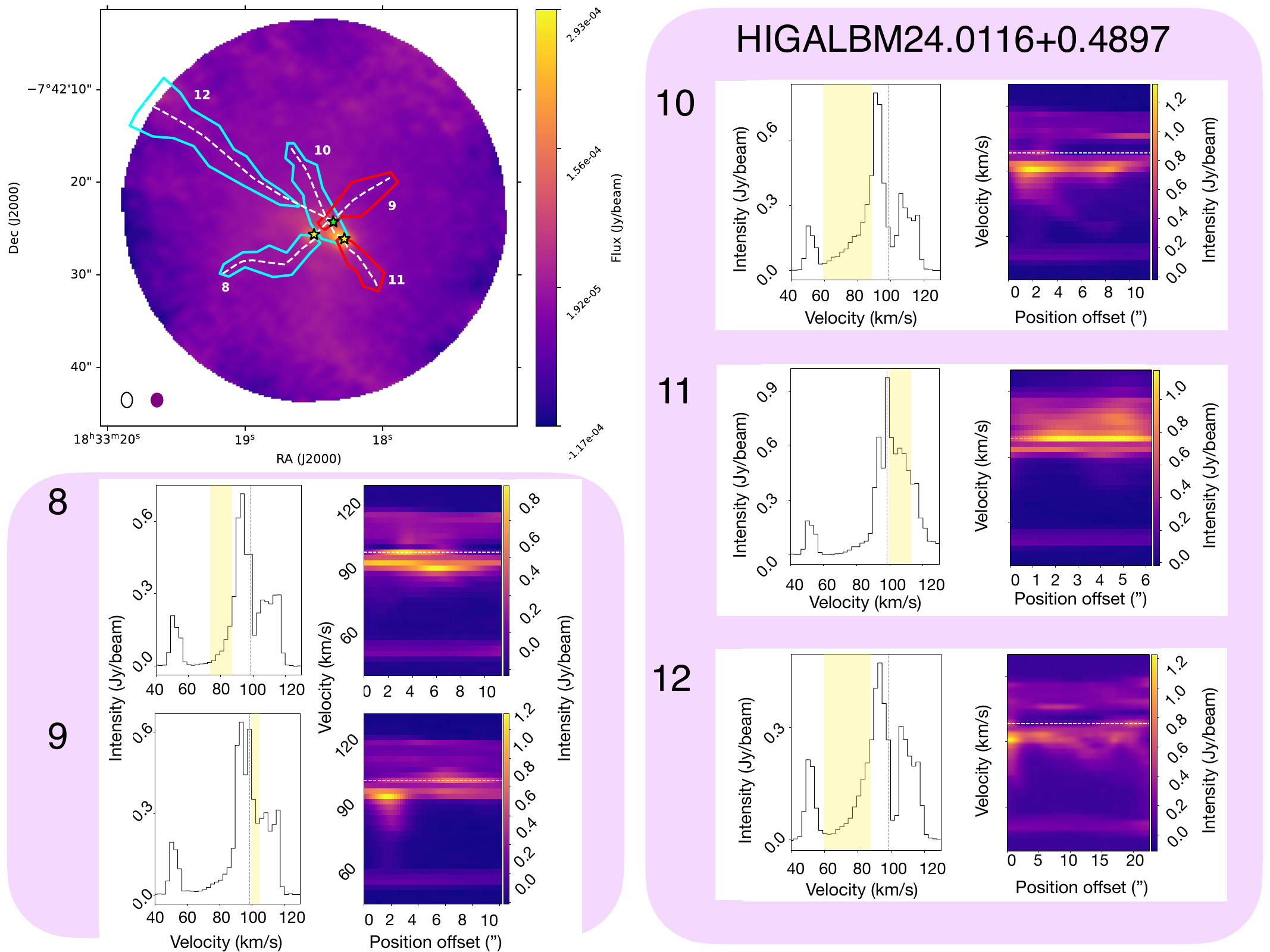}
    \caption{Example of the outflow identification procedure applied to
HIGALBM24.0116+0.4897. The left panel shows the Band~6
(1.3~mm) dust-continuum image, with red and cyan contours
indicating the integrated red- and blue-shifted CO
($J=2$--1) emission, respectively, detected above the
$5\sigma$ level with respect to the systemic velocity
($V_{\rm LSR}$). Black stars mark the continuum peaks, while
the numbered polygonal regions delineate the individual outflow
lobes used for the extraction of the integrated spectra and the
derivation of the outflow properties. The white and purple
ellipses in the lower-left corner indicate the synthesized beams
of the CO and continuum data, respectively. The surrounding
panels display the integrated spectra (left) and position--velocity
(PV) diagrams (right) corresponding to each numbered lobe. In
the spectra, the vertical dashed line indicates $V_{\rm LSR}$;
in the PV diagrams, the horizontal dashed line marks the same
velocity. The pale-yellow shaded regions in the spectral
panels indicate the lobe-specific velocity ranges adopted for
the calculation of the outflow properties, as listed in Table \ref{flows}.
Emission outside these shaded intervals is displayed to provide
kinematic context but was not included in the outflow calculations.
The white dashed lines within the polygonal regions show the
linear cuts used to extract the PV diagrams; unlike the spectra,
the PV diagrams are not averaged over the full polygonal
apertures.}
    \label{fig:SDC24_PV}
\end{figure}
\end{landscape}

Figure \ref{fig:SDC24_PV} presents the Band 6 (1.3 mm) dust-continuum image of HIGALBM24.0116+0.4897. Red and cyan contours with respect to the systemic velocity ($V_{\rm LSR}$) denote the integrated red- and blue-shifted CO ($J=2$--1) emission, respectively, detected above the 5$\sigma$ level. The numbered polygonal regions identify the individual outflow lobes used for the extraction of integrated spectra and the derivation of outflow properties. The spectrum of each lobe was obtained by spatially integrating the CO emission within its numbered polygonal aperture. By contrast, the corresponding position–velocity diagram was extracted along a linear cut following the longest axis of the lobe, as indicated by the white dashed line in the left panel, following an approach similar to that of \citet{Towner2024}; it therefore does not represent an average over the full polygon. For this example, the strong emission close to \(V_{\rm LSR}=94.8\ {\rm km\,s^{-1}}\) is attributed primarily to the ambient molecular cloud and was excluded from the outflow calculations. Specifically, the adopted intervals terminate at \(88.30\ {\rm km\,s^{-1}}\) for the blue-shifted lobes and begin at \(98.46\ {\rm km\,s^{-1}}\) for the red-shifted lobes, while their lobe-specific outer limits are listed in Table \ref{flows} as $\Delta_{int}$. For each lobe, the corresponding integrated spectrum and position--velocity (PV) diagram are shown in the surrounding panels. The vertical dashed line in the spectra and the horizontal dashed line in the PV diagrams indicate $V_{\rm LSR}$. 
An additional emission component is visible near \(50\ \mathrm{km\,s^{-1}}\) in some spectra and PV diagrams. It lies outside the adopted velocity intervals for HIGALBM24.0116+0.4897: the most blueshifted boundaries are \(60.36\ \mathrm{km\,s^{-1}}\) for lobes 10 and 12, and \(75.60\ \mathrm{km\,s^{-1}}\) for lobe 8 (Table \ref{flows}). Inspection of the channel maps and PV diagrams revealed no spatial or kinematic continuity linking this component to the identified lobes or a candidate driving core, so it was excluded from the outflow calculations. Although a detached high-velocity knot cannot be ruled out completely, the available data do not provide evidence to associate this component with the source.

The spectra and PV diagrams show the full velocity coverage of the extracted CO emission, including components outside the intervals adopted for the outflow calculations. The derived values span $5.19 \leq |V_{\max}| \leq 116.56~{\rm km\,s^{-1}}$. The highest value was retained because the emission remains spatially coherent across successive channels and is morphologically associated with a candidate driving source. Conversely, lobes with $|V_{\max}| < 10~{\rm km\,s^{-1}}$ were retained only when the emission formed a coherent structure above $5\sigma_{\rm ch}$ and could be associated with a continuum source. Nevertheless, these low-velocity cases are particularly susceptible to confusion with ambient-cloud emission, and their identification and derived properties should therefore be regarded as more uncertain.

The identified outflow lobes form the basis of the quantitative analysis presented in the following section. Geometrical parameters are derived directly from the observed outflow morphology, while physical and dynamical parameters are estimated from the measured CO emission. Figures showing the PV-diagram analyses for the remaining sources in the sample are provided in the online supplementary material.

\subsection{Outflow parameters}

The identified outflow lobes provide the basis for a quantitative characterization of the molecular outflows in the SQUALO sample. In this section, we derive the geometrical, physical, and dynamical properties of the outflows from the integrated CO ($J=2$--1) emission. We first determine the geometry-driven parameters, including the outflow size, opening angle, inclination constraints, and dynamical timescale, and then derive the corresponding physical and dynamical quantities, such as mass, momentum, kinetic energy, outflow force, and mass accretion rate.

\subsubsection{Geometry-driven parameters}
\label{parameters}
The geometrical parameters of each CO outflow lobe were measured manually using the Cube Analysis and Rendering Tool for Astronomy (CARTA)\footnote{https://cartavis.org} and the integrated high-velocity CO maps. The lobe boundaries were defined by the outermost spatially coherent 5$\sigma$ CO contour within the adopted polygonal aperture. For bipolar systems, the outflow axis was defined by the line connecting the intensity-weighted centroids of the blue- and red-shifted lobes; for monopolar systems, it was estimated from the elongation of the high-velocity emission relative to the candidate driving source. Following the biconical description of \citet{Cabrit1986}, each lobe was characterized by its projected length, maximum width, position of maximum width, half-opening angle, and indicative inclination limits.

\begin{table*}
\centering
     \begin{tabular}{llccccccccccc}
     \hline
Outflow ID & Core ID& Lobe & $\Delta_{int}$ & $x\rm_{O} $ & $L\rm_{O}$ & $W$ & $i_1$--$i_2$ & $\theta_{\rm half}$ & $l\rm_{proj}$ &|V$\rm_{max}$| &$t\rm_{dyn}$ &Flux\\
 &ID  &  & (km s$^{-1}$) & (\arcsec) & (\arcsec) & (\arcsec) & ($\rm ^{\circ})$ & ($\rm ^{\circ})$ &(pc)& (km s$^{-1}$)& (yr) & (Jy km s$^{-1}$)\\
\hline
\hline
\multicolumn{13}{c}{HIGALBM31.9462+0.0759 (94.7 km s$^{-1}$)} \\
\hline
1 & S1 &Blue  & [59.82, 69.98] & 3.87 & 6.90   &1.98   & 14--76 &14 &0.18 &34.88 &5158 &2.3\\
2& S1 &Red  & [103.0, 136.02] &6.72  & 9.71   &3.80   & 16--74 &16 &0.26 &41.32 &6127&61.1 \\
3& S1 & Blue &[80.14, 90.30]  &17.92  & 23.46   &6.99   & 11--79 &11 &0.63 &14.56 &42010&111.7 \\
4&S1  & Red & [103.0, 156.34] & 18.30 &  18.37  &6.36  & 10--80 &10 &0.49 &61.64 &7770&133.7\\
5&S2  &Red  & [103.0, 128.40] & 6.75 & 13.17   &2.24  & 9--81 & 9& 0.35&33.7 &10189&20.3\\
6&S3  &Blue  & [87.76, 90.30] & 1.38 & 2.06   &1.74   & 32--58 &32 & 0.05&6.94 &7739&2.4\\
7&S3  & Red & [103.0, 113.16] & 3.05 & 7.02   &8.46  & 54--36 &54 & 0.19&18.46 &9915&33.2\\
\hline
\multicolumn{13}{c}{HIGALBM24.0116+0.4897 (94.8 km s$^{-1}$)} \\
\hline
8&SA1 & Blue & [75.60, 88.30]  & 6.08 & 9.56 & 4.29  & 19--71  &19 &0.24 &19.2 & 12297&17.7\\
9& SA1   & Red  & [98.46, 103.54] &  4.16 & 7.83 & 4.16  & 27--63 &27 &0.20 &8.74 &22126&28.7\\
10& SA2 & Blue &  [60.36, 88.30] & 3.73 & 11.95 & 3.49   &  25--65&25 &0.30 &34.44 &8570&53.1\\
11&  SA2  & Red  & [98.46, 111.16] & 5.95 & 7.00 & 2.71  &  13--77&13 & 0.18&16.36 &10568&53.5\\
12 &A3 &Blue & [60.36, 88.30] &  24.10 & 24.10 & 5.48  & 6--84 &6 &0.61 &34.44 &17283&66.7\\

\hline
\multicolumn{13}{c}{HIGALBM28.1957--0.0724 (98.8 km s$^{-1}$)} \\
\hline
13& S1&Blue  &[66.90, 69.44]  &2.32  &4.55  & 2.30   & 26--64 &26 &0.12 &31.6 &3651&0.2\\ 
14& S1& Red &[115.16, 143.10]  & 2.05 &  7.10&2.34   & 30--60 &30 &0.19 & 44.3&4103&2.7\\
15& S2& Red & [115.16, 130.40] & 1.20 & 2.87 &1.82   & 37--53 &37 &0.08 &31.6 &2325&0.4\\
\hline
\multicolumn{13}{c}{HIGALBM338.9260+0.6340 (--65.1 km s$^{-1}$)} \\
\hline
16&SA1 & Blue &[--118.50, --77.86]  & 8.96 & 11.21 & 3.49  & 11--79 &11 &0.23 &53.4&4140 &100.2\\ 
17&SA1 & Red &[--39.76, --27.06]  & 2.05 & 5.50 & 2.60   & 32--58 &32 &0.11 &38.04&2851 &2.9\\ 
18&SA2 & Blue &[--98.18, --67.70]  & 3.63 & 3.63 & 6.48  & 42--48 &42 &0.07 &32.2&2223 &112.1\\ 
19&SA3 & Blue &[--80.40, --67.70]  & 4.89 & 4.89 & 3.74   & 21--69 &21 &0.10 &14.7 &6560&54.9\\ 
20&SA4 & Blue &[--80.40, --67.70]  & 6.75 & 8.61 & 7.39  & 29--61 &29 & 0.17& 14.7& 11550&138.5\\ 
21&SA4 & Red &[--55.00, --19.44]  & 7.58 & 10.55 & 3.84   & 14--76 &14 &0.21 &10.16&20477 &138.0\\ 
22&A5 & Blue &[--128.0, --119.84]  & 1.84 & 4.55 & 2.00  & 29--61 &29 &0.09 &116.9&768 &0.2\\ 
23&A5 & Red &[--55.00, --49.92]  & 3.24 & 4.91 & 2.80   & 23--67 &23 & 0.10&15.18 &6379&20.0\\ 
\hline
\multicolumn{13}{c}{HIGALBM332.6045--0.1674 (--48.34 km s$^{-1}$)} \\ 
\hline
24&S1 & Blue &[--78.10, --50.16]  & 10.4 & 14.66 & 26.7  & 38--52 &52 &0.22 &29.76 &7239&925.1\\
25 & S1 & Red & [--37.46, --19.68] &10.37 & 12.69&19.08  &43--47  &43 &0.19 &28.66&6507 &199.3\\
\hline
\multicolumn{13}{c}{HIGALBM341.2149--0.2359 (--45.72 km s$^{-1}$)} \\
\hline
26&SA1 & Blue & [--62.40, --54.78] & 8.43 & 10.74 &1.54  & 5--85 &5 &0.18 &16.68&10378 &17.0\\
27&SA1 & Red & [--24.30, --11.60] & 3.42 & 3.45 &6.44  & 43--47 & 43&0.06 &34.12&1630 &8.4\\
28&SA2 & Blue & [--59.86, --52.24] & 1.88 & 3.34 &2.18  &  30--60&30 &0.06 &14.14& 3807&5.5\\
\hline
\multicolumn{13}{c}{HIGALBM327.3918+0.1996 (--90.44 km s$^{-1}$)} \\
\hline
29&SA1 & Blue & [--104.86, --92.16] & 2.94 & 5.79 &4.34  & 36--54 &36 & 0.15&14.42&9898&166.7\\
30&SA1 & Red & [--84.54, --56.60] & 2.84 & 6.20 &4.65   & 39--51 & 39&0.16 &33.84 & 4516&95.5\\
31&SA2 & Blue & [--97.24, --94.70] & 4.45 & 9.87 &3.97  & 24--66 &24 & 0.25&6.2 &39242 &30.3\\
32&S3 & Blue & [--104.86, --92.16] & 6.92 & 9.76 &9.76  & 35--55 & 35&0.25 &14.42 & 16684&173.2\\
33&S4 & Red & [--59.14, --56.60] & 4.54 & 10.96 &5.86  & 33--57 &33 &0.28 & 33.84&7984 &8.2\\
\hline
\multicolumn{13}{c}{HIGALBM343.7560--0.1629 (--26.19 km s$^{-1}$)} \\
\hline
34&S1 & Blue &[--56.56, --38.78]  &7.83  & 10.69  & 1.48  & 5--85 &5 &0.10 &30.37 &7984&20.8\\
35 &S1& Blue& [--51.48, --36.24] & 3.55 & 15.00 & 1.46   & 12--78 &12 &0.15 &25.29 &3337 &40.0\\
36&S1 & Red & [--21.00] & 14.24 & 14.24 &12.56  & 24--66 &24 &0.14 &5.19&5623 &95.2\\
\hline
\multicolumn{13}{c}{HIGALBM344.1032--0.6609 (--27.90 km s$^{-1}$)} \\
\hline
& & & &   &   &None   &  &\\
\hline

\end{tabular}
\caption{Geometrical parameters and integrated CO
($J=2$--$1$) fluxes of the molecular outflow lobes identified
toward the SQUALO clumps. For each lobe, the table reports the LSR velocity interval used to integrate the CO emission, $v_{\rm int}=[v_{\min},v_{\max}]$. Projected offset from
the associated core ($x_{\rm O}$), lobe length ($L_{\rm O}$),
lobe width ($W$), the two geometrical inclination limits
($i_1=\theta_{\rm half}$ and
$i_2=90^\circ-\theta_{\rm half}$), half-opening angle
($\theta_{\rm half}$), projected lobe length ($l_{\rm proj}$),
maximum velocity relative to the systemic velocity
($|V_{\max}|$), dynamical timescale ($t_{\rm dyn}$), and
integrated CO flux. For $\theta_{\rm half}>45^\circ$, the first
geometrical limit exceeds the second, indicating that the
idealised non-overlapping conical geometry does not yield a
conventionally ordered inclination interval. Blue- and
red-shifted lobes are listed separately. The interval endpoints are the LSR-frame velocities of the selected channel centres, reported to two decimal places; the boundary uncertainty is at least one channel (2.54 km s$^{-1}$).}

\label{flows}
\end{table*}    

\addtocounter{table}{-1}
\begin{table*}
\centering
     \begin{tabular}{llccccccccccc}
     \hline
Name &Core & Lobe & $\Delta_{int}$ & $x\rm_{O} $ & $L\rm_{O}$ & $W$ & $i_1$--$i_2$ & $\theta_{\rm half}$ & $l\rm_{proj}$ &|V$\rm_{max}$| &$t\rm_{dyn}$ &Flux\\
 &ID  &  & (km s$^{-1}$) & (\arcsec) & (\arcsec) & (\arcsec) & ($\rm ^{\circ})$ & ($\rm ^{\circ})$ &(pc)& (km s$^{-1}$)& (yr) & (Jy km s$^{-1}$)\\
\hline
\hline
\multicolumn{13}{c}{HIGALBM343.5212--0.5172 (--34.62 km s$^{-1}$)} \\
\hline
37&SA1 & Blue & [--68.64, --50.86] & 6.51 & 7.79 &2.95  &13--77  &13 &0.11 &34.02 &3256&6.7\\
38&SA1 & Red & [--10.22, --2.60] & 2.99 & 5.61 &2.53  & 23--67 &23 &0.08 &8.62&9255 &5.9\\
39&SA2 & Blue & [--66.10, --53.40] & 7.30 & 8.83 &1.74  & 7--83 &7 &0.13 &26.72 &4700&1.4\\
\hline
\multicolumn{13}{c}{HIGALBM331.1314--0.2438 (--89.44 km s$^{-1}$)} \\
\hline
40&S1 & Blue & [--206.00, --94.24] & 3.90 & 10.73 &10.73  & 54--36 &54 &0.26 & 116.56&2182&6342.1 \\
41&A1 & Red & [--84.08, --56.14] & 7.87 & 16.26 &12.73  & 39--51 &39 &0.39 &33.3 &11574&61163.9 \\
42&S2 & Blue & [--119.64, --94.24] &5.29 & 10.34 &9.93  & 43--47 &43 & 0.25& 30.2&8115&6223.6 \\
43&S2 & Red & [--79.00, --56.14] & 2.30 & 6.28 &3.04   & 33-57 & 33&0.15 & 33.3&4470&697.2 \\
44&A2 & Blue & [--109.48, --94.24] & 3.35 & 10.04 &2.42   & 20--70 &20 &0.24 &20.04 &11875&683.4 \\
45&A3 & Blue & [--142.50, --99.32] & 2.11 & 5.12 &1.70   & 22--68 &22 &0.12 & 53.06&2287&688.8\\
46&A3 & Red & [--73.92, --56.14] & 2.49 & 4.59 &4.02  & 39--51 & 39&0.11 &33.3 &3267&690.5 \\
\hline
\multicolumn{13}{c}{HIGALBM327.4022+0.4449 (--80.80 km s$^{-1}$)} \\
\hline
47&A1 & Blue & [--111.10, --90.78] & 2.96 & 7.03 &2.62   & 24--66 & 24& 0.16&30.3 &5059&625.2\\
48&A1 & Red & [--67.92, --52.68] &  3.20 & 5.63 &3.20   & 27--63 & 27&0.13 &28.12 &4366&631.9\\
49&A2 & Blue & [--103.48, --88.24] & 2.14 & 3.90 &5.29   & 51--39 &51 &0.09 & 22.68&3750&670.5\\
50&A2 & Red & [--70.46, --60.30 ] & 3.74 & 8.55 &3.80   & 27--63 & 27&0.19 &20.5 &9095&6137.8\\

\hline
\multicolumn{13}{c}{HIGALBM344.2210--0.5932 (--27.90 km s$^{-1}$)} \\
\hline
51&S1 & Red & [--10.38, --7.84] & 11.92 & 12.31 &3.32   & 8--82 &8 &0.12 &20.06 &5818&615.5\\
\hline
\end{tabular}
\caption{Continued: Geometrical parameters and integrated CO
($J=2$--$1$) fluxes of the molecular outflow lobes identified
toward the SQUALO clumps. For each lobe, the table reports the
velocity integration interval ($\Delta v$), projected offset from
the associated core ($x_{\rm O}$), lobe length ($L_{\rm O}$),
lobe width ($W$), the two geometrical inclination limits
($i_1=\theta_{\rm half}$ and
$i_2=90^\circ-\theta_{\rm half}$), half-opening angle
($\theta_{\rm half}$), projected lobe length ($l_{\rm proj}$),
maximum velocity relative to the systemic velocity
($|V_{\max}|$), dynamical timescale ($t_{\rm dyn}$), and
integrated CO flux. For $\theta_{\rm half}>45^\circ$, the first
geometrical limit exceeds the second, indicating that the
idealised non-overlapping conical geometry does not yield a
conventionally ordered inclination interval. Blue- and
red-shifted lobes are listed separately. The interval endpoints are the LSR-frame velocities of the selected channel centres, reported to two decimal places; the boundary uncertainty is at least one channel (2.54 km s$^{-1}$).}

\label{flows}
\end{table*}  

The resulting measurements are reported in Table \ref{flows}. The core labels listed in Table \ref{flows} indicate the origin of the candidate driving source. The prefix \texttt{S} denotes cores identified only in the SQUALO survey \citep{Traficante2023}, \texttt{A} denotes cores identified only in the ALMAGAL continuum catalogue \citep{Coletta2025}, and \texttt{SA} denotes cores independently identified in both datasets. Numerical suffixes are assigned sequentially within each clump and serve only as identifiers.

For 8 of the 51 identified outflow lobes, marked with a dagger in Table \ref{flows}, the high-velocity CO emission extends to the edge of the mapped field of view.
For each lobe, we define $L_{\rm o}$ as the projected angular length of the outflow, measured from the candidate driving source to the outermost significant high-velocity CO emission. The lobe width, $W$, is measured perpendicular to the outflow axis, while $x_{\rm o}$ is the projected distance from the driving source to the position of maximum width. The half-opening angle is then estimated as
\begin{equation}
\theta_{\rm half} = \tan^{-1}\left(\frac{W/2}{x_{\rm o}}\right).
\end{equation}
For a small number of outflows, the high-velocity CO emission extends to the edge of the mapped field of view. In these cases, the measured values of $L_{\rm O}$, $W$, and $x_{\rm O}$ should be regarded as lower limits, since the full spatial extent and morphology of the lobe may not be completely sampled. Consequently, the derived opening angles and projected lengths for these outflows may also be underestimated. We nevertheless retain the measured values in the analysis, noting that the affected sources represent only a small fraction of the sample.
The full opening angle is $\theta = 2\theta_{\rm half}$.
This definition is consistent with the biconical geometry in which the opening angle is measured between the outflow axis and the edge of the lobe.

The projected physical length of each lobe $l_{\rm proj}$, in parsecs, is calculated from the observed angular length and the source distance as
\begin{equation}
l_{\rm proj} = \frac{L_{\rm o}}{206265}D,
\end{equation}
where $L_{\rm o}$ is in arcsec and $D$ is the distance to the source in pc. The maximum velocity offset is defined as
\begin{equation}
|V_{\rm max}| = |V_{\rm edge} - V_{\rm LSR}|,
\end{equation}
where $V_{\rm edge}$ is the most extreme velocity channel included in the adopted outflow velocity interval.

We estimate geometrical inclination limits for each lobe, where
$i$ is defined as the angle between the outflow axis and the line
of sight. Thus, $i=0^\circ$ corresponds to an outflow directed
along the line of sight, while $i=90^\circ$ corresponds to an
outflow lying in the plane of the sky. For clear bipolar systems exhibiting the non-overlapping red- and
blue-shifted morphology corresponding to Case 2 of
\citet{Cabrit1986}, we adopt the idealised conical outflow geometry
described by \citet{Cabrit1992}. Under this assumption, the two
geometrical inclination limits are given by
\begin{equation}
i_1 = \theta_{\rm half}, \qquad
i_2 = 90^\circ-\theta_{\rm half}.
\end{equation}
For lobes with $\theta_{\rm half}\leq45^\circ$, these limits define
the conventionally ordered interval
$\theta_{\rm half}\leq i\leq90^\circ-\theta_{\rm half}$.
For broader lobes with $\theta_{\rm half}>45^\circ$, the order of
the two limits is reversed. In these cases, both values are retained
because they provide a direct indication that the observed morphology
is not fully described by the idealised non-overlapping conical
geometry.

These limits should be regarded as indicative rather than definitive, since real molecular outflows frequently exhibit more complex morphologies than the idealised conical geometry assumed here. Among the 51 identified outflow lobes, 36 are associated with
bipolar systems and therefore permit the estimation of geometrical
inclination limits. For monopolar or morphologically complex outflows, the inclination is substantially more uncertain owing to projection effects and the absence of a clearly identified counter-lobe. Consequently, we report the two geometrical inclination limits
rather than a single preferred inclination angle whenever possible.
These values are intended only to provide a qualitative indication
of the likely outflow orientation.

For reference, the inclination-corrected length and velocity would be
\begin{equation}
l_{\rm corr} = \frac{l_{\rm proj}}{\sin i},
\end{equation}
and
\begin{equation}
V_{\rm corr} = \frac{|V_{\rm max}|}{\cos i},
\end{equation}
respectively. However, these corrected quantities are not adopted in the main parameter table and are not used in any subsequent statistical analyses. Instead, all physical and dynamical parameters presented in this work are derived from the observed projected quantities. The dynamical timescale listed in Table~\ref{flows} is therefore calculated as
\begin{equation}
t_{\rm dyn} = \frac{l_{\rm proj}}{|V_{\rm max}|}.
\end{equation}

Table~\ref{flows} lists the velocity interval $\Delta_{int}$, $x_{\rm O}$, $L_{\rm O}$, $W$, the geometrical inclination limits $i_1$ and $i_2$, half-opening angle $\theta_{\rm half}$, projected physical length $l$, maximum velocity offset $|V_{\rm max}|$, dynamical timescale $t_{\rm dyn}$, and integrated CO flux for each identified outflow lobe.

\subsubsection{Physical and dynamical parameters}
\label{parameters-2}
The physical properties of the outflows were derived from the integrated CO ($J=2$--1) emission under the assumption of local thermodynamic equilibrium (LTE). The outflow mass was calculated as

\begin{equation}
M_{\rm out}
=
\mu_{\rm H_2}\,m_{\rm H}\,N_{\rm H_2},
\end{equation}
where $\mu_{\rm H_2}=2.8$ is the mean molecular weight per hydrogen molecule, $m_{\rm H}$ is the hydrogen mass, and $N_{\rm H_2}$ is the total number of molecular hydrogen molecules associated with the outflow. Assuming a CO abundance relative to H$_2$ of $X_{\rm CO}=10^{-4}$ \citep{Frerking1982},
\begin{equation}
N_{\rm H_2}
=
\frac{N_{\rm CO}}{X_{\rm CO}},
\end{equation}
where $N_{\rm CO}$ is the total number of CO molecules derived from the observed CO ($J=2$--1) emission. To account for uncertainties in the excitation conditions of the outflowing gas, the calculations were performed assuming excitation temperatures of $T_{\rm ex}=20$ and 80 K. These values bracket the range of excitation temperatures commonly adopted for molecular outflows associated with massive protostars \citep[e.g.][]{Avison2021,Towner2024}. The CO column densities used to derive the outflow masses were corrected for optical-depth effects by applying fixed opacity correction factors for each adopted excitation temperature. We adopted correction factors of 3.5 for $T_{\rm ex}=20$ \citep{Cabrit1992} and 8.2 for $T_{\rm ex}=80$, following the opacity-correction prescription adopted by \citet{Avison2021}. The outflow mass was calculated independently for both temperature assumptions, and the average value was adopted as the fiducial mass estimate. The quoted uncertainties correspond only to the range spanned by the two excitation-temperature solutions and do not include systematic uncertainties associated with the adopted CO abundance or opacity correction. No explicit inclination correction is applied to the outflow mass because it is derived from the velocity-integrated CO emission. Nevertheless, inclination may introduce an indirect bias: for outflows oriented close to the plane of the sky, a larger fraction of the gas has low line-of-sight velocities and may be confused with ambient-cloud emission and excluded from the adopted integration intervals. The reported masses may therefore be underestimated for such outflows. Accordingly, the quoted uncertainties represent only the effect of the assumed excitation temperature and not the total uncertainty in the derived outflow masses.

Following the standard molecular outflow formalism \citep[e.g.][]{Cabrit1986,Avison2021}, the momentum and kinetic energy were calculated as
\begin{equation}
P_{\rm CO}=M_{\rm out}|V_{\rm max}|,
\end{equation}
and
\begin{equation}
E_{\rm CO}=\frac{1}{2}M_{\rm out}|V_{\rm max}|^2,
\end{equation}
respectively.

The outflow force was then derived from
\begin{equation}
F_{\rm CO}=\frac{P_{\rm CO}}{t_{\rm dyn}},
\end{equation}
where $t_{\rm dyn}$ is the dynamical timescale defined in Sect. \ref{parameters}.

Following the momentum-conservation formalism commonly adopted for molecular outflows \citep{Beuther2002a}, we assume that the observed molecular outflow is driven by an underlying protostellar wind, such that the outflow momentum flux, $F_{\rm CO}$, is equal to the wind momentum flux scaled by the entrainment efficiency, $f_{\rm ent}$. Under this assumption, the mass accretion rate ($\dot{M}_{\rm acc}$) can be estimated from the outflow force according to
\begin{equation}
F_{\rm CO} = f_{\rm ent}\,\dot{M}_{\rm acc}\,v_{\rm w},
\end{equation}
which can be rearranged to give
\begin{equation}
\dot{M}_{\rm acc} =
\frac{F_{\rm CO}}{f_{\rm ent}\,v_{\rm w}},
\label{Macc_equ}
\end{equation}
where $f_{\rm ent}=0.25$ is the entrainment efficiency, representing the fraction of the wind momentum transferred to the entrained molecular gas \citep{Bontemps1996}, and $v_{\rm w}=500\,{\rm km\,s^{-1}}$ is the assumed jet/wind velocity \citep{Lamers1995}.

Table~\ref{dynamics} lists the derived outflow masses, momenta, kinetic energies, outflow forces, and mass accretion rates for all identified outflow lobes. No inclination correction is applied to the physical and dynamical quantities reported in this work. Consequently, the values presented in Table~\ref{dynamics} should be regarded as projected estimates subject to potentially
substantial systematic uncertainties. The derived outflow properties span a wide range across the SQUALO sample. Outflow masses range from $2.5\times10^{-3}$ to $2.09\times10^{1} M_\odot$ with corresponding momenta of $2.9\times10^{-1}$ to $7.16\times10^{2} M_\odot {\rm km s^{-1}}$ and kinetic energies of $2.8\times10^{36}$ to $8.3\times10^{40}$ J. The inferred outflow forces cover more than four orders of magnitude, from $3.6\times10^{-5}$ to $3.3\times10^{-1} M_\odot {\rm km s^{-1} yr^{-1}}$, corresponding to mass accretion rates between $4.6\times10^{-7}$ and $7.6\times10^{-3} M_\odot {\rm yr^{-1}}$. These values are broadly consistent with those reported for molecular outflows associated with intermediate- and high-mass star-forming regions in previous studies \citep[e.g.][]{Beuther2002a,Maud2015,Avison2021, Towner2024}.
\subsubsection{Systematic uncertainties}
An additional source of systematic uncertainty arises from the choice
of characteristic outflow velocity. In this work, the momentum,
kinetic energy, and dynamical timescale are derived using the maximum
observed velocity offset ($|V_{\rm max}|$), following a commonly
adopted approach in studies of massive molecular outflows.
Alternative methods compute these quantities by summing over the
velocity distribution on a channel-by-channel basis or by using
intensity-weighted velocities, generally yielding lower
characteristic velocities and longer dynamical timescales
\citep[e.g.][]{Downes2007}. As discussed by
\citet{Downes2007}, different analysis methods can therefore produce
systematic differences in the derived outflow parameters. The adopted velocity intervals were selected individually for each source to exclude emission associated with the ambient cloud (Sect. \ref{identification}). Their precise boundaries are therefore source dependent, and residual contamination from ambient emission or exclusion of low-velocity outflow material may introduce additional systematic uncertainty, particularly in the derived masses and momenta.

The conversion from outflow momentum flux to protostellar accretion rate assumes fixed values of the entrainment efficiency and wind velocity for all sources. These quantities may vary among protostars and over time, particularly if the accretion and ejection processes are episodic. Consequently, the adopted conversion introduces source-dependent uncertainties that may affect not only the absolute accretion rates but also their relative values and the correlations discussed in this work. The inferred accretion rates should therefore be regarded as approximate, time-averaged estimates rather than direct measurements of the instantaneous protostellar accretion rates.

Importantly, all SQUALO clumps were observed, reduced, and analysed
using a homogeneous methodology. Consequently, although the absolute
values of the derived outflow parameters remain uncertain, the
relative comparisons within the SQUALO sample are expected to be less
sensitive to these systematic effects.

\section{Discussion}
\label{4}
The derived outflow properties span more than two orders of magnitude in mass, momentum, and kinetic energy, well above the uncertainties arising from the approximations made in deriving the parameters themselves and are indicative of  the diversity of protostellar activity within the SQUALO sample. In the following section, we examine how these properties relate to the evolutionary stage of the host clumps and to the larger-scale physical conditions of their natal environments.
\subsection{Molecular outflows as tracers of accretion in the SQUALO sample
and maximum outflow velocity and driving core mass}

Molecular outflows are widely regarded as one of the most direct observational tracers of ongoing protostellar accretion, owing to their physical connection with magneto-centrifugally launched disc winds and jets \citep{Blandford1982, Arce2007}. The high detection rate of CO outflows within the SQUALO sample therefore indicates that active mass accretion occurs across clumps spanning a broad range of masses and \(L/M\). Although \(L/M\) is commonly used as an approximate evolutionary indicator, variations among individual clumps cannot necessarily be attributed solely to evolutionary stage, because \(L/M\) is also sensitive to dust temperature and to the presence of a small number of luminous massive stars \citep{Ma2013}. This result is particularly noteworthy given that all SQUALO clumps were selected independently on the basis of large-scale infall signatures, suggesting that parsec-scale accretion and protostellar-scale accretion are simultaneously present throughout the sample.

The derived outflow properties span more than an order of magnitude in mass, momentum, energy, and momentum flux, indicating a substantial diversity in the strength of protostellar feedback among the observed clumps. Nevertheless, the measured values fall within the range reported by previous studies of high-mass star-forming regions \citep[e.g.][]{Beuther2002a, Duarte-Cabral2013, Maud2015, Yang2018, Li2020, Towner2024}. The SQUALO outflows therefore appear representative of the broader population of massive protostellar outflows despite being selected from a sample of clumps known to exhibit large-scale accretion. The derived outflow masses and energetics are comparable to those observed in other massive protostellar environments and significantly exceed the typical values reported for low-mass protostellar outflows \citep[e.g.][]{Bontemps1996,vanderMarel2013}, consistent with the broad literature compilation presented by \citet{Avison2021}.

The inferred protostellar accretion rates span $4.6\times10^{-7}$--$7.6\times10^{-3} M_\odot yr^{-1}$, with most values lying between $10^{-5}$ and $10^{-3}$ $M_\odot yr^{-1}$. Since the accretion rates are derived directly from the outflow momentum flux through Eq. \ref{Macc_equ}, the two quantities are expected to scale linearly. The inferred accretion rates are consistent with those required for the formation of massive stars and are comparable to values reported in previous studies of massive protostellar outflows (e.g. \citealt{Beuther2002a}; \citealt{Maud2015}; \citealt{Avison2021}; \citealt{Towner2024}). These results are consistent with the widely adopted picture in which molecular outflows are driven by accretion-powered jets or winds that transfer momentum to the surrounding molecular gas \citep[e.g.][]{Bontemps1996, Duarte-Cabral2013}. Within this framework, the observed outflow momentum flux provides an indirect probe of the ongoing accretion process, allowing accretion activity to be investigated even in deeply embedded massive protostars where direct measurements remain challenging.

The widespread detection of accretion-powered outflows therefore
motivates an examination of whether their properties depend on
the physical and dynamical conditions of the parent clumps.

\begin{figure}
    \centering
    \includegraphics[width=1.1\columnwidth]{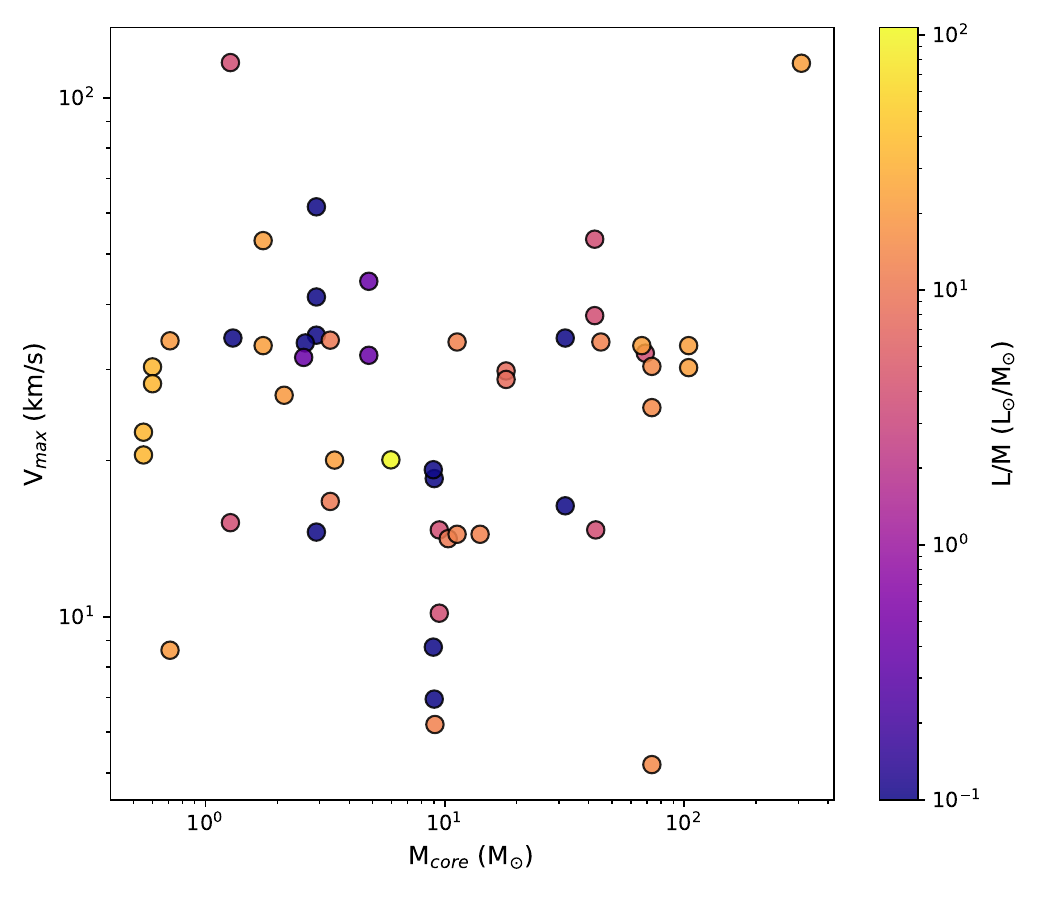}
    \caption{
    Maximum outflow velocity as a function of the mass of the
candidate driving core. Points are coloured according to the clump
$L/M$ ratio. A two-sided Pearson analysis in logarithmic space finds
no statistically significant correlation ($r=0.021$, $p=0.885$,
$N=51$).
    }
    \label{fig:vmax_mcore}
\end{figure}

One of the simplest expectations for accretion-driven outflows is that more massive protostellar cores should be capable of launching faster molecular outflows, since they are expected to host deeper gravitational potentials and potentially more energetic accretion. To investigate this possibility, we compare the maximum observed outflow velocity, $V_{max}$, with the mass of the candidate driving core identified from the SQUALO or ALMAGAL continuum data in Fig. \ref{fig:vmax_mcore}.
A two-sided Pearson analysis in logarithmic space finds no
statistically significant correlation between $V_{\max}$ and
$M_{\rm core}$ ($r=0.021$, $p=0.885$, $N=51$). The corresponding
power-law fit is nearly flat ($R^2=0.0004$). Outflows with comparable maximum velocities are associated with cores spanning more than two orders of magnitude in mass, while both low- and high-mass cores exhibit a broad range of observed velocities. The best-fitting relation is nearly flat, indicating that the maximum velocity of the molecular outflow is largely independent of the mass of the identified driving core within the SQUALO sample. 

The absence of a clear trend is not entirely unexpected. The observed $V_{max}$ represents only the line-of-sight velocity of entrained molecular gas rather than the intrinsic velocity of the underlying protostellar jet or wind. It is therefore strongly affected by projection effects, interaction with the surrounding cloud, and the adopted velocity range used to define the outflow emission. In clustered massive star-forming regions, additional uncertainty arises from the difficulty of uniquely associating each outflow with its driving source, particularly where multiple embedded cores are present.
Furthermore, the mass of the continuum core does not necessarily reflect the instantaneous mass of the central protostar responsible for launching the outflow. Instead, it represents the remaining reservoir of dense gas, which depends on the evolutionary stage, fragmentation, and local accretion history. Consequently, a simple one-to-one relationship between core mass and outflow velocity is not expected. Similar behaviour has been reported in previous studies of massive molecular outflows, where stronger correlations are generally found for integrated quantities such as outflow momentum, momentum flux, or mechanical luminosity than for the maximum velocity alone. Thus, within the SQUALO sample, the measured line-of-sight $V_{\max}$ does not provide a robust proxy for candidate driving-core mass. Its interpretation is limited by projection, uncertainty in the adopted velocity boundaries, possible contamination or exclusion of emission, and uncertainty in associating outflows with their driving sources. These effects may weaken or obscure an intrinsic relation. The integrated outflow quantities discussed in the following sections may provide more informative indicators of accretion activity.

\subsection{Outflow activity and the parent clump environment}
\label{outflow_activity}
Unless otherwise stated, correlations in the following analysis
were quantified using the two-sided Pearson product-moment
correlation coefficient, $r$. For power-law relations, the analysis
was performed in base-10 logarithmic space. We report the
corresponding $p$-value and sample size, $N$, and adopt $p<0.05$
as the threshold for nominal statistical significance. The
coefficient of determination, $R^2$, describes the goodness of the
fitted relation and is not used independently to assess statistical
significance.
\subsubsection{Clump surface density and outflow energetics}
\label{surfacedensity}

\begin{figure}
    \centering
    \includegraphics[width=1.1\columnwidth]{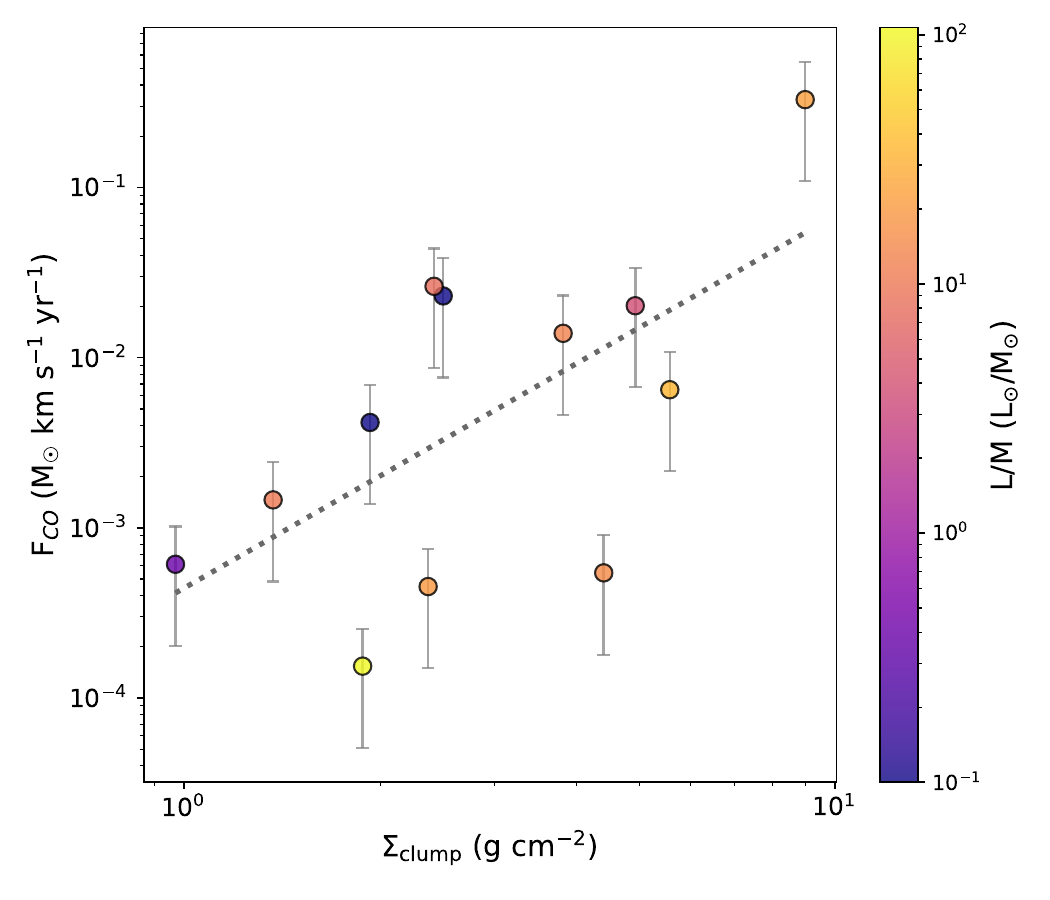}
    \caption{
    Figure~4. Outflow momentum flux ($F_{\rm CO}$) as a function of
clump surface density ($\Sigma_{\rm clump}$) for the 12 representative
outflows. Colours indicate the clump $L/M$. The dashed line shows the
best-fitting power-law relation. A two-sided Pearson analysis in
logarithmic space gives $r=0.627$, $p=0.029$, and $N=12$.
    }
    \label{fig:Fco_surface_density}
\end{figure}

To investigate whether the large-scale properties of the parent clumps influence the observed outflow activity, we compared the derived outflow parameters with the physical characteristics of the SQUALO clumps. The strongest correlation is found between the outflow momentum flux and the clump surface density, $\Sigma_{\rm cl}$ (Fig.~\ref{fig:Fco_surface_density}). This behaviour suggests that the physical conditions of the parent clump play an important role in regulating the accretion process occurring on protostellar scales.

The clump surface densities were taken from the SQUALO I analysis of \citet{Traficante2023}, where they were derived from the dust continuum properties. Since several SQUALO clumps host multiple outflows, we consider only the outflow with the highest inferred accretion rate in each clump, resulting in a sample of 12 independent measurements. This approach provides a single representative outflow for each clump while avoiding multiple non-independent measurements associated with the same parent structure.
Over nearly three orders of magnitude in $F_{\rm CO}$, a moderate
positive Pearson correlation is found in logarithmic space
($r=0.627$, $p=0.029$, $N=12$). This correlation is nominally
statistically significant at the adopted $p<0.05$ threshold. The
corresponding power-law fit yields

\[
F_{\rm CO}\propto\Sigma_{\rm cl}^{2.19},
\]

with $R^2=0.39$. Although substantial scatter is present, the result
indicates that clumps with higher surface densities tend to host
outflows with larger momentum fluxes. The colour coding shows that
sources spanning a broad range of $L/M$ values are distributed
throughout the relation.

 Notably, SQUALO~I found a strong positive correlation between the surface density of the parent clumps and that of the densest embedded cores (their Fig.~8; \citealt{Traficante2023}). Together with the correlation presented here between $F_{\rm CO}$ and $\Sigma_{\rm cl}$, these results suggest that the properties of the molecular outflows may ultimately be linked to the surface density of the dense cores from which they originate. Although the present dataset does not allow this possibility to be tested directly, the combined results are consistent with a scenario in which the dense core population reflects the physical properties of its parent clump, while the molecular outflows trace the subsequent protostellar accretion occurring within those cores.

The observed positive correlation between outflow momentum flux and clump surface density indicates that stronger outflows tend to occur in clumps with higher surface densities. This trend is consistent with a possible connection between protostellar accretion activity and the conditions within the parent clump. However, the correlation alone does not establish that the large-scale environment directly regulates accretion, because both \(F_{\rm CO}\) and \(\Sigma_{\rm cl}\) may depend on other properties, including clump mass, luminosity, evolutionary stage, and the number and masses of the embedded protostars. We therefore regard this relationship as suggestive rather than conclusive evidence of an environmental influence.

These results complement the findings of the SQUALO I survey
\citep{Traficante2023}, which revealed strong links between the properties
of dense fragments and those of their parent clumps. Similar relationships
have been reported in other interferometric studies, with more massive and
denser clumps generally hosting more massive cores or fragments, and with
the number of detected cores also increasing with clump surface density
\citep{Csengeri2017,Morii2024,Xu2024}. Our observed correlation between
$F_{\rm CO}$ and $\Sigma_{\rm cl}$ is consistent with this broader picture
and suggests a possible association between outflow activity and the
conditions in the parent clump. However, this correlation alone does not
establish a causal influence of the clump-scale environment, because both
quantities may depend on other clump and source properties. Whether the
outflow properties correlate more directly with the surface densities of
their candidate driving cores remains an interesting question beyond the
scope of the present work.

\subsubsection{Virial state and accretion activity}
\label{virial_state}

In addition to the dependence on clump surface density, the outflow properties also show a tendency to vary with the virial state of the parent clumps. The virial parameter, $\alpha_{\rm vir}$, is widely used as an indicator of the relative importance of self-gravity and internal support, although its physical interpretation depends on assumptions regarding the density structure, cloud geometry, magnetic fields, external pressure, and the origin of the observed motions \citep[e.g.][]{Ballesteros-Paredes2006,Kauffmann2013,Traficante2018c}. In particular, several theoretical studies have shown that collapsing molecular clouds do not necessarily exhibit very low virial parameters, since gravitational contraction itself contributes to the observed kinetic energy and may maintain $\alpha_{\rm vir}\sim1$ \citep[e.g.][]{Ballesteros-Paredes2011}. Nevertheless, values well below unity are generally interpreted as indicating that self-gravity plays an increasingly important role in the cloud dynamics \citep[e.g.][]{Tan2014}. The virial parameters adopted here were derived by \citet{Traficante2023} from the physical properties of the SQUALO clumps.

Following the approach adopted in Sect.~\ref{surfacedensity}, we consider only the outflow with the highest inferred accretion rate in each clump to provide a one-to-one comparison with the parent-clump properties. 

\begin{figure}
  \includegraphics[width=1.1\columnwidth]{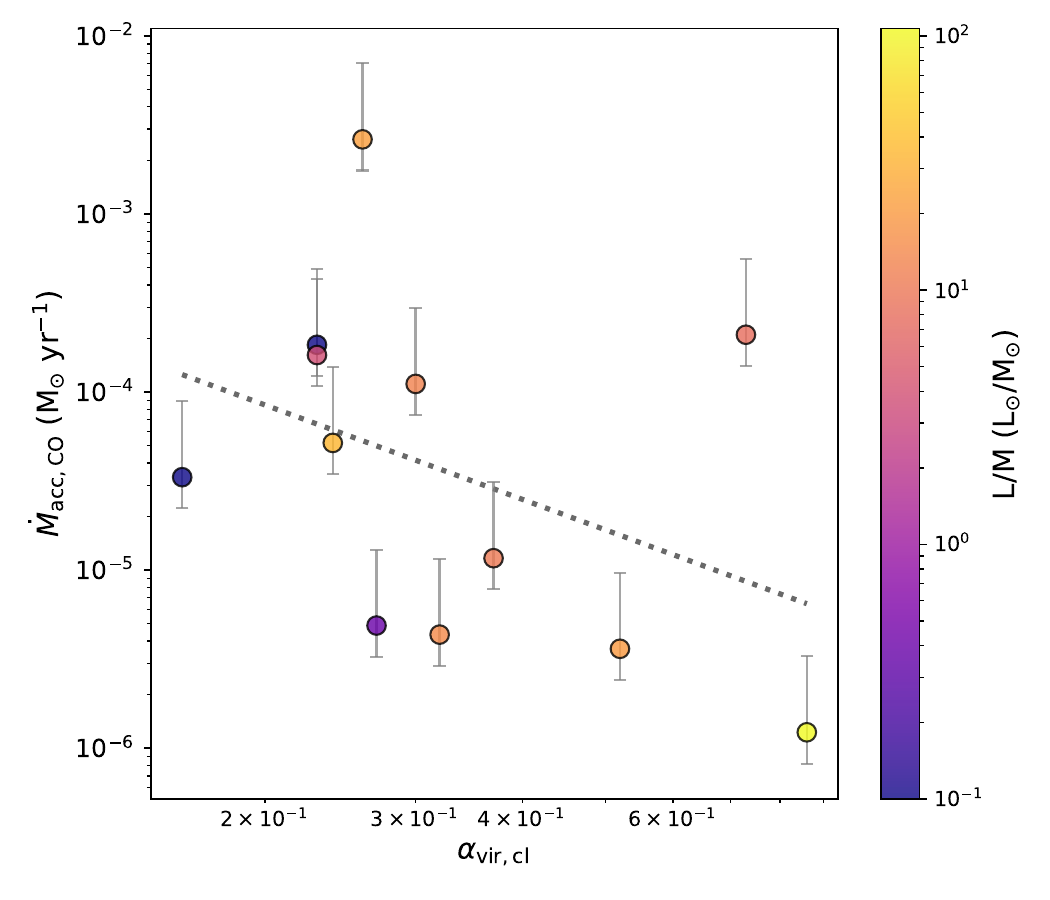}
    \caption{Inferred accretion rate ($\dot{M}_{\rm acc,CO}$) as a
function of the clump virial parameter ($\alpha_{\rm vir}$) for the
12 representative outflows. Colours indicate the clump $L/M$ ratio.
The dashed line shows the best-fitting relation. The tentative
negative tendency is not statistically significant
($r=-0.397$, $p=0.201$, $N=12$).}
    \label{fig:Virial_Macc}
\end{figure}

Figure ~\ref{fig:Virial_Macc} presents the inferred accretion rate as a function of the
clump virial parameter. A two-sided Pearson analysis gives a weak-to-
moderate negative coefficient, but the relation is not statistically
significant ($r=-0.397$, $p=0.201$, $N=12$). The corresponding fit
has $R^2=0.16$. The data therefore show only a tentative tendency for
clumps with lower virial parameters to host outflows with higher
inferred accretion rates. Considerable scatter is present, and similar
qualitative tendencies in the other outflow properties have not been
assigned formal statistical significance.

If lower values of the virial parameter are interpreted as indicating
that self-gravity plays a greater dynamical role within the parent
clumps, the tentative tendency in Fig.\ref{fig:Virial_Macc} would be qualitatively
consistent with enhanced protostellar accretion in such environments.
Since molecular outflows are believed to be powered by
accretion-driven winds, increased accretion activity would naturally
be expected to produce larger momentum fluxes and more energetic
outflows. However, because the observed relation is not statistically
significant, the present data do not establish that the virial state
of the parent clump regulates the accretion activity of its embedded
protostellar population.

The observed tendency is also broadly consistent with hierarchical or global-collapse scenarios for massive star formation, in which the parent clump represents a dynamically evolving structure that channels material from parsec scales towards dense cores and embedded protostars \citep[e.g.][]{Ballesteros-Paredes2011,Peretto2013,Motte2018}. However, the interpretation of the virial parameter within these models is not unique, and collapsing regions may still exhibit virial parameters close to unity. The present results should therefore not be interpreted as demonstrating that lower values of $\alpha_{\rm vir}$ directly trace more rapid gravitational collapse. Rather, the data may indicate a tentative association between outflow
activity and the dynamical state of the parent clump, subject to the
assumptions involved in interpreting the virial parameter.

Given the modest sample size and the substantial scatter in Fig.~\ref{fig:Virial_Macc}, the observed relation should be regarded as suggestive rather than conclusive. The virial parameter is unlikely to be the sole quantity regulating outflow activity, with evolutionary stage, multiplicity, local environmental conditions, magnetic support, and uncertainties in both the outflow and clump properties likely contributing to the observed dispersion.

\subsubsection{Outflows, embedded cores, and the larger-scale environment}
\label{outflow_environment}

The detected molecular outflows are spatially associated with
continuum cores embedded within the infrared-dark structures of
the SQUALO clumps. In several cases, the candidate driving sources
are more clearly associated with compact ALMAGAL cores than with
the lower-resolution SQUALO fragments. For example, in
HIGALBM331.1314$-$0.2438, the northern bipolar outflow and an
additional blue-shifted monopolar component can be associated with
compact ALMAGAL cores located away from the SQUALO core positions
(Fig.~\ref{CO_MAPS}). This illustrates the importance of
high-angular-resolution continuum observations for identifying
candidate driving sources in crowded massive star-forming regions.

Many of these cores are projected against infrared-dark filaments or
elongated dust structures. This demonstrates that active protostellar
accretion is occurring while the sources remain embedded within their
larger-scale natal environment, but it does not by itself establish
that material is transferred continuously from the observed filaments
to the individual protostars. The commonly expected perpendicularity
applies most directly between an outflow axis and its launching disc
plane, and need not extend to the spine of a parsec-scale parent
filament. The angular-momentum direction may change between filament,
core, envelope, streamer, and disc scales, particularly in fragmented
and dynamically complex high-mass environments. In nearby low-mass
systems, spatially and kinematically resolved streamers can sometimes
be traced toward disc scales \citep[e.g.][]{Pineda2020,Garufi2022},
whereas the structures considered here are observed on substantially
larger scales and cannot be connected directly to unresolved
protostellar discs. Because the identified lobes extend to high velocity offsets and
remain coherent across successive channels, their highest-velocity
emission is unlikely to originate from the ambient filamentary
structure; nevertheless, residual contamination near the low-velocity
boundaries cannot be excluded.

The moderate positive relation between the clump-scale infall rate
and the summed outflow-inferred protostellar accretion rate
(Sect.~\ref{clump_infall}; Fig.~\ref{Minf_Macc}) is consistent with
the possibility that the larger-scale gas reservoir contributes to
the ongoing accretion activity of the embedded protostars. However,
the relation is only tentative ($r=0.55$, $p=0.064$, $N=12$), and the
two quantities probe different spatial scales and characteristic
timescales. It therefore does not demonstrate continuous mass
transfer from filaments to individual protostars. The individual
sources showing apparent projected correspondences between CO lobes
and infrared-dark structures are discussed separately in
Sect.~\ref{outflow_filament_interactions}, where possible ambient
contamination and the limitations of a morphological interpretation
are considered.

\subsubsection{Clump-scale infall and protostellar accretion}
\label{clump_infall}
To investigate whether the accretion activity traced by the molecular outflows remains connected to the large-scale collapse of the parent clumps, we compared the cumulative outflow-derived protostellar accretion rate within each clump,
$\sum \dot{M}_{\rm acc}$,
with the clump-scale infall rate,
$\dot{M}_{\rm cl}$. For each clump, we calculated $\sum \dot{M}_{\rm acc}$ by summing the outflow-inferred accretion rates of all identified outflows associated with that clump. We adopted the summed rate because the clump-scale infall rate is compared with the collective accretion activity of the embedded protostellar population.

The clump-scale infall rates were taken from \citet{Traficante2017,Traficante2018a}, where they were derived from blue-asymmetric molecular line profiles tracing large-scale collapse. 

\begin{figure}
  \includegraphics[width=1.0\columnwidth]{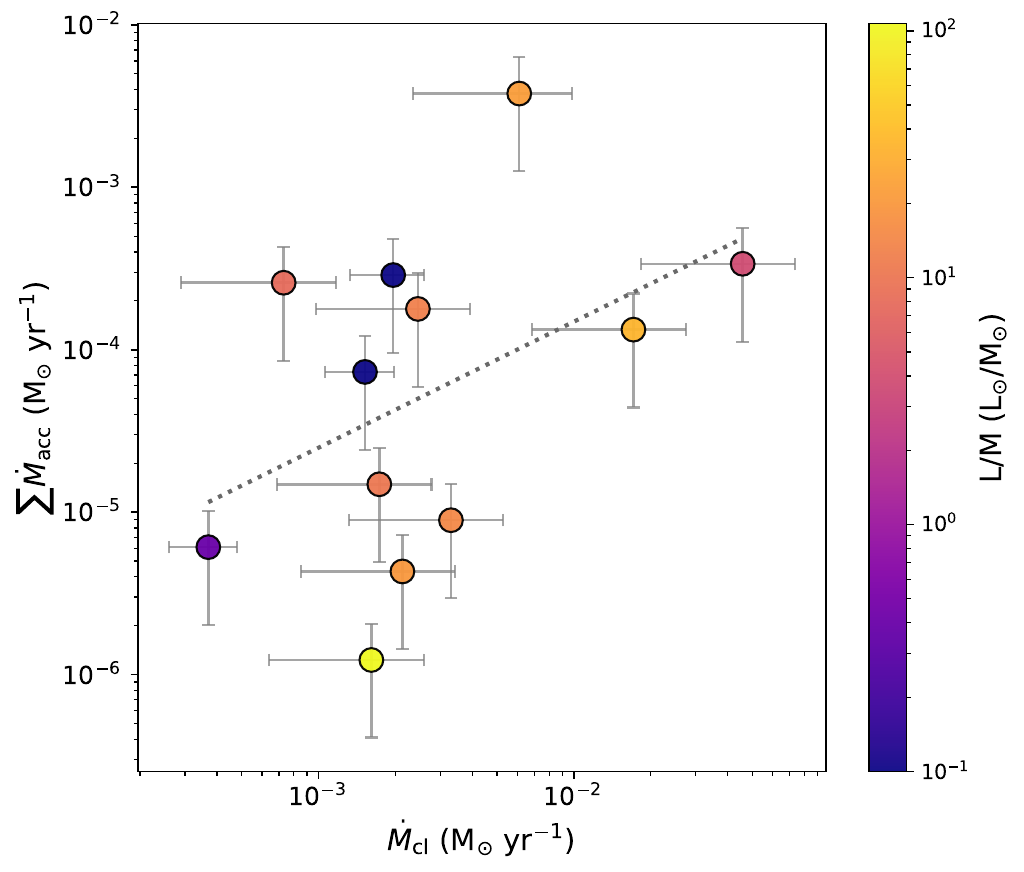}
    \caption{Total protostellar accretion rate inferred from the CO outflows, $\sum \dot{M}_{\rm acc}$, as a function of the clump-scale infall rate, $\dot{M}_{\rm cl}$, for the SQUALO sample. The infall rates are taken from \citet{Traficante2017,Traficante2018a}. Colours indicate the clump luminosity-to-mass ratio ($L/M$). The dashed line shows the best-fitting power-law relation. A moderate positive Pearson correlation is obtained, but it does not
meet the adopted threshold for statistical significance
($r=0.55$, $p=0.064$, $N=12$).
}
    \label{Minf_Macc}
\end{figure}

While $\dot{M}_{\rm cl}$ characterises the rate at which material is
transported into the clump on parsec scales,
$\sum\dot{M}_{\rm acc}$ traces the cumulative accretion activity of
the embedded protostellar population inferred from the observed
molecular outflows. Figure~\ref{Minf_Macc} compares the summed
outflow-inferred protostellar accretion rate with the clump-scale
infall rate for the 12 SQUALO clumps with detected outflows.
A moderate positive Pearson correlation is obtained between the two
rates ($r=0.55$, $p=0.064$, $N=12$). However, the relation does not
meet the adopted $p<0.05$ threshold and is therefore regarded as
tentative. The substantial scatter indicates that a larger clump-scale
infall rate does not translate directly into a proportionally larger
outflow-inferred protostellar accretion rate.

The nominal summed outflow-inferred accretion rate is lower than
the clump-scale infall rate in every SQUALO clump with detected
outflows. The ratio
$\sum\dot{M}_{\rm acc}/\dot{M}_{\rm cl}$ ranges from approximately
$7\times10^{-4}$ to 0.32, with a median of
$7.6\times10^{-3}$. Thus, the clump-scale infall rates exceed the
summed outflow-inferred protostellar accretion rates by factors of
approximately 3--1400, with a median factor of about 135.

This difference should not be interpreted as an equivalent excess
mass-deposition rate onto the identified cores. The clump-scale infall
rate measures the transport of material into the clump on parsec
scales, whereas $\sum\dot{M}_{\rm acc}$ represents the time-averaged
accretion activity traced by the currently detected molecular
outflows. The two quantities therefore probe different spatial scales
and characteristic timescales. Infalling material may remain
temporarily within the clump, filaments, dense cores, or
circumstellar envelopes; it may be distributed among unresolved or
outflow-undetected protostars; or it may be dispersed or otherwise
fail to reach a protostar. In addition, both rate estimates carry
substantial systematic uncertainties. The measured ratios therefore
indicate that only a fraction of the mass flux inferred on clump
scales is represented by the currently detected outflow-driving
protostars, rather than directly measuring the accumulation of excess
mass on individual cores.

\subsubsection{Candidate outflow--filament interactions in individual sources}
\label{outflow_filament_interactions}
\begin{figure*}
\centering
\includegraphics[width=16cm]{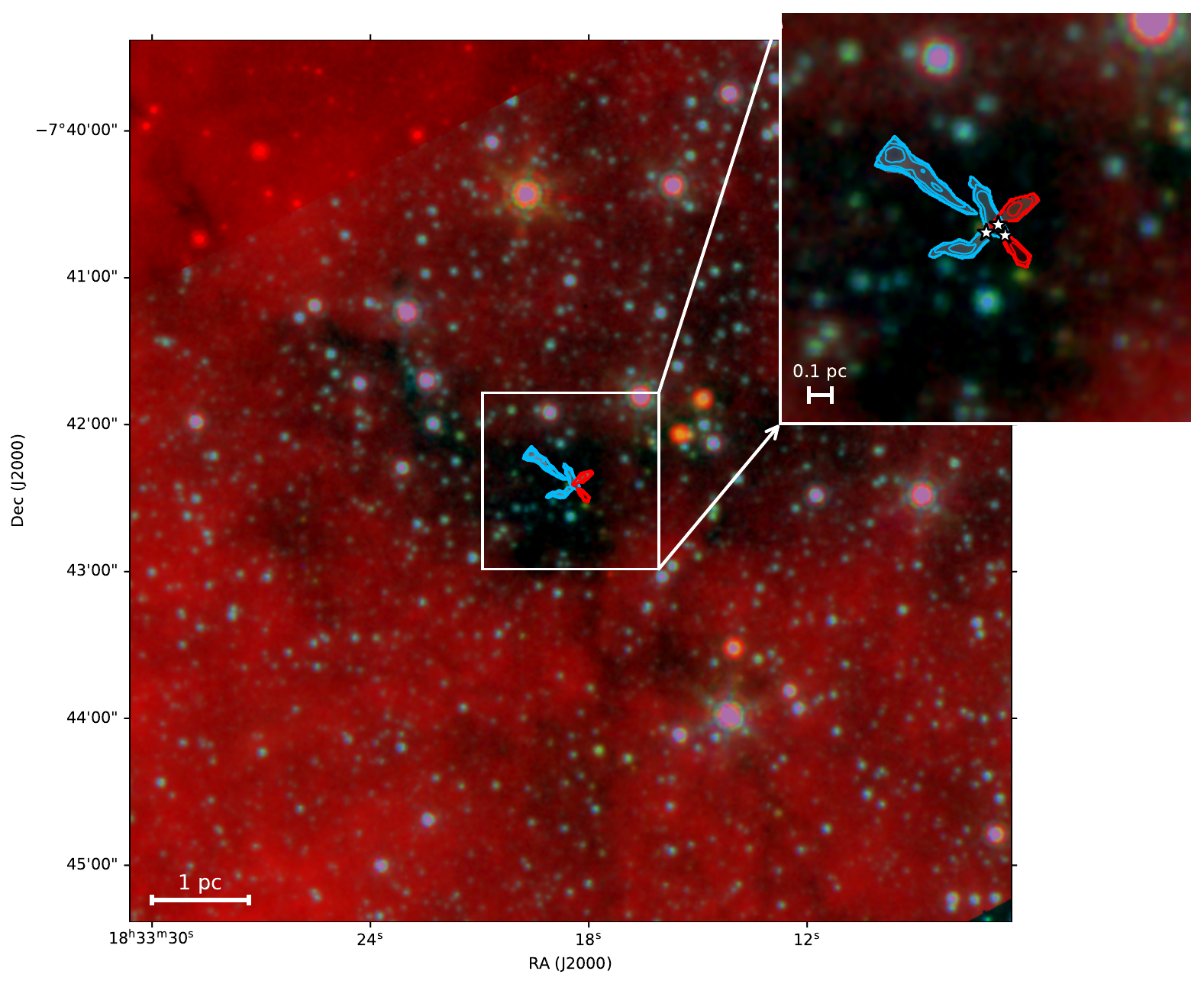}
\caption{The large panel shows a three-color \textit{Spitzer} image constructed from the IRAC 8.0 $\mu$m (red), 5.8 $\mu$m (green), and 4.5 $\mu$m (blue) bands towards HIGALBM24.0116+0.4897. The white box marks the region shown in the zoomed inset. Blue- and red-shifted CO (2--1) outflow emission, integrated within velocity ranges defined for each lobe and masked using polygon apertures enclosing significant emission above a $5\sigma$ threshold, are overlaid as blue and red contours, respectively. The semi-transparent blue and red shaded regions indicate the corresponding integrated intensity distributions. White star symbols mark the positions of compact continuum fragments identified from the high-resolution continuum observations. The inset highlights the immediate vicinity of the protostellar system, revealing multiple outflow lobes and their spatial relationship to the embedded fragments. Black arrows connect the large-scale view to the zoomed region. Scale bars corresponding to 1 pc and 0.1 pc are shown in the lower-left corners of the main panel and inset, respectively. $V_{\rm LSR}=94.8$~km~s$^{-1}$; the extended northern blue-shifted
component is integrated over 60.36--88.30~km~s$^{-1}$.}
\label{Spitzer_ALMA_24}
\end{figure*}

The three-colour \textit{Spitzer} images presented in Figs.~\ref{Spitzer_ALMA_31}, \ref{Spitzer_ALMA_24}, and \ref{Spitzer_ALMA_343} reveal several sources in which one of the CO outflow lobes closely follows the morphology of the surrounding infrared-dark filamentary structures. Similar projected configurations have been reported by \citet{Avison2021}, who identified outflows aligned with large-scale accretion filaments in the massive star-forming region SDC335. Although projected morphology alone cannot demonstrate a physical interaction, the close correspondence between the high-velocity CO emission and the infrared filaments makes these sources particularly interesting candidates for investigating the relationship between protostellar outflows and filamentary environments.

The clearest example is HIGALBM24.0116+0.4897 (Fig. \ref{Spitzer_ALMA_24}),
where the extended blue-shifted lobe, integrated between 60.36
and 88.30~km~s$^{-1}$, overlaps in projection with the northern
infrared-dark filament over much of its extent. Relative to the
systemic velocity, $V_{\rm LSR}=+94.8$~km~s$^{-1}$, this interval
corresponds to velocity offsets of approximately $-34$ to
$-6.5$~km~s$^{-1}$. Although the inner boundary lies relatively
close to $V_{\rm LSR}$, it was not selected using a fixed minimum
velocity offset. Inspection of the channel maps showed that, at
velocities of $\leq88.30$~km~s$^{-1}$, the emission becomes
spatially distinguishable from the extended ambient-cloud
component, remains coherent across successive channels, and is
morphologically connected to the higher-velocity emission
associated with the candidate driving source. Channels closer to
$V_{\rm LSR}$ were excluded because their emission is dominated
by the ambient cloud. Nevertheless, because velocity gradients
of comparable magnitude can occur in parsec-scale structures,
some residual contamination at the low-velocity boundary cannot
be excluded. The apparent correspondence with the filament is
therefore treated as a projected morphological association rather
than evidence of a physical outflow-filament interaction.

A similar configuration is observed towards HIGALBM31.9462+0.0759 (Fig.~\ref{Spitzer_ALMA_31}), although in this case the red-shifted lobe exhibits the strongest correspondence with the filamentary structure. The red-shifted emission, integrated between 103.0 and 128.4 km\,s$^{-1}$ ($+8$ to $+34$ km\,s$^{-1}$ relative to $V_{\rm LSR}=+94.7$ km\,s$^{-1}$), extends along the prominent infrared-dark filament emerging from the central protostellar system. Its morphology resembles the filament-aligned outflow components reported by \citet{Avison2021}, making it another interesting candidate for future studies of possible outflow--filament interactions.

HIGALBM343.7560$-$0.1629 (Fig.~\ref{Spitzer_ALMA_343}) is particularly intriguing. This source was identified by \citet{Traficante2023} as a peculiar object because it remains unfragmented despite containing a massive ($\sim73\,M_\odot$) continuum core, leading those authors to suggest that an additional support mechanism, such as magnetic fields, may suppress fragmentation. Our CO observations reveal multiple outflow lobes originating
from this dominant core. Relative to the systemic velocity of
$V_{\rm LSR}=-26.19$~km~s$^{-1}$, the northern blue-shifted
component integrated between $-51.48$ and
$-36.24$~km~s$^{-1}$ corresponds to velocity offsets of
approximately $-25.3$ to $-10.1$~km~s$^{-1}$ and shows an
apparent spatial correspondence in projection with the
infrared-dark filament north of the source. Although this alignment alone cannot establish a physical interaction, it identifies HIGALBM343.7560$-$0.1629 as another promising target for future investigations of the interplay between protostellar feedback and filamentary structures.

In summary, the three sources presented in Figs.~\ref{Spitzer_ALMA_31}, \ref{Spitzer_ALMA_24}, and \ref{Spitzer_ALMA_343} exhibit unusually close morphological correspondences between their largest CO outflow lobes and the surrounding infrared-dark filaments. While these projected alignments do not by themselves demonstrate a physical interaction, they identify promising targets for future studies of the interplay between protostellar feedback and filamentary accretion. High-resolution observations of dense-gas tracers and shock diagnostics, together with detailed kinematic analyses of the filaments, will be required to determine whether these systems represent genuine outflow--filament interactions or chance superpositions along the line of sight. Additional RGB maps for the remaining sources in the sample are provided in the online supplementary material.

\subsection{Evolution of outflow properties along the SQUALO sequence}

\subsubsection{Outflow extent as a function of evolutionary stage}

To investigate how the outflow properties evolve across the SQUALO sequence, we compared the clump luminosity-to-mass ratios ($L/M$) derived by \citet{Traficante2023} with the properties of the associated molecular outflows.

\begin{figure}
    \centering
    \includegraphics[width=1.0\columnwidth]{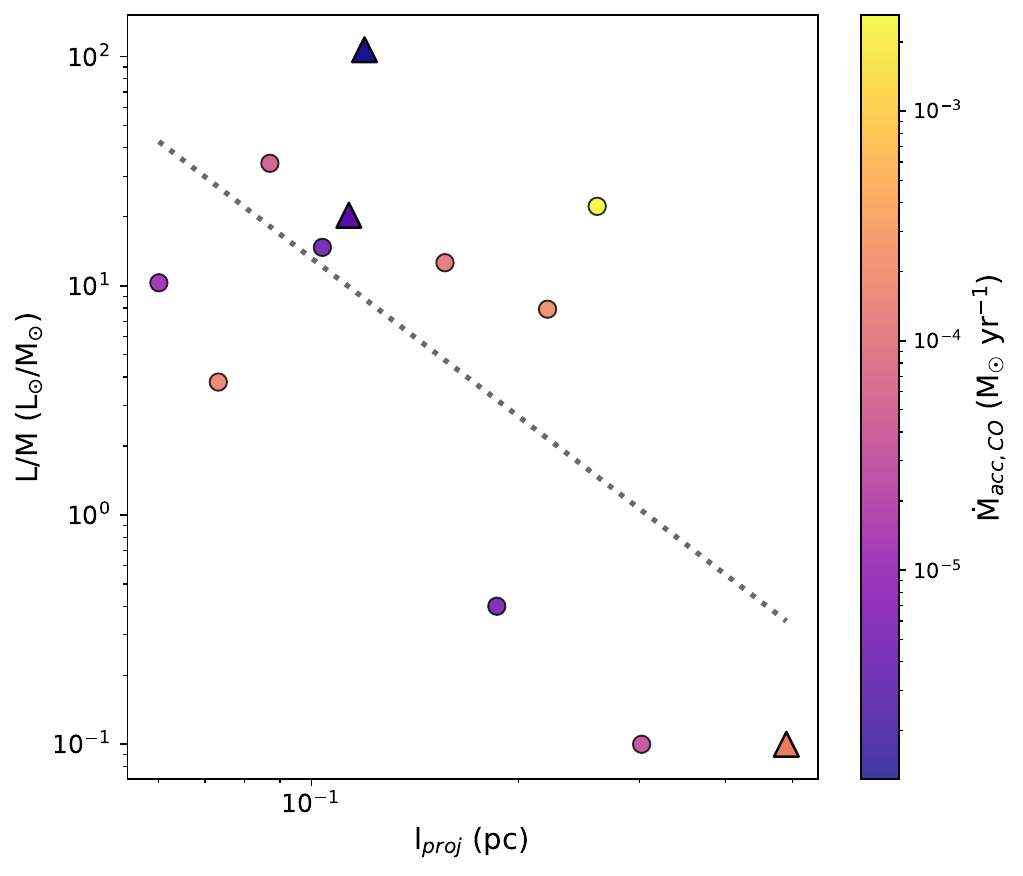}
    \caption{Relation between the luminosity-to-mass ratio ($L/M$) and the projected outflow length ($l_{\rm proj}$) for the representative outflow (the outflow with the highest inferred accretion rate) in each SQUALO clump. The $L/M$ values are taken from \citet{Traficante2023}. Colours indicate the inferred accretion rate derived from the CO outflow properties. The triangular symbols indicate lobes whose projected lengths,
$l_{\rm proj}$, are lower limits because their CO emission extends beyond
the region of reliable primary-beam sensitivity. The dashed line shows the best-fitting power-law relation. A two-sided Pearson analysis in logarithmic space gives
$r=-0.630$, $p=0.028$, and $N=12$.}
    \label{fig:LM-l}
\end{figure}

Figure~\ref{fig:LM-l} shows the relation between $L/M$ and the projected outflow length, $l_{\rm proj}$, using the representative outflow adopted throughout this work, namely the outflow with the highest inferred accretion rate in each SQUALO clump. A two-sided Pearson analysis in logarithmic space reveals a moderate
negative correlation between the two quantities ($r=-0.630$,
$p=0.028$, $N=12$). The corresponding power-law fit has
$R^2=0.40$. The anticorrelation is therefore nominally statistically
significant at the adopted $p<0.05$ threshold, although substantial
scatter is present. The lobes whose emission extends beyond the region
of reliable primary-beam sensitivity are shown with triangular
symbols in Fig. \ref{fig:LM-l}, indicating that their measured $l_{\rm proj}$
values are lower limits.

The interpretation of this anticorrelation is not unique. Although $L/M$ is frequently used as an approximate evolutionary indicator, it also depends on the mass and luminosity of the most luminous embedded source, the sampling of the protostellar population, dust temperature, and the adopted clump mass. In particular, a high $L/M$
value may reflect the presence of a more massive central object rather than evolutionary stage alone. The projected outflow length is also affected by inclination, source distance, sensitivity, confusion with ambient emission, interaction with the surrounding cloud, and the
finite map coverage. Indeed, several of the measured lengths are lower limits because the CO emission extends beyond the region of reliable
primary-beam sensitivity.

Consequently, the observed anticorrelation cannot by itself
demonstrate that younger clumps intrinsically drive more extended
outflows. Nor do the present data establish that outflows become
shorter because more evolved environments are increasingly disrupted
by feedback. We therefore regard the relation as an empirical
association between $L/M$ and projected outflow length within the
SQUALO sample, rather than evidence for a unique evolutionary
sequence. A larger sample, together with independent constraints on
protostellar masses, evolutionary stages, inclination angles, and
outflow ages, would be required to determine its physical origin.

When all identified outflows within each clump are displayed rather
than only the representative outflow, the distribution shows
considerably greater scatter and no comparably clear monotonic
relation. We therefore do not assign formal statistical significance
to this expanded, non-independent sample.

\subsubsection{Connection with fragmentation properties from SQUALO I}

One of the primary goals of the SQUALO project is to investigate how the internal structure of massive clumps evolves alongside their accretion activity. In SQUALO I, \citet{Traficante2023} analysed the 1.3 mm continuum emission and characterised the fragmentation properties of the sample, deriving the masses, sizes, and spatial distribution of the embedded cores. Their analysis revealed systematic variations in the characteristic fragment spacing across the SQUALO evolutionary sequence, with more evolved clumps generally exhibiting smaller separations between fragments.

When considered together with the outflow properties derived in this work, an interesting evolutionary picture emerges. When considered together with the outflow properties derived in this
work, the fragmentation analysis reveals a tentative correspondence:
clumps with lower $L/M$ values tend to show both larger characteristic
fragment separations and longer projected representative outflows,
whereas clumps with higher $L/M$ values tend to show smaller fragment
separations and shorter projected outflows. However, we do not
interpret this correspondence as a direct evolutionary sequence.
Because $L/M$ depends on both evolutionary state and the mass and
luminosity of the embedded stellar population, the observed behaviour
may partly reflect differences in source content among the clumps.
Furthermore, projected outflow lengths are affected by inclination,
sensitivity, confusion, and map coverage.

The fragmentation and outflow measurements therefore describe
different aspects of the internal structure and protostellar activity
of the SQUALO clumps, but the present sample does not establish a
causal connection between them. In particular, we cannot determine
whether the apparent correspondence arises from evolution, differences
in the masses of the dominant embedded sources, environmental
conditions, or observational effects. A homogeneous comparison using
independent evolutionary diagnostics and direct estimates of
protostellar masses will be required to distinguish among these
possibilities.

Jointly, the fragmentation properties derived in SQUALO I and the molecular outflow analysis presented here support a hierarchical picture of massive star formation. The fragmentation of the clump, the growth of the embedded core population, and the evolution of the associated molecular outflows appear to be interconnected manifestations of the same underlying process, namely the continued transfer of mass from larger than clump scales to protostellar scales. In this framework, molecular outflows provide a direct probe of ongoing accretion, while the fragmentation properties reveal how the available mass reservoir is distributed among the forming stellar population.

\subsection{Comparison with literature scaling relations}
\label{comparison}
\subsubsection{Outflow force versus clump luminosity}
\label{FcoVSL}
The outflow momentum flux, $F_{\rm CO}$, is commonly regarded as one of the most robust tracers of the underlying protostellar accretion activity because it is less sensitive than other outflow parameters to uncertainties in the dynamical timescale \citep[e.g.,][]{Bontemps1996, Duarte-Cabral2013}. Numerous studies have reported a positive correlation between outflow force and source luminosity over a wide range of stellar masses, suggesting a fundamental link between the accretion process and the feedback generated by molecular outflows \citep[e.g.,][]{Beuther2002a, Duarte-Cabral2013, Maud2015, Yang2018}.

\begin{figure*}
\centering
\includegraphics[width=16cm]{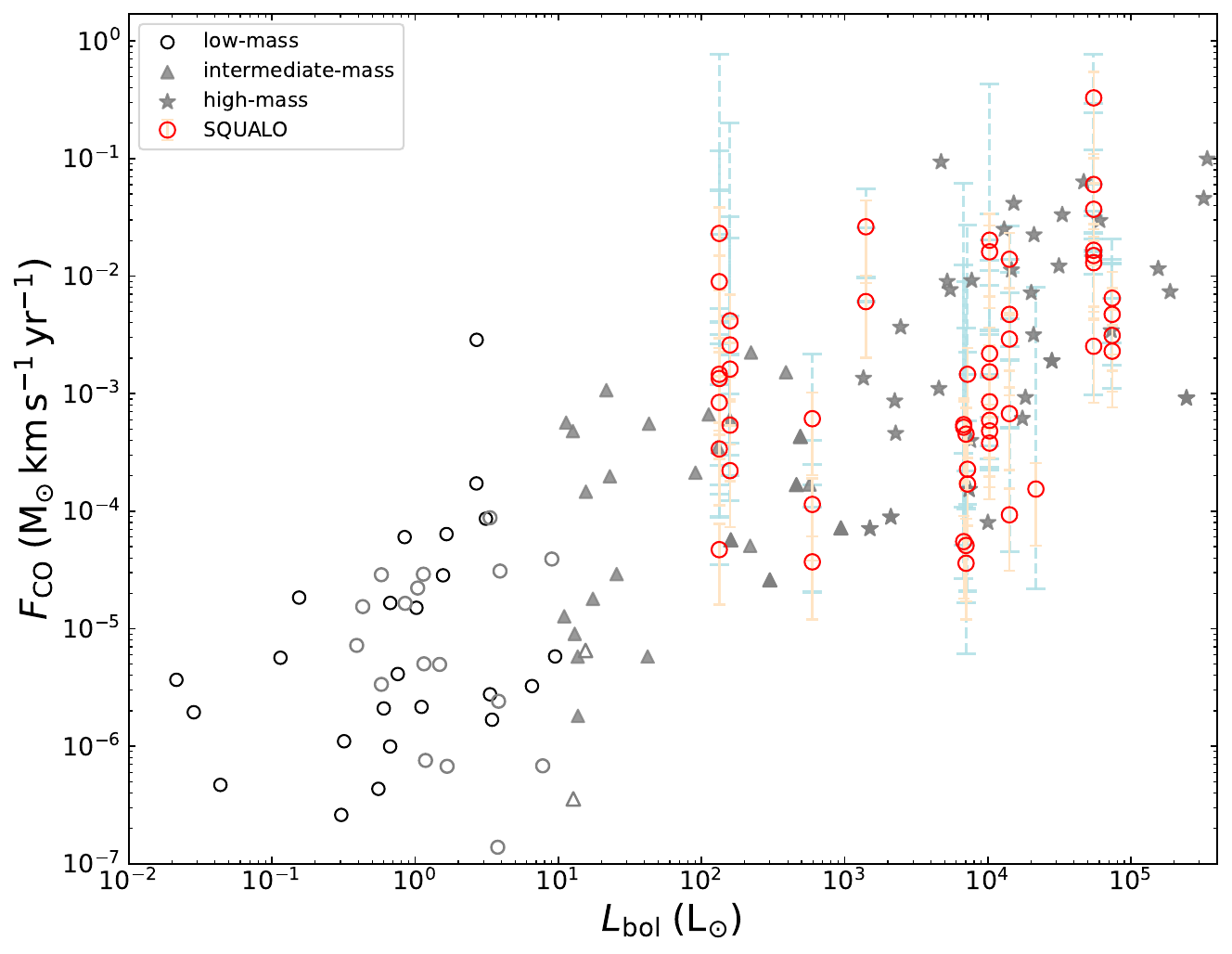}
\caption{Outflow momentum flux ($F_{\rm CO}$) as a function of clump bolometric luminosity ($L_{\rm clump}$). The red open circles show the SQUALO outflows identified in this work. The blue error bars illustrate the range that the derived values could span if the geometrically estimated inclination corrections were applied, while the orange error bars represent the uncertainty associated with the adopted excitation-temperature range of $T_{\rm ex}=20$--$80$ K. The inclination corrections are shown only to illustrate their potential systematic effect and are not applied to the values adopted in the main analysis. The literature samples include low-, intermediate-, and high-mass protostellar systems cited in the text.}
\label{Fco_L}
\end{figure*}

Figure \ref{Fco_L} compares the outflow momentum fluxes derived from CO emission for the SQUALO sample with the compilation presented by \citet{Avison2021}, which includes low-, intermediate-, and high-mass protostellar systems spanning several orders of magnitude in luminosity. The comparison sample combines literature measurements of low-mass protostars \citep{Bontemps1996, vanderMarel2013}, high-mass Class 0 analogues \citep{Duarte-Cabral2013}, and massive protostellar objects \citep{Maud2015}, together with the outflows identified in the SDC335 protocluster \citep{Avison2021}. We have updated this comparison by including two more recent CO outflow
studies. At low luminosities, we include 19 protostellar systems from the
Orion~A sample of \citet{Hsieh2023}, observed in
$^{12}\mathrm{CO}(2\mbox{--}1)$. Although their complete sample contains
21 targets, HOPS-200 and HOPS-408 were excluded because reliable momentum
rates were not reported. For the remaining targets, we adopted a
representative momentum rate obtained by averaging the summed red- and
blueshifted, inclination-corrected rates over the available radial
annuli. We also include the ten Cygnus-X massive protostars studied by
\citet{Skretas2022} using
$^{12}\mathrm{CO}(3\mbox{--}2)$ observations.

Using this compilation provides a broad reference framework for placing the SQUALO sources in the broader context of protostellar outflow activity. The SQUALO sources occupy the intermediate- to high-luminosity regime of the distribution and occupy the region of parameter space defined by the broad increase
of outflow momentum flux with luminosity seen in the heterogeneous
literature compilation of increasing outflow momentum flux with increasing clump luminosity established by previous studies. Although the comparison sample spans a range of source masses, evolutionary stages, and observing strategies, the SQUALO measurements lie within the scatter defined by the literature, indicating that the outflow forces derived here are consistent with those observed in other massive star-forming regions.

The agreement with the compiled literature sample suggests that the molecular outflows identified in the SQUALO clumps are representative of those associated with massive protostellar systems. In particular, the overlap with the high-mass sources in the compilation demonstrates that the outflow momentum fluxes measured in SQUALO are comparable to those found in actively accreting clustered environments. At the same time, the SQUALO survey extends this comparison to a sample of clumps selected specifically on the basis of large-scale infall signatures, providing a unique opportunity to investigate molecular outflow activity in a uniformly selected population of globally collapsing massive clumps.

Thus, while the SQUALO outflows are consistent with established
outflow-force--luminosity scaling relations, they extend those
comparisons to a uniformly selected sample of clumps exhibiting
large-scale infall.

\subsubsection{Accretion rate versus clump luminosity}

\begin{figure*}
\centering
\includegraphics[width=16cm]{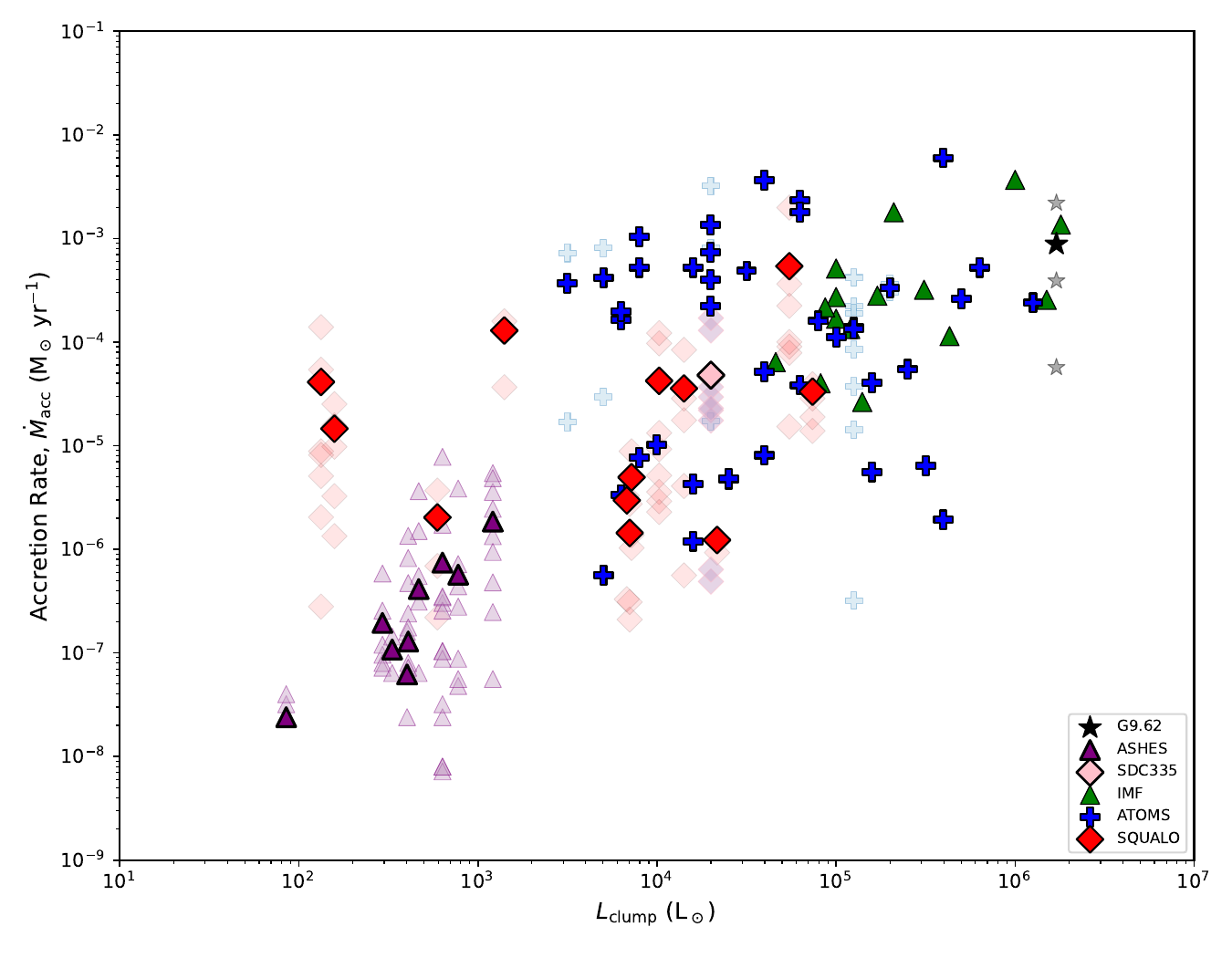}
\caption{Accretion rate inferred from the molecular outflows
($\dot{M}_{\rm acc}$) as a function of clump bolometric luminosity
($L_{\rm clump}$). The pale cyan diamonds show the individual SQUALO
outflows identified in this work, while the larger darker symbols show
the mean value for each SQUALO clump. The comparison samples
include G9.62+0.19 \citep{Liu2017}, ALMA-ASHES \citep{Li2020}, SDC335 \citep{Avison2021}, ALMA-IMF \citep{Towner2024}, and ALMA-ATOMS \citep{Hoque2025}. The SQUALO sources occupy an intermediate region of parameter space between the infrared-dark ALMA-ASHES clumps and the more luminous ALMA-IMF regions, while exhibiting accretion rates comparable to those observed in actively forming massive protostars. For consistency, the literature accretion rates were recalculated from the published outflow momentum fluxes using the same conversion adopted in this work.}
\label{Macc_L}
\end{figure*}

The accretion rate is one of the key quantities governing the growth of massive protostars and is expected to be closely linked to the luminosity of the parent star-forming region. Figure \ref{Macc_L} compares the accretion rates inferred from the SQUALO molecular outflows with those reported for a range of intermediate- and high-mass protostellar systems from \citet{Liu2017, Li2020, Avison2021, Towner2024, Hoque2025}.

A positive trend between accretion rate and clump luminosity is observed across the combined dataset, consistent with the general expectation that more luminous regions host more actively accreting protostellar populations. The SQUALO sources occupy an intermediate regime between the infrared-dark ALMA-ASHES clumps and the more luminous ALMA-IMF regions, extending the observed relation across a broad range of luminosities. A broadly similar trend is also observed for the CO outflow force (Sect. \ref{FcoVSL}). Although the SQUALO accretion rates are inferred exclusively from the CO ($J=2$--1) outflows presented in this work, the literature compilation shown in Fig.~\ref{Macc_L} includes values derived from both CO- and SiO-based outflow studies. To facilitate a homogeneous comparison, the literature accretion rates were recalculated from the published outflow momentum fluxes using Eq.~(\ref{Macc_equ}), adopting the same assumptions for the entrainment efficiency, jet-to-outflow velocity ratio, and jet mass-loss efficiency as used throughout this work. Consequently, the comparison is not simply a rescaling of a uniform CO-only dataset, but instead places the SQUALO results within the broader context of outflow studies employing different molecular tracers. The broadly similar behaviour obtained with the two tracers suggests that the broad qualitative trend between accretion activity and clump
luminosity is not driven solely by the choice of outflow tracer.

Among the comparison samples, the ALMA-IMF field G338.93 corresponds to the SQUALO clump HIGALBM338.9260+0.6340. However, a direct comparison of their individual outflow parameters is not possible. The SQUALO measurements are based on CO (2--1) observations combining the 12-m, 7-m, and Total Power arrays, whereas \citet{Towner2024} identified outflow candidates using 12-m SiO (5--4) data without systematically associating them with individual driving sources. Consequently, differences in molecular tracer, flux recovery, abundance assumptions, velocity selection, and dynamical-timescale definition may affect the derived outflow parameters. Although the literature accretion rates were recalculated using the same conversion from momentum flux, the underlying outflow masses, momenta, dynamical timescales, and momentum fluxes remain methodologically heterogeneous. The comparison should therefore be regarded as qualitative.

At a given luminosity, however, several SQUALO sources exhibit accretion rates that are systematically higher than the average values reported for the ALMA-ASHES sample. This offset is particularly evident in the luminosity range $L_{\rm clump}\sim10^2$--$10^4 L_\odot$, where many of the SQUALO clumps occupy the upper envelope of the distribution. A likely explanation is related to the source-selection criteria of the SQUALO survey. Unlike the ALMA-ASHES sample, which targeted infrared-dark clumps representative of an earlier evolutionary stage, the SQUALO sample was specifically selected to exhibit clear signatures of ongoing clump-scale infall \citep{Traficante2017,Traficante2023}. The elevated accretion rates inferred from the outflows are therefore consistent with the presence of particularly active mass assembly within these systems.

It is important to note, however, that observational methodology may also contribute to part of the observed offset. The SQUALO observations combine the 12-m, 7-m, and Total Power arrays, enabling recovery of extended outflow emission that may be partially filtered out in interferometer-only datasets. Consequently, part of the elevated outflow momentum fluxes and inferred accretion rates relative to some literature samples may reflect differences in flux recovery in addition to intrinsic physical differences between the source populations. Nevertheless, the broad qualitative increase of accretion rate with
clump luminosity is consistent with the behaviour seen across the
comparison samples. Because these measurements are methodologically
heterogeneous, we do not assign a formal correlation coefficient or
statistical significance to the combined dataset.

The uncertainties in the SQUALO measurements are dominated by the adopted excitation temperature and, particularly, the unknown outflow inclination. Despite these systematic uncertainties, the inferred accretion rates broadly overlap those reported for massive protostars in other high-angular-resolution studies.

Consequently, the results presented in Figs. \ref{Fco_L} and \ref{Macc_L} demonstrate that the SQUALO clumps follow the same general scaling relations observed in other massive star-forming regions while preferentially populating the upper range of accretion activity at a given luminosity. 

\subsubsection{Distribution of accretion among embedded driving sources}
\label{sec:acc_partition}

While the previous sections examined how the global outflow properties correlate with the characteristics of their parent clumps, the presence of multiple embedded protostellar sources within most SQUALO regions also allows us to investigate how the total accretion budget is partitioned among the individual driving sources. This provides a direct measure of whether the outflow-derived accretion is dominated by a single embedded source or distributed more evenly among the protostellar population.

For each candidate driving source, the contributions from all associated molecular outflow systems were combined. The total accretion rate associated with an individual source was therefore calculated as

\begin{equation}
\dot{M}_{{\rm acc},i}
=
\sum_{j}
\dot{M}_{{\rm acc},ij},
\end{equation}

where $i$ denotes the driving source and $j$ represents each molecular outflow system associated with that source. For bipolar outflows, the blue- and red-shifted lobes were combined, while sources associated with multiple outflow systems were assigned the sum of all corresponding outflow-derived accretion rates.

The total accretion rate within each clump is then

\begin{equation}
\dot{M}_{\rm acc,clump}
=
\sum_i
\dot{M}_{{\rm acc},i},
\end{equation}

from which we define the dominant-source accretion fraction as

\begin{equation}
f_{\rm dom}
=
\frac{\dot{M}_{\rm acc,max}}
{\sum_i \dot{M}_{{\rm acc},i}}
=
\frac{\dot{M}_{\rm acc,max}}
{\dot{M}_{\rm acc,clump}},
\label{eq:fdom}
\end{equation}

where $\dot{M}_{\rm acc,max}$ is the largest outflow-derived accretion rate among all candidate driving sources within the clump. The quantity $f_{\rm dom}$ therefore represents the fraction of the total inferred accretion budget contributed by the dominant embedded source.

\begin{figure*}
\centering
\includegraphics[width=16cm]{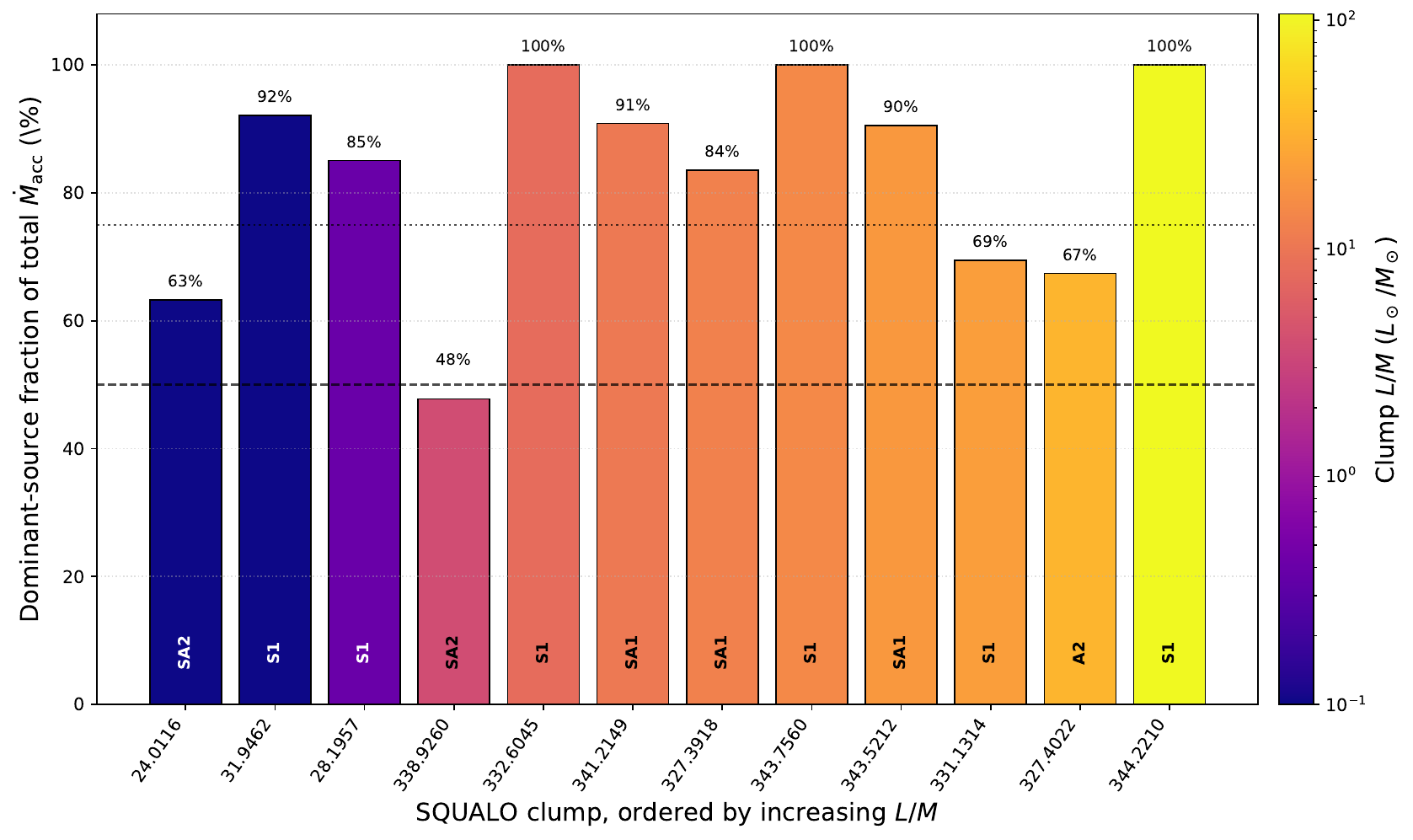}
\caption{Fraction of the total outflow-derived accretion rate within each SQUALO clump contributed by the dominant embedded driving source, defined as $f_{\rm dom}=\dot{M}_{\rm acc,max}/\dot{M}_{\rm acc,clump}$ (Eq.~\ref{eq:fdom}). The clumps are ordered by increasing $L/M$, and the bar colours indicate the corresponding $L/M$ values, shown by the colour bar. The labels within each bar identify the dominant driving source, while the percentages denote the fraction of the total clump accretion budget associated with that source. The horizontal dashed lines at 50\% and 75\% mark equal partition between two dominant sources and a strongly concentrated accretion budget, respectively. Most SQUALO clumps are dominated by a single embedded source contributing more than 60\% of the total inferred accretion rate, although one clump exhibits a substantially more even distribution of the accretion budget among multiple protostars.} 

\label{fig:acc_partition}
\end{figure*}

Figure~\ref{fig:acc_partition} presents the resulting dominant-source accretion fractions for the SQUALO sample, ordered by increasing clump evolutionary indicator ($L/M$). The bars are colour-coded according to the clump $L/M$ ratio, while the labels identify the dominant driving source in each region. The results show that the outflow-derived accretion budget is generally highly concentrated within a single embedded object. In ten of the twelve analysed clumps, the dominant source contributes more than 60\% of the total inferred accretion, while six clumps exhibit dominant-source fractions exceeding 90\%. The only clear exception is HIGALBM338.9260+0.6340, where the dominant source contributes only $\sim$48\% of the total accretion, indicating that the accretion budget is shared more evenly among multiple embedded protostars. No obvious trend is observed between $f_{\rm dom}$ and the evolutionary indicator $L/M$, suggesting that the degree of accretion concentration does not evolve monotonically over the evolutionary range sampled by the SQUALO clumps.

These results indicate that, although massive star-forming clumps commonly host several actively accreting protostars, the global accretion budget inferred from the molecular outflows is frequently dominated by a single embedded source. Such behaviour is consistent with the expectation that one source may dominate the observable outflow activity within a clustered environment, while neighbouring protostars continue to accrete at lower rates. However, the present analysis alone does not distinguish between competing massive star formation scenarios, since a dominant accreting source could arise in both competitive accretion and core-fed models. Instead, the principal result of this analysis is that the partitioning of the outflow-derived accretion budget varies substantially between clumps, ranging from systems in which a single source contributes essentially all of the inferred accretion to regions where the accretion is distributed much more evenly among several embedded protostars.

Several caveats should be considered when interpreting these results. First, the inferred accretion rates are derived from molecular outflow properties and therefore represent time-averaged accretion over the dynamical lifetime of the outflows rather than instantaneous protostellar accretion rates. Secondly, the identification of candidate driving sources is based on the spatial association between the CO outflows and the continuum emission and is therefore subject to uncertainties in crowded regions and projection effects. Finally, the driving sources were identified using both the SQUALO and ALMAGAL continuum catalogues, which employ different angular resolutions and source-identification methodologies. Although the calculation of $f_{\rm dom}$ is independent of the adopted continuum catalogue, a direct comparison between outflow-derived accretion rates and continuum-derived core masses requires a homogeneous core catalogue. Such an analysis will enable us to determine whether the dominant accreting source also corresponds to the most massive embedded core, providing a more direct test of how the clump-scale accretion reservoir is partitioned within clustered massive star-forming regions, and will be presented in a future study.

\section{Conclusions}
\label{5}
We have presented an ALMA study of CO ($J=2$--1) molecular outflows toward the 13 clumps of the SQUALO sample, extending the SQUALO project from the continuum analysis of embedded cores presented in SQUALO I \citep{Traficante2023} to the investigation of accretion and feedback processes traced by molecular outflows. The sample was specifically selected to span a wide range of evolutionary stages while exhibiting clear signatures of clump-scale infall. Our main findings can be summarised as follows:

\begin{enumerate}

\item We identified 51 high-velocity CO outflow lobes, corresponding
to 31 outflow systems, across 12 of the 13 SQUALO clumps. These comprise 15 bipolar, 14 monopolar, and two complex systems. The outflows span a broad range of masses, momenta, energies, and momentum fluxes, with values comparable to those reported in other studies of massive star-forming regions.

\item The outflow-derived accretion rates are typically of order $10^{-5}$--$10^{-3} M_\odot {\rm yr}^{-1}$, consistent with the high accretion rates required for the formation of massive stars. Comparison with literature samples shows that the SQUALO sources follow the same general scaling relations between outflow activity and luminosity observed in other clustered star-forming environments.

\item A nominally significant positive correlation is found
between outflow momentum flux and clump surface density
($r=0.627$, $p=0.029$, $N=12$). In contrast, the tendency for
clumps with lower virial parameters to host higher inferred
accretion rates is not statistically significant
($r=-0.397$, $p=0.201$, $N=12$). The latter should therefore be
regarded as tentative rather than evidence of a firm connection.

\item The summed outflow-inferred protostellar accretion rate increases tentatively with the clump-scale infall rate but remains below the inferred clump inflow in every source. The nominal ratios $\Sigma\dot{M}_{\rm acc}/\dot{M}_{\rm cl}$ range from approximately 7$\times 10^{-4}$ to 0.32, indicating that only a fraction of the clump-scale mass flux is represented by the currently detected protostellar outflows. Because these quantities probe different spatial scales and timescales and are subject to substantial systematic uncertainties, the difference cannot be interpreted directly as mass deposited onto the identified cores.

\item We find a nominally significant anticorrelation between
the clump $L/M$ ratio and the projected length of the representative
outflow ($r=-0.630$, $p=0.028$, $N=12$). However, $L/M$ depends on
both evolutionary state and the mass and luminosity of the embedded
source population, while the projected outflow lengths are affected
by inclination, sensitivity, confusion, and map coverage. We
therefore interpret this result as an empirical association and not
as evidence that younger clumps intrinsically drive more extended
outflows.

\item Comparison with the fragmentation properties derived in
SQUALO I and with higher-resolution ALMAGAL continuum data shows that
the molecular outflows are associated with compact cores embedded
within larger-scale infrared-dark structures. Several outflows are
more clearly associated with compact ALMAGAL cores than with the
lower-resolution SQUALO continuum sources, highlighting the
importance of high-angular-resolution observations for identifying
their candidate driving sources. However, this spatial embedding does
not by itself demonstrate continuous mass transfer from the
large-scale filaments to individual protostars.

\item We quantified how the outflow-derived accretion budget is partitioned among the embedded driving sources within each SQUALO clump. In most cases, a single source dominates the inferred accretion, contributing more than 60\% of the total clump accretion rate, while only one clump exhibits a nearly equal partition of the accretion budget among multiple protostars. This result indicates that, although clustered massive star-forming regions commonly host several actively accreting objects, the observable accretion traced by molecular outflows is frequently concentrated in a single dominant source.

\end{enumerate}

In conclusion, these results support a hierarchical clump-fed scenario of massive star formation in which material is transferred from large-scale filamentary structures to collapsing clumps, from clumps to dense cores, and ultimately onto accreting protostars that power the observed molecular outflows. The close connection between clump-scale infall, protostellar accretion, and outflow feedback suggests that molecular outflows retain signatures of the large-scale mass assembly process throughout the early evolution of massive star-forming regions. 

\begin{table*}
\centering
\caption{Outflow parameters with uncertainties derived from $T_{\rm ex}=20$--80\,K bracket.}
\begin{tabular}{lccccc}
\hline
Name & $M_{\rm out}$ & $P_{\rm CO}$ & $E_{\rm CO}$ & $F_{\rm CO}$ & $\dot{M}_{\rm acc}$ \\
 & ($M_\odot$) & ($M_\odot$ km s$^{-1}$) & ($10^{38}$ J) & ($10^{-5}$ $M_\odot$ km s$^{-1}$ yr$^{-1}$) & ($10^{-5}$ $M_\odot$ yr$^{-1}$) \\
\hline
1 & $0.05 \pm 0.0334$ & $1.743 \pm 1.165$ & $0.6045 \pm 0.404$ & $33.8 \pm 22.6$ & $0.270 \pm 0.1805$ \\
2 & $1.328 \pm 0.8873$ & $54.86 \pm 36.66$ & $22.54 \pm 15.06$ & $895.3 \pm 598.5$ & $7.163 \pm 4.787$ \\
3 & $2.427 \pm 1.622$ & $35.34 \pm 23.62$ & $5.116 \pm 3.419$ & $84.1 \pm 56.2$ & $0.673 \pm 0.450$ \\
4 & $2.905 \pm 1.942$ & $179.1 \pm 119.7$ & $109.7 \pm 73.35$ & $2305 \pm 1540$ & $18.44 \pm 12.32$ \\
5 & $0.4411 \pm 0.2948$ & $14.86 \pm 9.935$ & $4.981 \pm 3.329$ & $145.9 \pm 97.5$ & $1.167 \pm 0.780$ \\
6 & $0.0521 \pm 0.03485$ & $0.3619 \pm 0.2419$ & $0.025 \pm 0.0167$ & $4.7 \pm 3.1$ & $0.037 \pm 0.025$ \\
7 & $0.7214 \pm 0.4821$ & $13.32 \pm 8.9$ & $2.444 \pm 1.634$ & $134.3 \pm 89.8$ & $1.074 \pm 0.7185$ \\
8 & $0.3451 \pm 0.2306$ & $6.626 \pm 4.429$ & $1.265 \pm 0.8454$ & $53.9 \pm 36$ & $0.431 \pm 0.288$ \\
9 & $0.5596 \pm 0.374$ & $4.891 \pm 3.269$ & $0.425 \pm 0.284$ & $22.1 \pm 14.8$ & $0.177 \pm 0.118$ \\
10 & $1.035 \pm 0.692$ & $35.66 \pm 23.83$ & $12.21 \pm 8.16$ & $416.1 \pm 278.1$ & $3.329 \pm 2.224$ \\
11 & $1.043 \pm 0.6972$ & $17.07 \pm 11.41$ & $2.776 \pm 1.855$ & $161.5 \pm 107.9$ & $1.292 \pm 0.8635$ \\
12 & $1.3 \pm 0.8692$ & $44.79 \pm 29.93$ & $15.34 \pm 10.25$ & $259.2 \pm 173.2$ & $2.073 \pm 1.385$ \\
13 & $0.0042 \pm 0.0028$ & $0.1336 \pm 0.0893$ & $0.0424 \pm 0.0283$ & $3.7 \pm 2.45$ & $0.029 \pm 0.0195$ \\
14 & $0.0566 \pm 0.0378$ & $2.505 \pm 1.674$ & $1.104 \pm 0.7375$ & $61.1 \pm 40.8$ & $0.489 \pm 0.3265$ \\
15 & $0.0084 \pm 0.0056$ & $0.2648 \pm 0.177$ & $0.0832 \pm 0.0556$ & $11.4 \pm 7.6$ & $0.091 \pm 0.061$ \\
16 & $1.246 \pm 0.8324$ & $66.51 \pm 44.45$ & $35.31 \pm 23.6$ & $1607 \pm 1074$ & $12.85 \pm 8.591$ \\
17 & $0.036 \pm 0.02405$ & $1.371 \pm 0.9165$ & $0.5186 \pm 0.3467$ & $48.1 \pm 32.1$ & $0.385 \pm 0.257$ \\
18 & $1.393 \pm 0.9314$ & $44.87 \pm 29.99$ & $14.36 \pm 9.601$ & $2018 \pm 1349$ & $16.15 \pm 10.79$ \\
19 & $0.6824 \pm 0.4562$ & $10.03 \pm 6.705$ & $1.466 \pm 0.9799$ & $152.9 \pm 102.2$ & $1.223 \pm 0.8175$ \\
20 & $1.722 \pm 1.151$ & $25.31 \pm 16.91$ & $3.699 \pm 2.472$ & $219.1 \pm 146.5$ & $1.753 \pm 1.171$ \\
21 & $1.715 \pm 1.147$ & $17.43 \pm 11.65$ & $1.761 \pm 1.177$ & $85.1 \pm 56.9$ & $0.681 \pm 0.455$ \\
22 & $0.0025 \pm 0.00165$ & $0.2906 \pm 0.1943$ & $0.3378 \pm 0.2258$ & $37.9 \pm 25.3$ & $0.303 \pm 0.2025$ \\
23 & $0.2486 \pm 0.1661$ & $3.774 \pm 2.522$ & $0.5696 \pm 0.3807$ & $59.2 \pm 39.55$ & $0.473 \pm 0.3165$ \\
24 & $6.386 \pm 4.268$ & $190 \pm 127$ & $56.23 \pm 37.58$ & $2625 \pm 1755$ & $21.00 \pm 14.04$ \\
25 & $1.376 \pm 0.9194$ & $39.43 \pm 26.35$ & $11.24 \pm 7.509$ & $606 \pm 405$ & $4.848 \pm 3.240$ \\
26 & $0.1412 \pm 0.09435$ & $2.355 \pm 1.574$ & $0.3905 \pm 0.261$ & $22.7 \pm 15.2$ & $0.182 \pm 0.1215$ \\
27 & $0.0697 \pm 0.04665$ & $2.38 \pm 1.591$ & $0.8073 \pm 0.5396$ & $146 \pm 97.6$ & $1.168 \pm 0.781$ \\
28 & $0.0457 \pm 0.03055$ & $0.6458 \pm 0.4316$ & $0.0908 \pm 0.0607$ & $17 \pm 11.35$ & $0.136 \pm 0.0905$ \\
29 & $3.238 \pm 2.164$ & $46.69 \pm 31.2$ & $6.694 \pm 4.474$ & $471.7 \pm 315.3$ & $3.774 \pm 2.522$ \\
30 & $1.855 \pm 1.24$ & $62.77 \pm 41.95$ & $21.12 \pm 14.11$ & $1390 \pm 928.9$ & $11.12 \pm 7.431$ \\
31 & $0.5885 \pm 0.3933$ & $3.649 \pm 2.439$ & $0.2249 \pm 0.1503$ & $9.3 \pm 6.2$ & $0.074 \pm 0.0495$ \\
32 & $3.364 \pm 2.248$ & $48.51 \pm 32.42$ & $6.955 \pm 4.648$ & $290.7 \pm 194.3$ & $2.326 \pm 1.555$ \\
33 & $0.1593 \pm 0.1064$ & $5.39 \pm 3.602$ & $1.813 \pm 1.212$ & $67.5 \pm 45.1$ & $0.540 \pm 0.361$ \\
34 & $0.0598 \pm 0.03995$ & $1.815 \pm 1.213$ & $0.548 \pm 0.3662$ & $54.4 \pm 36.35$ & $0.435 \pm 0.291$ \\
35 & $0.1149 \pm 0.0768$ & $2.907 \pm 1.943$ & $0.7308 \pm 0.4885$ & $51.7 \pm 34.55$ & $0.413 \pm 0.2765$ \\
36 & $0.2735 \pm 0.1828$ & $1.42 \pm 0.9488$ & $0.0733 \pm 0.04895$ & $5.5 \pm 3.65$ & $0.044 \pm 0.0295$ \\
37 & $0.0433 \pm 0.02895$ & $1.474 \pm 0.9849$ & $0.4984 \pm 0.3331$ & $45.2 \pm 30.25$ & $0.362 \pm 0.242$ \\
38 & $0.0381 \pm 0.0255$ & $0.3288 \pm 0.2198$ & $0.0282 \pm 0.01885$ & $3.6 \pm 2.35$ & $0.028 \pm 0.019$ \\
39 & $0.0091 \pm 0.00605$ & $0.2418 \pm 0.1617$ & $0.0642 \pm 0.04295$ & $5.1 \pm 3.45$ & $0.041 \pm 0.0275$ \\
40 & $6.143 \pm 4.106$ & $716.1 \pm 478.6$ & $829.8 \pm 554.6$ & $32820 \pm 21930$ & $262.5 \pm 175.5$ \\
41 & $20.9 \pm 13.97$ & $696 \pm 465.2$ & $230.4 \pm 154$ & $6014 \pm 4019$ & $48.11 \pm 32.15$ \\
42 & $4.015 \pm 2.684$ & $121.3 \pm 81.05$ & $36.41 \pm 24.33$ & $1494 \pm 998.7$ & $11.95 \pm 7.99$ \\
43 & $1.746 \pm 1.167$ & $58.12 \pm 38.85$ & $19.24 \pm 12.86$ & $1300 \pm 869.1$ & $10.40 \pm 6.953$ \\
44 & $1.498 \pm 1.001$ & $30.01 \pm 20.06$ & $5.98 \pm 3.997$ & $252.7 \pm 168.9$ & $2.022 \pm 1.351$ \\
45 & $1.595 \pm 1.066$ & $84.61 \pm 56.55$ & $44.64 \pm 29.83$ & $3699 \pm 2472$ & $29.60 \pm 19.78$ \\
46 & $1.625 \pm 1.086$ & $54.12 \pm 36.17$ & $17.92 \pm 11.98$ & $1656 \pm 1107$ & $13.25 \pm 8.857$ \\
47 & $0.383 \pm 0.256$ & $11.61 \pm 7.757$ & $3.496 \pm 2.337$ & $229.4 \pm 153.3$ & $1.835 \pm 1.227$ \\
48 & $0.4849 \pm 0.3241$ & $13.63 \pm 9.113$ & $3.812 \pm 2.548$ & $312.3 \pm 208.7$ & $2.498 \pm 1.669$ \\
49 & $1.072 \pm 0.7162$ & $24.3 \pm 16.24$ & $5.48 \pm 3.663$ & $648.1 \pm 433.2$ & $5.185 \pm 3.465$ \\
50 & $2.094 \pm 1.4$ & $42.94 \pm 28.7$ & $8.751 \pm 5.849$ & $472.1 \pm 315.5$ & $3.777 \pm 2.524$ \\
51 & $0.0445 \pm 0.02975$ & $0.8934 \pm 0.5971$ & $0.1782 \pm 0.1191$ & $15.4 \pm 10.25$ & $0.123 \pm 0.082$ \\
\hline
\end{tabular}
\label{dynamics}
\end{table*} 

\section*{Acknowledgements}
We thank the anonymous referee for their constructive comments, which significantly improved the quality of this work. The project was funded by the European Research Council via the ERC
Synergy Grant “ECOGAL” (project ID 855130). SM and DW acknowledge support from STFC Grant ST/Y004108/1. SM, WC, and GAF acknowledge funding by the Deutsche Forschungsgemeinschaft (DFG, German Research Foundation) under Germany’s Excellence Strategy EXC 3037 -
533607693 - Our Dynamic Universe.
GAF acknowledges support from the Collaborative Research Centre 956, funded by the Deutsche
Forschungsgemeinschaft (DFG) with project ID 184018867.

\section*{Data Availability}
This work makes use of observations obtained as part of the ALMAGAL (project code 2019.1.00195.L) and the SQUALO programme (project code 2018.1.00443.S), both publicly available through the ALMA Science Archive. Mid-infrared data were obtained from the \textit{Spitzer Space Telescope} Heritage Archive. The derived outflow catalogues and physical parameters presented in this paper are available from the corresponding author upon reasonable request.



\bibliographystyle{mnras}
\bibliography{example} 

@ARTICLE{Kepley2020,
       author = {{Kepley}, Amanda A. and {Tsutsumi}, Takahiro and {Brogan}, Crystal L. and {Indebetouw}, Remy and {Yoon}, Ilsang and {Mason}, Brian and {Donovan Meyer}, Jennifer},
        title = "{Auto-multithresh: A General Purpose Automasking Algorithm}",
      journal = {\pasp},
         year = 2020,
        month = feb,
       volume = {132},
       number = {1008},
          eid = {024505},
        pages = {024505},
          doi = {10.1088/1538-3873/ab5e14},
archivePrefix = {arXiv},
       eprint = {1912.04970},
 primaryClass = {astro-ph.IM},
       adsurl = {https://ui.adsabs.harvard.edu/abs/2020PASP..132b4505K}
}

@ARTICLE{Zinnecker2007,
       author = {{Zinnecker}, Hans and {Yorke}, Harold W.},
        title = "{Toward Understanding Massive Star Formation}",
      journal = {\araa},
         year = 2007,
        month = sep,
       volume = {45},
       number = {1},
        pages = {481-563},
          doi = {10.1146/annurev.astro.44.051905.092549},
archivePrefix = {arXiv},
       eprint = {0707.1279},
 primaryClass = {astro-ph},
       adsurl = {https://ui.adsabs.harvard.edu/abs/2007ARA&A..45..481Z}
}

@INPROCEEDINGS{Tan2014,
       author = {{Tan}, J.~C. and {Beltr{\'a}n}, M.~T. and {Caselli}, P. and {Fontani}, F. and {Fuente}, A. and {Krumholz}, M.~R. and {McKee}, C.~F. and {Stolte}, A.},
        title = "{Massive Star Formation}",
    booktitle = {Protostars and Planets VI},
         year = 2014,
       editor = {{Beuther}, Henrik and {Klessen}, Ralf S. and {Dullemond}, Cornelis P. and {Henning}, Thomas},
        month = jan,
        pages = {149-172},
          doi = {10.2458/azu_uapress_9780816531240-ch007},
archivePrefix = {arXiv},
       eprint = {1402.0919},
 primaryClass = {astro-ph.GA},
       adsurl = {https://ui.adsabs.harvard.edu/abs/2014prpl.conf..149T}
}

@ARTICLE{Blandford1982,
       author = {{Blandford}, R.~D. and {Payne}, D.~G.},
        title = "{Hydromagnetic flows from accretion disks and the production of radio jets.}",
      journal = {\mnras},
         year = 1982,
        month = jun,
       volume = {199},
        pages = {883-903},
          doi = {10.1093/mnras/199.4.883},
       adsurl = {https://ui.adsabs.harvard.edu/abs/1982MNRAS.199..883B}
}

@INPROCEEDINGS{Arce2007,
       author = {{Arce}, H.~G. and {Shepherd}, D. and {Gueth}, F. and {Lee}, C.-F. and {Bachiller}, R. and {Rosen}, A. and {Beuther}, H.},
        title = "{Molecular Outflows in Low- and High-Mass Star-forming Regions}",
    booktitle = {Protostars and Planets V},
         year = 2007,
       editor = {{Reipurth}, Bo and {Jewitt}, David and {Keil}, Klaus},
        month = jan,
        pages = {245},
          doi = {10.48550/arXiv.astro-ph/0603071},
archivePrefix = {arXiv},
       eprint = {astro-ph/0603071},
 primaryClass = {astro-ph},
       adsurl = {https://ui.adsabs.harvard.edu/abs/2007prpl.conf..245A}
}

@ARTICLE{Cabrit1992,
       author = {{Cabrit}, Sylvie and {Bertout}, Claude},
        title = "{CO line formation in bipolar flows. III. The energetics of molecular flows and ionized winds.}",
      journal = {\aap},
         year = 1992,
        month = jul,
       volume = {261},
        pages = {274-284},
       adsurl = {https://ui.adsabs.harvard.edu/abs/1992A&A...261..274C}
}

@ARTICLE{Bontemps1996,
       author = {{Bontemps}, S. and {Andre}, P. and {Terebey}, S. and {Cabrit}, S.},
        title = "{Evolution of outflow activity around low-mass embedded young stellar objects}",
      journal = {\aap},
         year = 1996,
        month = jul,
       volume = {311},
        pages = {858-872},
       adsurl = {https://ui.adsabs.harvard.edu/abs/1996A&A...311..858B}
}

@ARTICLE{Maud2015,
       author = {{Maud}, L.~T. and {Moore}, T.~J.~T. and {Lumsden}, S.~L. and {Mottram}, J.~C. and {Urquhart}, J.~S. and {Hoare}, M.~G.},
        title = "{A distance-limited sample of massive molecular outflows}",
      journal = {\mnras},
         year = 2015,
        month = oct,
       volume = {453},
       number = {1},
        pages = {645-665},
          doi = {10.1093/mnras/stv1635},
archivePrefix = {arXiv},
       eprint = {1509.00199},
 primaryClass = {astro-ph.SR},
       adsurl = {https://ui.adsabs.harvard.edu/abs/2015MNRAS.453..645M}
}

@ARTICLE{Yang2018,
       author = {{Yang}, A.~Y. and {Thompson}, M.~A. and {Urquhart}, J.~S. and {Tian}, W.~W.},
        title = "{Massive Outflows Associated with ATLASGAL Clumps}",
      journal = {\apjs},
         year = 2018,
        month = mar,
       volume = {235},
       number = {1},
          eid = {3},
        pages = {3},
          doi = {10.3847/1538-4365/aaa297},
archivePrefix = {arXiv},
       eprint = {1712.04599},
 primaryClass = {astro-ph.GA},
       adsurl = {https://ui.adsabs.harvard.edu/abs/2018ApJS..235....3Y}
}

@ARTICLE{Li2018,
       author = {{Li}, Qiang and {Zhou}, Jianjun and {Esimbek}, Jarken and {He}, Yuxin and {Baan}, W.~A. and {Li}, Dalei and {Wu}, Gang and {Tang}, Xindi and {Ji}, Weiguang and {Zhexeray}, Dauren},
        title = "{High-mass Outflows Identified from COHRS CO (3-2) Survey}",
      journal = {\apj},
         year = 2018,
        month = nov,
       volume = {867},
       number = {2},
          eid = {167},
        pages = {167},
          doi = {10.3847/1538-4357/aae2b8},
archivePrefix = {arXiv},
       eprint = {1809.08739},
 primaryClass = {astro-ph.GA},
       adsurl = {https://ui.adsabs.harvard.edu/abs/2018ApJ...867..167L}
}

@ARTICLE{Beuther2002b,
       author = {{Beuther}, H. and {Schilke}, P. and {Menten}, K.~M. and {Motte}, F. and {Sridharan}, T.~K. and {Wyrowski}, F.},
        title = "{High-Mass Protostellar Candidates. II. Density Structure from Dust Continuum and CS Emission}",
      journal = {\apj},
         year = 2002,
        month = feb,
       volume = {566},
       number = {2},
        pages = {945-965},
          doi = {10.1086/338334},
archivePrefix = {arXiv},
       eprint = {astro-ph/0110370},
 primaryClass = {astro-ph},
       adsurl = {https://ui.adsabs.harvard.edu/abs/2002ApJ...566..945B}
}

@ARTICLE{Beuther2002a,
       author = {{Beuther}, H. and {Schilke}, P. and {Gueth}, F. and {McCaughrean}, M. and {Andersen}, M. and {Sridharan}, T.~K. and {Menten}, K.~M.},
        title = "{IRAS 05358+3543: Multiple outflows at the earliest stages of massive star formation}",
      journal = {\aap},
         year = 2002,
        month = jun,
       volume = {387},
        pages = {931-943},
          doi = {10.1051/0004-6361:20020319},
archivePrefix = {arXiv},
       eprint = {astro-ph/0203010},
 primaryClass = {astro-ph},
       adsurl = {https://ui.adsabs.harvard.edu/abs/2002A&A...387..931B}
}

@ARTICLE{Duarte-Cabral2013,
       author = {{Duarte-Cabral}, A. and {Bontemps}, S. and {Motte}, F. and {Hennemann}, M. and {Schneider}, N. and {Andr{\'e}}, Ph.},
        title = "{CO outflows from high-mass Class 0 protostars in Cygnus-X}",
      journal = {\aap},
         year = 2013,
        month = oct,
       volume = {558},
          eid = {A125},
        pages = {A125},
          doi = {10.1051/0004-6361/201321393},
archivePrefix = {arXiv},
       eprint = {1308.6490},
 primaryClass = {astro-ph.GA},
       adsurl = {https://ui.adsabs.harvard.edu/abs/2013A&A...558A.125D}
}

@ARTICLE{Yang2022,
       author = {{Yang}, A.~Y. and {Urquhart}, J.~S. and {Wyrowski}, F. and {Thompson}, M.~A. and {K{\"o}nig}, C. and {Colombo}, D. and {Menten}, K.~M. and {Duarte-Cabral}, A. and {Schuller}, F. and {Csengeri}, T. and {Eden}, D. and {Barnes}, P. and {Traficante}, A. and {Bronfman}, L. and {Sanchez-Monge}, A. and {Ginsburg}, A. and {Cesaroni}, R. and {Lee}, M.-Y. and {Beuther}, H. and {Medina}, S.-N.~X. and {Mazumdar}, P. and {Henning}, T.},
        title = "{The SEDIGISM survey: A search for molecular outflows}",
      journal = {\aap},
         year = 2022,
        month = feb,
       volume = {658},
          eid = {A160},
        pages = {A160},
          doi = {10.1051/0004-6361/202142039},
archivePrefix = {arXiv},
       eprint = {2111.10850},
 primaryClass = {astro-ph.GA},
       adsurl = {https://ui.adsabs.harvard.edu/abs/2022A&A...658A.160Y}
}

@ARTICLE{Traficante2023,
       author = {{Traficante}, A. and {Jones}, B.~M. and {Avison}, A. and {Fuller}, G.~A. and {Benedettini}, M. and {Elia}, D. and {Molinari}, S. and {Peretto}, N. and {Pezzuto}, S. and {Pillai}, T. and {Rygl}, K.~L.~J. and {Schisano}, E. and {Smith}, R.~J.},
        title = "{The SQUALO project (Star formation in QUiescent And Luminous Objects) I: clump-fed accretion mechanism in high-mass star-forming objects}",
      journal = {\mnras},
         year = 2023,
        month = apr,
       volume = {520},
       number = {2},
        pages = {2306-2327},
          doi = {10.1093/mnras/stad272},
archivePrefix = {arXiv},
       eprint = {2301.09917},
 primaryClass = {astro-ph.GA},
       adsurl = {https://ui.adsabs.harvard.edu/abs/2023MNRAS.520.2306T}
}

@ARTICLE{Li2020,
       author = {{Li}, Shanghuo and {Sanhueza}, Patricio and {Zhang}, Qizhou and {Nakamura}, Fumitaka and {Lu}, Xing and {Wang}, Junzhi and {Liu}, Tie and {Tatematsu}, Ken'ichi and {Jackson}, James M. and {Silva}, Andrea and {Guzm{\'a}n}, Andr{\'e}s E. and {Sakai}, Takeshi and {Izumi}, Natsuko and {Tafoya}, Daniel and {Li}, Fei and {Contreras}, Yanett and {Morii}, Kaho and {Kim}, Kee-Tae},
        title = "{The ALMA Survey of 70 {\ensuremath{\mu}}m Dark High-mass Clumps in Early Stages (ASHES). II. Molecular Outflows in the Extreme Early Stages of Protocluster Formation}",
      journal = {\apj},
         year = 2020,
        month = nov,
       volume = {903},
       number = {2},
          eid = {119},
        pages = {119},
          doi = {10.3847/1538-4357/abb81f},
archivePrefix = {arXiv},
       eprint = {2009.05506},
 primaryClass = {astro-ph.GA},
       adsurl = {https://ui.adsabs.harvard.edu/abs/2020ApJ...903..119L}
}

@ARTICLE{Csengeri2014,
       author = {{Csengeri}, T. and {Urquhart}, J.~S. and {Schuller}, F. and {Motte}, F. and {Bontemps}, S. and {Wyrowski}, F. and {Menten}, K.~M. and {Bronfman}, L. and {Beuther}, H. and {Henning}, Th. and {Testi}, L. and {Zavagno}, A. and {Walmsley}, M.},
        title = "{The ATLASGAL survey: a catalog of dust condensations in the Galactic plane}",
      journal = {\aap},
         year = 2014,
        month = may,
       volume = {565},
          eid = {A75},
        pages = {A75},
          doi = {10.1051/0004-6361/201322434},
archivePrefix = {arXiv},
       eprint = {1312.0937},
 primaryClass = {astro-ph.GA},
       adsurl = {https://ui.adsabs.harvard.edu/abs/2014A&A...565A..75C}
}

@ARTICLE{Traficante2018a,
       author = {{Traficante}, A. and {Fuller}, G.~A. and {Smith}, R.~J. and {Billot}, N. and {Duarte-Cabral}, A. and {Peretto}, N. and {Molinari}, S. and {Pineda}, J.~E.},
        title = "{Massive 70 {\ensuremath{\mu}}m quiet clumps - II. Non-thermal motions driven by gravity in massive star formation?}",
      journal = {\mnras},
         year = 2018,
        month = feb,
       volume = {473},
       number = {4},
        pages = {4975-4985},
          doi = {10.1093/mnras/stx2672},
archivePrefix = {arXiv},
       eprint = {1710.04904},
 primaryClass = {astro-ph.GA},
       adsurl = {https://ui.adsabs.harvard.edu/abs/2018MNRAS.473.4975T}
}

@ARTICLE{Towner2024,
       author = {{Towner}, A.~P.~M. and {Ginsburg}, A. and {Dell'Ova}, P. and {Gusdorf}, A. and {Bontemps}, S. and {Csengeri}, T. and {Galv{\'a}n-Madrid}, R. and {Louvet}, F.~K. and {Motte}, F. and {Sanhueza}, P. and {Stutz}, A.~M. and {Bally}, J. and {Baug}, T. and {Chen}, H.-R.~V. and {Cunningham}, N. and {Fern{\'a}ndez-L{\'o}pez}, M. and {Liu}, H.-L. and {Lu}, X. and {Nony}, T. and {Valeille-Manet}, M. and {Wu}, B. and {{\'A}lvarez-Guti{\'e}rrez}, R.~H. and {Bonfand}, M. and {Di Francesco}, J. and {Nguyen-Luong}, Q. and {Olguin}, F. and {Whitworth}, A.~P.},
        title = "{ALMA-IMF. IX. Catalog and Physical Properties of 315 SiO Outflow Candidates in 15 Massive Protoclusters}",
      journal = {\apj},
         year = 2024,
        month = jan,
       volume = {960},
       number = {1},
          eid = {48},
        pages = {48},
          doi = {10.3847/1538-4357/ad0786},
archivePrefix = {arXiv},
       eprint = {2310.13125},
 primaryClass = {astro-ph.GA},
       adsurl = {https://ui.adsabs.harvard.edu/abs/2024ApJ...960...48T}
}

@ARTICLE{Schneider2010,
       author = {{Schneider}, N. and {Csengeri}, T. and {Bontemps}, S. and {Motte}, F. and {Simon}, R. and {Hennebelle}, P. and {Federrath}, C. and {Klessen}, R.},
        title = "{Dynamic star formation in the massive DR21 filament}",
      journal = {\aap},
         year = 2010,
        month = sep,
       volume = {520},
          eid = {A49},
        pages = {A49},
          doi = {10.1051/0004-6361/201014481},
archivePrefix = {arXiv},
       eprint = {1003.4198},
 primaryClass = {astro-ph.GA},
       adsurl = {https://ui.adsabs.harvard.edu/abs/2010A&A...520A..49S}
}

@ARTICLE{Peretto2013,
       author = {{Peretto}, N. and {Fuller}, G.~A. and {Duarte-Cabral}, A. and {Avison}, A. and {Hennebelle}, P. and {Pineda}, J.~E. and {Andr{\'e}}, Ph. and {Bontemps}, S. and {Motte}, F. and {Schneider}, N. and {Molinari}, S.},
        title = "{Global collapse of molecular clouds as a formation mechanism for the most massive stars}",
      journal = {\aap},
         year = 2013,
        month = jul,
       volume = {555},
          eid = {A112},
        pages = {A112},
          doi = {10.1051/0004-6361/201321318},
archivePrefix = {arXiv},
       eprint = {1307.2590},
 primaryClass = {astro-ph.GA},
       adsurl = {https://ui.adsabs.harvard.edu/abs/2013A&A...555A.112P}
}

@ARTICLE{Sanhueza2019,
       author = {{Sanhueza}, Patricio and {Contreras}, Yanett and {Wu}, Benjamin and {Jackson}, James M. and {Guzm{\'a}n}, Andr{\'e}s E. and {Zhang}, Qizhou and {Li}, Shanghuo and {Lu}, Xing and {Silva}, Andrea and {Izumi}, Natsuko and {Liu}, Tie and {Miura}, Rie E. and {Tatematsu}, Ken'ichi and {Sakai}, Takeshi and {Beuther}, Henrik and {Garay}, Guido and {Ohashi}, Satoshi and {Saito}, Masao and {Nakamura}, Fumitaka and {Saigo}, Kazuya and {Veena}, V.~S. and {Nguyen-Luong}, Quang and {Tafoya}, Daniel},
        title = "{The ALMA Survey of 70 {\ensuremath{\mu}}m Dark High-mass Clumps in Early Stages (ASHES). I. Pilot Survey: Clump Fragmentation}",
      journal = {\apj},
         year = 2019,
        month = dec,
       volume = {886},
       number = {2},
          eid = {102},
        pages = {102},
          doi = {10.3847/1538-4357/ab45e9},
archivePrefix = {arXiv},
       eprint = {1909.07985},
 primaryClass = {astro-ph.GA},
       adsurl = {https://ui.adsabs.harvard.edu/abs/2019ApJ...886..102S}
}

@ARTICLE{Wyrowski2016,
       author = {{Wyrowski}, F. and {G{\"u}sten}, R. and {Menten}, K.~M. and {Wiesemeyer}, H. and {Csengeri}, T. and {Heyminck}, S. and {Klein}, B. and {K{\"o}nig}, C. and {Urquhart}, J.~S.},
        title = "{Infall through the evolution of high-mass star-forming clumps}",
      journal = {\aap},
         year = 2016,
        month = jan,
       volume = {585},
          eid = {A149},
        pages = {A149},
          doi = {10.1051/0004-6361/201526361},
archivePrefix = {arXiv},
       eprint = {1510.08374},
 primaryClass = {astro-ph.SR},
       adsurl = {https://ui.adsabs.harvard.edu/abs/2016A&A...585A.149W}
}

@ARTICLE{Ballesteros-Paredes2011,
       author = {{Ballesteros-Paredes}, Javier and {Hartmann}, Lee W. and {V{\'a}zquez-Semadeni}, Enrique and {Heitsch}, Fabian and {Zamora-Avil{\'e}s}, Manuel A.},
        title = "{Gravity or turbulence? Velocity dispersion-size relation}",
      journal = {\mnras},
         year = 2011,
        month = feb,
       volume = {411},
       number = {1},
        pages = {65-70},
          doi = {10.1111/j.1365-2966.2010.17657.x},
archivePrefix = {arXiv},
       eprint = {1009.1583},
 primaryClass = {astro-ph.GA},
       adsurl = {https://ui.adsabs.harvard.edu/abs/2011MNRAS.411...65B}
}

@ARTICLE{Wang2020,
       author = {{Wang}, Jia-Wei and {Koch}, Patrick M. and {Galv{\'a}n-Madrid}, Roberto and {Lai}, Shih-Ping and {Liu}, Hauyu Baobab and {Lin}, Sheng-Jun and {Pattle}, Kate},
        title = "{Formation of the Hub-Filament System G33.92+0.11: Local Interplay between Gravity, Velocity, and Magnetic Field}",
      journal = {\apj},
         year = 2020,
        month = dec,
       volume = {905},
       number = {2},
          eid = {158},
        pages = {158},
          doi = {10.3847/1538-4357/abc74e},
archivePrefix = {arXiv},
       eprint = {2011.01555},
 primaryClass = {astro-ph.GA},
       adsurl = {https://ui.adsabs.harvard.edu/abs/2020ApJ...905..158W}
}

@ARTICLE{Motte2018,
       author = {{Motte}, Fr{\'e}d{\'e}rique and {Bontemps}, Sylvain and {Louvet}, Fabien},
        title = "{High-Mass Star and Massive Cluster Formation in the Milky Way}",
      journal = {\araa},
         year = 2018,
        month = sep,
       volume = {56},
        pages = {41-82},
          doi = {10.1146/annurev-astro-091916-055235},
archivePrefix = {arXiv},
       eprint = {1706.00118},
 primaryClass = {astro-ph.GA},
       adsurl = {https://ui.adsabs.harvard.edu/abs/2018ARA&A..56...41M}
}

@ARTICLE{Avison2021,
       author = {{Avison}, A. and {Fuller}, G.~A. and {Peretto}, N. and {Duarte-Cabral}, A. and {Rosen}, A.~L. and {Traficante}, A. and {Pineda}, J.~E. and {G{\"u}sten}, R. and {Cunningham}, N.},
        title = "{Continuity of accretion from clumps to Class 0 high-mass protostars in SDC335}",
      journal = {\aap},
         year = 2021,
        month = jan,
       volume = {645},
          eid = {A142},
        pages = {A142},
          doi = {10.1051/0004-6361/201936043},
archivePrefix = {arXiv},
       eprint = {2012.08948},
 primaryClass = {astro-ph.GA},
       adsurl = {https://ui.adsabs.harvard.edu/abs/2021A&A...645A.142A}
}

@ARTICLE{Elia2021,
       author = {{Elia}, Davide and {Merello}, M. and {Molinari}, S. and {Schisano}, E. and {Zavagno}, A. and {Russeil}, D. and {M{\`e}ge}, P. and {Martin}, P.~G. and {Olmi}, L. and {Pestalozzi}, M. and {Plume}, R. and {Ragan}, S.~E. and {Benedettini}, M. and {Eden}, D.~J. and {Moore}, T.~J.~T. and {Noriega-Crespo}, A. and {Paladini}, R. and {Palmeirim}, P. and {Pezzuto}, S. and {Pilbratt}, G.~L. and {Rygl}, K.~L.~J. and {Schilke}, P. and {Strafella}, F. and {Tan}, J.~C. and {Traficante}, A. and {Baldeschi}, A. and {Bally}, J. and {di Giorgio}, A.~M. and {Fiorellino}, E. and {Liu}, S.~J. and {Piazzo}, L. and {Polychroni}, D.},
        title = "{The Hi-GAL compact source catalogue - II. The 360{\textdegree} catalogue of clump physical properties}",
      journal = {\mnras},
         year = 2021,
        month = jun,
       volume = {504},
       number = {2},
        pages = {2742-2766},
          doi = {10.1093/mnras/stab1038},
archivePrefix = {arXiv},
       eprint = {2104.04807},
 primaryClass = {astro-ph.GA},
       adsurl = {https://ui.adsabs.harvard.edu/abs/2021MNRAS.504.2742E}
}

@ARTICLE{Molinari2025,
       author = {{Molinari}, S. and {Schilke}, P. and {Battersby}, C. and {Ho}, P.~T.~P. and {S{\'a}nchez-Monge}, {\'A}. and {Traficante}, A. and {Jones}, B. and {Beltr{\'a}n}, M.~T. and {Beuther}, H. and {Fuller}, G.~A. and {Zhang}, Q. and {Klessen}, R.~S. and {Walch}, S. and {Tang}, Y.-W. and {Benedettini}, M. and {Elia}, D. and {Coletta}, A. and {Mininni}, C. and {Schisano}, E. and {Avison}, A. and {Law}, C.~Y. and {Nucara}, A. and {Soler}, J.~D. and {Stroud}, G. and {Wallace}, J. and {Wells}, M.~R.~A. and {Ahmadi}, A. and {Brogan}, C.~L. and {Hunter}, T.~R. and {Liu}, S.-Y. and {Pezzuto}, S. and {Su}, Y.-N. and {Zimmermann}, B. and {Zhang}, T. and {Wyrowski}, F. and {De Angelis}, F. and {Liu}, S. and {Clarke}, S.~D. and {Fontani}, F. and {Klaassen}, P.~D. and {Koch}, P. and {Johnston}, K.~G. and {Lebreuilly}, U. and {Liu}, T. and {Lumsden}, S.~L. and {Moeller}, T. and {Moscadelli}, L. and {Kuiper}, R. and {Lis}, D. and {Peretto}, N. and {Pfalzner}, S. and {Rigby}, A.~J. and {Sanhueza}, P. and {Rygl}, K.~L.~J. and {van der Tak}, F. and {Zinnecker}, H. and {Amaral}, F. and {Bally}, J. and {Bronfman}, L. and {Cesaroni}, R. and {Goh}, K. and {Hoare}, M.~G. and {Hatchfield}, P. and {Hennebelle}, P. and {Henning}, T. and {Kim}, K.-T. and {Kim}, W.-J. and {Maud}, L. and {Merello}, M. and {Nakamura}, F. and {Plume}, R. and {Qin}, S.-L. and {Svoboda}, B. and {Testi}, L. and {Veena}, V.~S. and {Walker}, D.},
        title = "{ALMAGAL: I. The ALMA evolutionary study of high-mass protocluster formation in the Galaxy: Presentation of the survey and early results}",
      journal = {\aap},
         year = 2025,
        month = apr,
       volume = {696},
          eid = {A149},
        pages = {A149},
          doi = {10.1051/0004-6361/202452702},
archivePrefix = {arXiv},
       eprint = {2503.05555},
 primaryClass = {astro-ph.GA},
       adsurl = {https://ui.adsabs.harvard.edu/abs/2025A&A...696A.149M}
}

@ARTICLE{Fazio2004,
       author = {{Fazio}, G.~G. and {Hora}, J.~L. and {Allen}, L.~E. and {Ashby}, M.~L.~N. and {Barmby}, P. and {Deutsch}, L.~K. and {Huang}, J.-S. and {Kleiner}, S. and {Marengo}, M. and {Megeath}, S.~T. and {Melnick}, G.~J. and {Pahre}, M.~A. and {Patten}, B.~M. and {Polizotti}, J. and {Smith}, H.~A. and {Taylor}, R.~S. and {Wang}, Z. and {Willner}, S.~P. and {Hoffmann}, W.~F. and {Pipher}, J.~L. and {Forrest}, W.~J. and {McMurty}, C.~W. and {McCreight}, C.~R. and {McKelvey}, M.~E. and {McMurray}, R.~E. and {Koch}, D.~G. and {Moseley}, S.~H. and {Arendt}, R.~G. and {Mentzell}, J.~E. and {Marx}, C.~T. and {Losch}, P. and {Mayman}, P. and {Eichhorn}, W. and {Krebs}, D. and {Jhabvala}, M. and {Gezari}, D.~Y. and {Fixsen}, D.~J. and {Flores}, J. and {Shakoorzadeh}, K. and {Jungo}, R. and {Hakun}, C. and {Workman}, L. and {Karpati}, G. and {Kichak}, R. and {Whitley}, R. and {Mann}, S. and {Tollestrup}, E.~V. and {Eisenhardt}, P. and {Stern}, D. and {Gorjian}, V. and {Bhattacharya}, B. and {Carey}, S. and {Nelson}, B.~O. and {Glaccum}, W.~J. and {Lacy}, M. and {Lowrance}, P.~J. and {Laine}, S. and {Reach}, W.~T. and {Stauffer}, J.~A. and {Surace}, J.~A. and {Wilson}, G. and {Wright}, E.~L. and {Hoffman}, A. and {Domingo}, G. and {Cohen}, M.},
        title = "{The Infrared Array Camera (IRAC) for the Spitzer Space Telescope}",
      journal = {\apjs},
         year = 2004,
        month = sep,
       volume = {154},
       number = {1},
        pages = {10-17},
          doi = {10.1086/422843},
archivePrefix = {arXiv},
       eprint = {astro-ph/0405616},
 primaryClass = {astro-ph},
       adsurl = {https://ui.adsabs.harvard.edu/abs/2004ApJS..154...10F}
}

@ARTICLE{Benjamin2003,
       author = {{Benjamin}, Robert A. and {Churchwell}, E. and {Babler}, Brian L. and {Bania}, T.~M. and {Clemens}, Dan P. and {Cohen}, Martin and {Dickey}, John M. and {Indebetouw}, R{\'e}my and {Jackson}, James M. and {Kobulnicky}, Henry A. and {Lazarian}, Alex and {Marston}, A.~P. and {Mathis}, John S. and {Meade}, Marilyn R. and {Seager}, Sara and {Stolovy}, S.~R. and {Watson}, C. and {Whitney}, Barbara A. and {Wolff}, Michael J. and {Wolfire}, Mark G.},
        title = "{GLIMPSE. I. An SIRTF Legacy Project to Map the Inner Galaxy}",
      journal = {\pasp},
         year = 2003,
        month = aug,
       volume = {115},
       number = {810},
        pages = {953-964},
          doi = {10.1086/376696},
archivePrefix = {arXiv},
       eprint = {astro-ph/0306274},
 primaryClass = {astro-ph},
       adsurl = {https://ui.adsabs.harvard.edu/abs/2003PASP..115..953B}
}

@ARTICLE{Cabrit1986,
       author = {{Cabrit}, S. and {Bertout}, C.},
        title = "{CO Line Formation in Bipolar Flows. I. Accelerated Outflows}",
      journal = {\apj},
         year = 1986,
        month = aug,
       volume = {307},
        pages = {313},
          doi = {10.1086/164418},
       adsurl = {https://ui.adsabs.harvard.edu/abs/1986ApJ...307..313C}
}

@ARTICLE{Frerking1982,
       author = {{Frerking}, M.~A. and {Langer}, W.~D. and {Wilson}, R.~W.},
        title = "{The relationship between carbon monoxide abundance and visual extinction in interstellar clouds.}",
      journal = {\apj},
         year = 1982,
        month = nov,
       volume = {262},
        pages = {590-605},
          doi = {10.1086/160451},
       adsurl = {https://ui.adsabs.harvard.edu/abs/1982ApJ...262..590F}
}

@ARTICLE{Traficante2017,
       author = {{Traficante}, A. and {Fuller}, G.~A. and {Billot}, N. and {Duarte-Cabral}, A. and {Merello}, M. and {Molinari}, S. and {Peretto}, N. and {Schisano}, E.},
        title = "{Massive 70 {\ensuremath{\mu}}m quiet clumps I: evidence of embedded low/intermediate-mass star formation activity}",
      journal = {\mnras},
         year = 2017,
        month = oct,
       volume = {470},
       number = {4},
        pages = {3882-3923},
          doi = {10.1093/mnras/stx1375},
archivePrefix = {arXiv},
       eprint = {1706.00432},
 primaryClass = {astro-ph.GA},
       adsurl = {https://ui.adsabs.harvard.edu/abs/2017MNRAS.470.3882T}
}

@ARTICLE{Molinari2008,
       author = {{Molinari}, S. and {Pezzuto}, S. and {Cesaroni}, R. and {Brand}, J. and {Faustini}, F. and {Testi}, L.},
        title = "{The evolution of the spectral energy distribution in massive young stellar objects}",
      journal = {\aap},
         year = 2008,
        month = apr,
       volume = {481},
       number = {2},
        pages = {345-365},
          doi = {10.1051/0004-6361:20078661},
       adsurl = {https://ui.adsabs.harvard.edu/abs/2008A&A...481..345M}
}

@ARTICLE{Jackson2006,
author = {{Jackson}, J.~M. and {Rathborne}, J.~M. and {Shah}, R.~Y. and
{Simon}, R. and {Bania}, T.~M. and {Clemens}, D.~P. and
{Chambers}, E.~T. and {Johnson}, A.~M. and {Dormody}, M. and
{Lavoie}, R. and {Heyer}, M.},
title = "{The Boston University-FCRAO Galactic Ring Survey}",
journal = {\apjs},
year = 2006,
month = feb,
volume = {163},
number = {1},
pages = {145-159},
doi = {10.1086/500091},
adsurl = {https://ui.adsabs.harvard.edu/abs/2006ApJS..163..145J}
}

@ARTICLE{Jackson2013,
author = {{Jackson}, J.~M. and {Foster}, J.~B. and {Contreras}, Y. and
{et al.}},
title = "{MALT90: The Millimetre Astronomy Legacy Team 90 GHz Survey}",
journal = {\pasa},
year = 2013,
month = nov,
volume = {30},
eid = {e057},
pages = {e057},
doi = {10.1017/pasa.2013.44},
adsurl = {https://ui.adsabs.harvard.edu/abs/2013PASA...30...57J}
}

@ARTICLE{Burton2013,
author = {{Burton}, M.~G. and {Braiding}, C. and {Glueck}, C. and
{et al.}},
title = "{The Mopra Southern Galactic Plane CO Survey}",
journal = {\pasa},
year = 2013,
month = nov,
volume = {30},
eid = {e044},
pages = {e044},
doi = {10.1017/pasa.2013.35},
adsurl = {https://ui.adsabs.harvard.edu/abs/2013PASA...30...44B}
}

@ARTICLE{Barnes2015,
author = {{Barnes}, P.~J. and {Muller}, E. and {Indermuehle}, B. and
{Henshaw}, J.~D. and {Hernandez}, A.~K.},
title = "{THRUMMS: The Three-mm Ultimate Mopra Milky Way Survey. I. Overview and Data Release}",
journal = {\apj},
year = 2015,
month = may,
volume = {812},
number = {1},
eid = {6},
pages = {6},
doi = {10.1088/0004-637X/812/1/6},
adsurl = {https://ui.adsabs.harvard.edu/abs/2015ApJ...812....6B}
}

@ARTICLE{Coletta2025,
       author = {{Coletta}, A. and {Molinari}, S. and {Schisano}, E. and {Traficante}, A. and {Elia}, D. and {Benedettini}, M. and {Mininni}, C. and {Soler}, J.~D. and {S{\'a}nchez-Monge}, {\'A}. and {Schilke}, P. and {Battersby}, C. and {Fuller}, G.~A. and {Beuther}, H. and {Zhang}, Q. and {Beltr{\'a}n}, M.~T. and {Jones}, B. and {Klessen}, R.~S. and {Walch}, S. and {Fontani}, F. and {Avison}, A. and {Brogan}, C.~L. and {Clarke}, S.~D. and {Hatchfield}, P. and {Hennebelle}, P. and {Ho}, P.~T.~P. and {Hunter}, T.~R. and {Johnston}, K.~G. and {Klaassen}, P.~D. and {Koch}, P.~M. and {Kuiper}, R. and {Lis}, D.~C. and {Liu}, T. and {Lumsden}, S.~L. and {Maruccia}, Y. and {M{\"o}ller}, T. and {Moscadelli}, L. and {Nucara}, A. and {Rigby}, A.~J. and {Rygl}, K.~L.~J. and {Sanhueza}, P. and {van der Tak}, F. and {Wells}, M.~R.~A. and {Wyrowski}, F. and {De Angelis}, F. and {Liu}, S. and {Ahmadi}, A. and {Bronfman}, L. and {Liu}, S.-Y. and {Su}, Y.-N. and {Tang}, Y. and {Testi}, L. and {Zinnecker}, H.},
        title = "{ALMAGAL: III. Compact source catalog: Fragmentation statistics and physical evolution of the core population}",
      journal = {\aap},
         year = 2025,
        month = apr,
       volume = {696},
          eid = {A151},
        pages = {A151},
          doi = {10.1051/0004-6361/202452706},
archivePrefix = {arXiv},
       eprint = {2503.05663},
 primaryClass = {astro-ph.GA},
       adsurl = {https://ui.adsabs.harvard.edu/abs/2025A&A...696A.151C}
}

@ARTICLE{Liu2017,
       author = {{Liu}, Tie and {Lacy}, John and {Li}, Pak Shing and {Wang}, Ke and {Qin}, Sheng-Li and {Zhang}, Qizhou and {Kim}, Kee-Tae and {Garay}, Guido and {Wu}, Yuefang and {Mardones}, Diego and {Zhu}, Qingfeng and {Tatematsu}, Ken'ichi and {Hirota}, Tomoya and {Ren}, Zhiyuan and {Liu}, Sheng-Yuan and {Chen}, Huei-Ru and {Su}, Yu-Nung and {Li}, Di},
        title = "{ALMA Reveals Sequential High-mass Star Formation in the G9.62+0.19 Complex}",
      journal = {\apj},
         year = 2017,
        month = nov,
       volume = {849},
       number = {1},
          eid = {25},
        pages = {25},
          doi = {10.3847/1538-4357/aa8d73},
archivePrefix = {arXiv},
       eprint = {1705.04907},
 primaryClass = {astro-ph.GA},
       adsurl = {https://ui.adsabs.harvard.edu/abs/2017ApJ...849...25L}
}

@ARTICLE{Beuther2025,
       author = {{Beuther}, H. and {Kuiper}, R. and {Tafalla}, M.},
        title = "{Star Formation from Low to High Mass: A Comparative View}",
      journal = {\araa},
         year = 2025,
        month = aug,
       volume = {63},
       number = {1},
        pages = {1-44},
          doi = {10.1146/annurev-astro-013125-122023},
archivePrefix = {arXiv},
       eprint = {2501.16866},
 primaryClass = {astro-ph.GA},
       adsurl = {https://ui.adsabs.harvard.edu/abs/2025ARA&A..63....1B}
}

@ARTICLE{Molinari2016,
       author = {{Molinari}, S. and {Merello}, M. and {Elia}, D. and {Cesaroni}, R. and {Testi}, L. and {Robitaille}, T.},
        title = "{Calibration of Evolutionary Diagnostics in High-mass Star Formation}",
      journal = {\apjl},
         year = 2016,
        month = jul,
       volume = {826},
       number = {1},
          eid = {L8},
        pages = {L8},
          doi = {10.3847/2041-8205/826/1/L8},
archivePrefix = {arXiv},
       eprint = {1604.06192},
 primaryClass = {astro-ph.GA},
       adsurl = {https://ui.adsabs.harvard.edu/abs/2016ApJ...826L...8M}
}

@ARTICLE{Traficante2020,
       author = {{Traficante}, A. and {Fuller}, G.~A. and {Duarte-Cabral}, A. and {Elia}, D. and {Heyer}, M.~H. and {Molinari}, S. and {Peretto}, N. and {Schisano}, E.},
        title = "{Multiscale dynamics in star-forming regions: the interplay between gravity and turbulence}",
      journal = {\mnras},
         year = 2020,
        month = jan,
       volume = {491},
       number = {3},
        pages = {4310-4324},
          doi = {10.1093/mnras/stz3344},
archivePrefix = {arXiv},
       eprint = {1912.00031},
 primaryClass = {astro-ph.GA},
       adsurl = {https://ui.adsabs.harvard.edu/abs/2020MNRAS.491.4310T}
}

@ARTICLE{Ballesteros-Paredes2006,
       author = {{Ballesteros-Paredes}, Javier},
        title = "{Six myths on the virial theorem for interstellar clouds}",
      journal = {\mnras},
         year = 2006,
        month = oct,
       volume = {372},
       number = {1},
        pages = {443-449},
          doi = {10.1111/j.1365-2966.2006.10880.x},
archivePrefix = {arXiv},
       eprint = {astro-ph/0606071},
 primaryClass = {astro-ph},
       adsurl = {https://ui.adsabs.harvard.edu/abs/2006MNRAS.372..443B}
}

@ARTICLE{Kauffmann2013,
       author = {{Kauffmann}, Jens and {Pillai}, Thushara and {Goldsmith}, Paul F.},
        title = "{Low Virial Parameters in Molecular Clouds: Implications for High-mass Star Formation and Magnetic Fields}",
      journal = {\apj},
         year = 2013,
        month = dec,
       volume = {779},
       number = {2},
          eid = {185},
        pages = {185},
          doi = {10.1088/0004-637X/779/2/185},
archivePrefix = {arXiv},
       eprint = {1308.5679},
 primaryClass = {astro-ph.GA},
       adsurl = {https://ui.adsabs.harvard.edu/abs/2013ApJ...779..185K}
}

@ARTICLE{Traficante2018c,
       author = {{Traficante}, A. and {Lee}, Y.-N. and {Hennebelle}, P. and {Molinari}, S. and {Kauffmann}, J. and {Pillai}, T.},
        title = "{A possible observational bias in the estimation of the virial parameter in virialized clumps}",
      journal = {\aap},
         year = 2018,
        month = nov,
       volume = {619},
          eid = {L7},
        pages = {L7},
          doi = {10.1051/0004-6361/201833513},
archivePrefix = {arXiv},
       eprint = {1811.00561},
 primaryClass = {astro-ph.GA},
       adsurl = {https://ui.adsabs.harvard.edu/abs/2018A&A...619L...7T}
}

@ARTICLE{vanderMarel2013,
       author = {{van der Marel}, N. and {Kristensen}, L.~E. and {Visser}, R. and {Mottram}, J.~C. and {Y{\i}ld{\i}z}, U.~A. and {van Dishoeck}, E.~F.},
        title = "{Outflow forces of low-mass embedded objects in Ophiuchus: a quantitative comparison of analysis methods}",
      journal = {\aap},
         year = 2013,
        month = aug,
       volume = {556},
          eid = {A76},
        pages = {A76},
          doi = {10.1051/0004-6361/201220717},
archivePrefix = {arXiv},
       eprint = {1305.6453},
 primaryClass = {astro-ph.SR},
       adsurl = {https://ui.adsabs.harvard.edu/abs/2013A&A...556A..76V}
}

@ARTICLE{Lamers1995,
       author = {{Lamers}, Henny J.~G.~L.~M. and {Snow}, Theodore P. and {Lindholm}, Douglas M.},
        title = "{Terminal Velocities and the Bistability of Stellar Winds}",
      journal = {\apj},
         year = 1995,
        month = dec,
       volume = {455},
        pages = {269},
          doi = {10.1086/176575},
       adsurl = {https://ui.adsabs.harvard.edu/abs/1995ApJ...455..269L}
}

@ARTICLE{Downes2007,
       author = {{Downes}, T.~P. and {Cabrit}, S.},
        title = "{Jet-driven molecular outflows from class 0 sources: younger and stronger than they seem?}",
      journal = {\aap},
         year = 2007,
        month = sep,
       volume = {471},
       number = {3},
        pages = {873-884},
          doi = {10.1051/0004-6361:20066921},
       adsurl = {https://ui.adsabs.harvard.edu/abs/2007A&A...471..873D}
}

@ARTICLE{Zhang2005,
       author = {{Zhang}, Qizhou and {Hunter}, T.~R. and {Brand}, J. and {Sridharan}, T.~K. and {Cesaroni}, R. and {Molinari}, S. and {Wang}, J. and {Kramer}, M.},
        title = "{Search for CO Outflows toward a Sample of 69 High-Mass Protostellar Candidates. II. Outflow Properties}",
      journal = {\apj},
         year = 2005,
        month = jun,
       volume = {625},
       number = {2},
        pages = {864-882},
          doi = {10.1086/429660},
       adsurl = {https://ui.adsabs.harvard.edu/abs/2005ApJ...625..864Z}
}

@ARTICLE{Ma2013,
       author = {{Ma}, Bo and {Tan}, Jonathan C. and {Barnes}, Peter J.},
        title = "{The Galactic Census of High- and Medium-mass Protostars. II. Luminosities and Evolutionary States of a Complete Sample of Dense Gas Clumps}",
      journal = {\apj},
         year = 2013,
        month = dec,
       volume = {779},
       number = {1},
          eid = {79},
        pages = {79},
          doi = {10.1088/0004-637X/779/1/79},
archivePrefix = {arXiv},
       eprint = {1211.6492},
 primaryClass = {astro-ph.GA},
       adsurl = {https://ui.adsabs.harvard.edu/abs/2013ApJ...779...79M}
}

@ARTICLE{Csengeri2017,
       author = {{Csengeri}, T. and {Bontemps}, S. and {Wyrowski}, F. and {Motte}, F. and {Menten}, K.~M. and {Beuther}, H. and {Bronfman}, L. and {Commer{\c{c}}on}, B. and {Chapillon}, E. and {Duarte-Cabral}, A. and {Fuller}, G.~A. and {Henning}, Th. and {Leurini}, S. and {Longmore}, S. and {Palau}, A. and {Peretto}, N. and {Schuller}, F. and {Tan}, J.~C. and {Testi}, L. and {Traficante}, A. and {Urquhart}, J.~S.},
        title = "{ALMA survey of massive cluster progenitors from ATLASGAL. Limited fragmentation at the early evolutionary stage of massive clumps}",
      journal = {\aap},
         year = 2017,
        month = apr,
       volume = {600},
          eid = {L10},
        pages = {L10},
          doi = {10.1051/0004-6361/201629754},
archivePrefix = {arXiv},
       eprint = {1703.03273},
 primaryClass = {astro-ph.GA},
       adsurl = {https://ui.adsabs.harvard.edu/abs/2017A&A...600L..10C}
}

@ARTICLE{Morii2024,
       author = {{Morii}, Kaho and {Sanhueza}, Patricio and {Zhang}, Qizhou and {Nakamura}, Fumitaka and {Li}, Shanghuo and {Sabatini}, Giovanni and {Olguin}, Fernando A. and {Beuther}, Henrik and {Tafoya}, Daniel and {Izumi}, Natsuko and {Tatematsu}, Ken'ichi and {Sakai}, Takeshi},
        title = "{The ALMA Survey of 70 {\ensuremath{\mu}}m Dark High-mass Clumps in Early Stages (ASHES). XI. Statistical Study of Early Fragmentation}",
      journal = {\apj},
         year = 2024,
        month = may,
       volume = {966},
       number = {2},
          eid = {171},
        pages = {171},
          doi = {10.3847/1538-4357/ad32d0},
archivePrefix = {arXiv},
       eprint = {2403.07058},
 primaryClass = {astro-ph.GA},
       adsurl = {https://ui.adsabs.harvard.edu/abs/2024ApJ...966..171M}
}

@ARTICLE{Xu2024,
       author = {{Xu}, Fengwei and {Wang}, Ke and {Liu}, Tie and {Tang}, Mengyao and {Evans}, II, Neal J. and {Palau}, Aina and {Morii}, Kaho and {He}, Jinhua and {Sanhueza}, Patricio and {Liu}, Hong-Li and {Stutz}, Amelia and {Zhang}, Qizhou and {Chen}, Xi and {Li}, Pak Shing and {G{\'o}mez}, Gilberto C. and {V{\'a}zquez-Semadeni}, Enrique and {Li}, Shanghuo and {Mai}, Xiaofeng and {Lu}, Xing and {Liu}, Meizhu and {Chen}, Li and {Li}, Chuanshou and {Shi}, Hongqiong and {Ren}, Zhiyuan and {Li}, Di and {Garay}, Guido and {Bronfman}, Leonardo and {Dewangan}, Lokesh and {Juvela}, Mika and {Lee}, Chang Won and {Zhang}, S. and {Yue}, Nannan and {Wang}, Chao and {Ge}, Yifei and {Jiao}, Wenyu and {Luo}, Qiuyi and {Zhou}, J.-W. and {Tatematsu}, Ken'ichi and {Chibueze}, James O. and {Su}, Keyun and {Sun}, Shenglan and {Ristorcelli}, I. and {Toth}, L. Viktor},
        title = "{The ALMA Survey of Star Formation and Evolution in Massive Protoclusters with Blue Profiles (ASSEMBLE): Core Growth, Cluster Contraction, and Primordial Mass Segregation}",
      journal = {\apjs},
         year = 2024,
        month = jan,
       volume = {270},
       number = {1},
          eid = {9},
        pages = {9},
          doi = {10.3847/1538-4365/acfee5},
archivePrefix = {arXiv},
       eprint = {2309.14684},
 primaryClass = {astro-ph.GA},
       adsurl = {https://ui.adsabs.harvard.edu/abs/2024ApJS..270....9X}
}

@ARTICLE{Hsieh2023,
       author = {{Hsieh}, Cheng-Han and {Arce}, H{\'e}ctor G. and {Li}, Zhi-Yun and {Dunham}, Michael and {Offner}, Stella and {Stephens}, Ian W. and {Stutz}, Amelia and {Megeath}, Tom and {Kong}, Shuo and {Plunkett}, Adele and {Tobin}, John J. and {Zhang}, Yichen and {Mardones}, Diego and {Pineda}, Jaime E. and {Stanke}, Thomas and {Carpenter}, John},
        title = "{The Evolution of Protostellar Outflow Cavities, Kinematics, and Angular Distribution of Momentum and Energy in Orion A: Evidence for Dynamical Cores}",
      journal = {\apj},
         year = 2023,
        month = apr,
       volume = {947},
       number = {1},
          eid = {25},
        pages = {25},
          doi = {10.3847/1538-4357/acba13},
archivePrefix = {arXiv},
       eprint = {2302.03174},
 primaryClass = {astro-ph.SR},
       adsurl = {https://ui.adsabs.harvard.edu/abs/2023ApJ...947...25H}
}

@ARTICLE{Skretas2022,
       author = {{Skretas}, I.~M. and {Kristensen}, L.~E.},
        title = "{Connecting Galactic and extragalactic outflows: From the Cygnus-X cluster to active galaxies}",
      journal = {\aap},
         year = 2022,
        month = apr,
       volume = {660},
          eid = {A39},
        pages = {A39},
          doi = {10.1051/0004-6361/202141944},
archivePrefix = {arXiv},
       eprint = {2202.04676},
 primaryClass = {astro-ph.GA},
       adsurl = {https://ui.adsabs.harvard.edu/abs/2022A&A...660A..39S}
}

@ARTICLE{Hoque2025,
       author = {{Hoque}, Ariful and {Baug}, Tapas and {Dewangan}, Lokesh K. and {Juvela}, Mika and {Tej}, Anandmayee and {Goldsmith}, Paul F. and {Garc{\'\i}a}, Pablo and {Stutz}, Amelia M. and {Liu}, Tie and {Lee}, Chang Won and {Xu}, Fengwei and {Sanhueza}, Patricio and {Bhadari}, N.~K. and {Tatematsu}, K. and {Liu}, Xunchuan and {Liu}, Hong-Li and {Zhang}, Yong and {Tang}, Xindi and {Garay}, Guido and {Wang}, Ke and {Zhang}, Siju and {T{\'o}th}, L. Viktor and {Nazeer}, Hafiz and {Hwang}, Jihye and {Gorai}, Prasanta and {Bronfman}, Leonardo and {Das}, Swagat Ranjan and {Sinha}, Tirthendu},
        title = "{The ALMA-ATOMS Survey: Exploring Protostellar Outflows in HC$_{3}$N}",
      journal = {\apj},
         year = 2025,
        month = jul,
       volume = {987},
       number = {2},
          eid = {197},
        pages = {197},
          doi = {10.3847/1538-4357/add928},
archivePrefix = {arXiv},
       eprint = {2505.04164},
 primaryClass = {astro-ph.GA},
       adsurl = {https://ui.adsabs.harvard.edu/abs/2025ApJ...987..197H}
}

@ARTICLE{Pineda2020,
       author = {{Pineda}, Jaime E. and {Segura-Cox}, Dominique and {Caselli}, Paola and {Cunningham}, Nichol and {Zhao}, Bo and {Schmiedeke}, Anika and {Maureira}, Mar{\'\i}a Jos{\'e} and {Neri}, Roberto},
        title = "{A protostellar system fed by a streamer of 10,500 au length}",
      journal = {Nature Astronomy},
         year = 2020,
        month = jan,
       volume = {4},
        pages = {1158-1163},
          doi = {10.1038/s41550-020-1150-z},
archivePrefix = {arXiv},
       eprint = {2007.13430},
 primaryClass = {astro-ph.GA},
       adsurl = {https://ui.adsabs.harvard.edu/abs/2020NatAs...4.1158P}
}

@ARTICLE{Garufi2022,
       author = {{Garufi}, A. and {Podio}, L. and {Codella}, C. and {Segura-Cox}, D. and {Vander Donckt}, M. and {Mercimek}, S. and {Bacciotti}, F. and {Fedele}, D. and {Kasper}, M. and {Pineda}, J.~E. and {Humphreys}, E. and {Testi}, L.},
        title = "{ALMA chemical survey of disk-outflow sources in Taurus (ALMA-DOT). VI. Accretion shocks in the disk of DG Tau and HL Tau}",
      journal = {\aap},
         year = 2022,
        month = feb,
       volume = {658},
          eid = {A104},
        pages = {A104},
          doi = {10.1051/0004-6361/202141264},
archivePrefix = {arXiv},
       eprint = {2110.13820},
 primaryClass = {astro-ph.GA},
       adsurl = {https://ui.adsabs.harvard.edu/abs/2022A&A...658A.104G}
}


\appendix
\section{RGB Maps}
This appendix presents the \textit{Spitzer} RGB images of the two peculiar sources discussed in Sect.~\ref{outflow_filament_interactions}. The maps illustrate the spatial relationship between the large-scale infrared filamentary emission and the CO (2--1) molecular outflows identified in this work, providing additional visual context for the discussion in the main text. Similar RGB maps for the remaining SQUALO sources are available in the online supplementary material.
\begin{figure*}
\centering
\includegraphics[width=16cm]{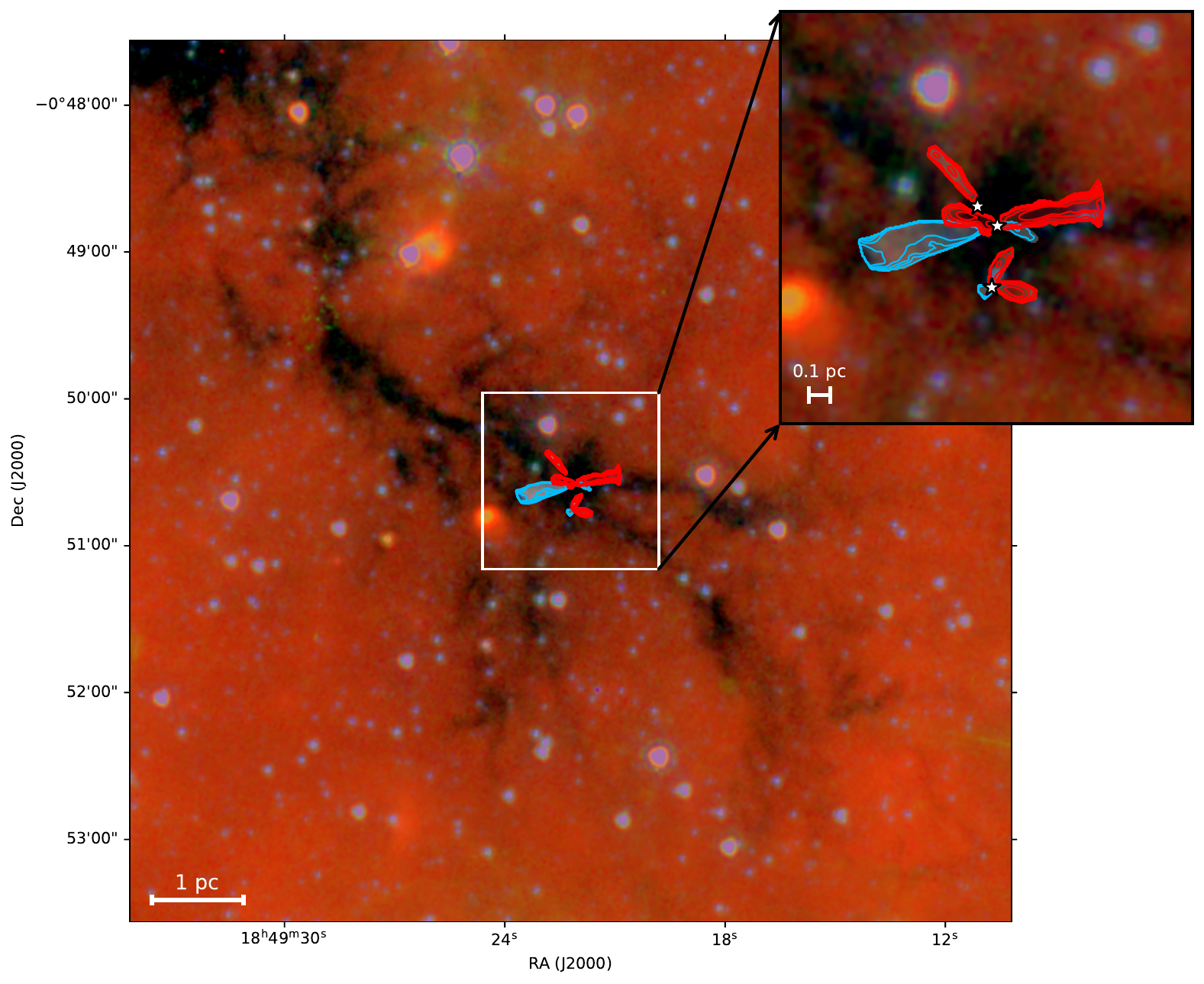}
\caption{The large panel shows a three-color \textit{Spitzer} image constructed from the IRAC 8.0 $\mu$m (red), 5.8 $\mu$m (green), and 4.5 $\mu$m (blue) bands towards HIGALBM31.9462+0.0759. The white box marks the region shown in the zoomed inset. Blue- and red-shifted CO (2--1) outflow emission, integrated within velocity ranges defined for each lobe and masked using polygon apertures enclosing significant emission above a $5\sigma$ threshold, are overlaid as blue and red contours, respectively. The semi-transparent blue and red shaded regions indicate the corresponding integrated intensity distributions. White star symbols mark the positions of compact continuum fragments identified from the high-resolution continuum observations. The inset highlights the immediate vicinity of the protostellar system, revealing multiple outflow lobes and their spatial relationship to the embedded fragments. Black arrows connect the large-scale view to the zoomed region. Scale bars corresponding to 1 pc and 0.1 pc are shown in the lower-left corners of the main panel and inset, respectively.}
\label{Spitzer_ALMA_31}
\end{figure*}
\begin{figure*}
\centering
\includegraphics[width=16cm]{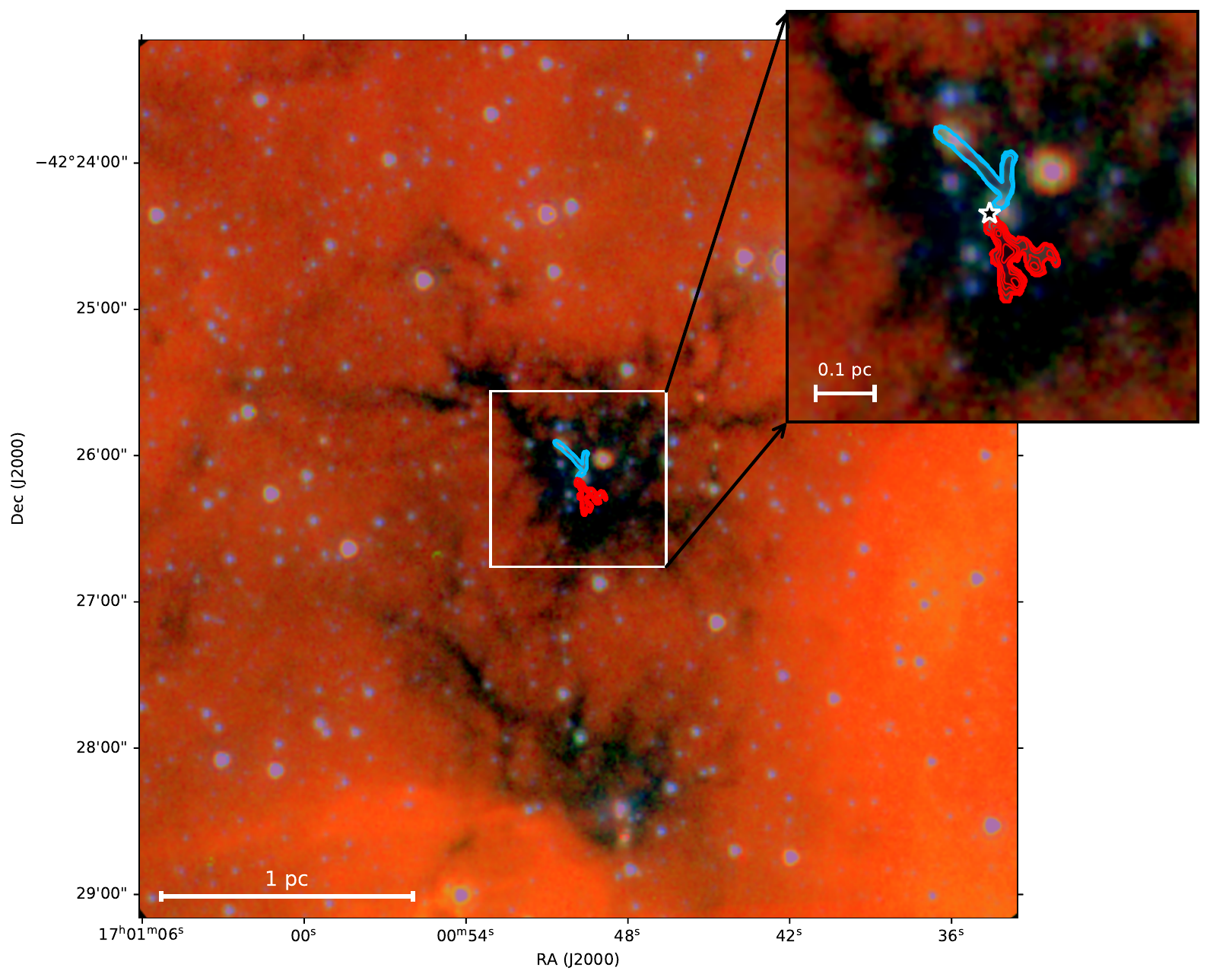}
\caption{The large panel shows a three-color \textit{Spitzer} image constructed from the IRAC 8.0 $\mu$m (red), 5.8 $\mu$m (green), and 4.5 $\mu$m (blue) bands towards HIGALBM343.7560--0.1629. The white box marks the region shown in the zoomed inset. Blue- and red-shifted CO (2--1) outflow emission, integrated within velocity ranges defined for each lobe and masked using polygon apertures enclosing significant emission above a $5\sigma$ threshold, are overlaid as blue and red contours, respectively. The semi-transparent blue and red shaded regions indicate the corresponding integrated intensity distributions. White star symbols mark the positions of compact continuum fragments identified from the high-resolution continuum observations. The inset highlights the immediate vicinity of the protostellar system, revealing multiple outflow lobes and their spatial relationship to the embedded fragments. Black arrows connect the large-scale view to the zoomed region. Scale bars corresponding to 1 pc and 0.1 pc are shown in the lower-left corners of the main panel and inset, respectively.}
\label{Spitzer_ALMA_343}
\end{figure*}

\bsp	
\label{lastpage}
\end{document}